\documentclass[acmsmall]{acmart}

\usepackage{amsmath,amsthm,amsfonts}

\usepackage{float}    

\usepackage{graphicx}
\usepackage{subcaption}
\usepackage{bcprules}

\usepackage{enumerate}
\usepackage{multicol}

\usepackage{booktabs}
\usepackage{etoolbox}
\AtBeginEnvironment{tabular}{\footnotesize}

\usepackage{csquotes}

\usepackage{fancyvrb}

\usepackage{todonotes}

\usepackage{setspace} 

\usepackage{hyperref}
\usepackage[noend,linesnumbered]{algorithm2e}

\theoremstyle{remark} 
\usepackage[inline]{enumitem}
\usepackage{stmaryrd}
\usepackage{url}
\usepackage{textcomp}

\usepackage{mathtools}

\usepackage{xcolor,colortbl}
\definecolor{Gray}{gray}{0.9}
\definecolor{LightCyan}{rgb}{0.88,1,1}

\usepackage{booktabs}

\colorlet{punct}{red!60!black}
\definecolor{background}{HTML}{EEEEEE}
\definecolor{delim}{RGB}{20,105,176}
\colorlet{numb}{magenta!60!black}

\usepackage{tikz}
\usepackage{ulem}
\newif\iflon
\lontrue                        
\iflon
\newcommand{\iflong}[1]{#1}
\newcommand{\ifshort}[1]{}
\else
\newcommand{\iflong}[1]{}
\newcommand{\ifshort}[1]{#1}
\fi

\newtheorem{remark}{Remark}

\newtheorem{example}{Example}

\iflon
\usepackage[appendix=inline]{apxproof} 
\else
\usepackage[appendix=strip,bibliography=common]{apxproof}
\fi

\newtheoremrep{theorem}{Theorem}
\newtheoremrep{lemma}[theorem]{Lemma}
\newtheoremrep{corollary}[theorem]{Corollary}

  {\small \em \tt ccc\begin{table} aaa bbb}%
  {\end{table} ddd}

\newcommand{\json}{JSON}
\newcommand{\kw}[1]{\textbf{#1}}
\renewcommand{\kw}[1]{\ensuremath{\mathtt{#1}}}
\newcommand{\qkw}[1]{\ensuremath{\mathtt{\QQ{#1}\QQ}}}

\renewcommand{\kw}[1]{\akw{#1}}
\renewcommand{\qkw}[1]{\qakw{#1}}
\newcommand{\qkwfn}[1]{\qkw{\footnotesize #1}}
\newcommand{\akwfn}[1]{\akw{\footnotesize #1}}

\newcommand{\key}[1]{\ensuremath{\mathit{#1}}}

\newcommand{\akey}[1]{\ensuremath{\mathsf{#1}}}

\newcommand{\rkw}[1]{\ensuremath{\mathsf{#1}}}
\renewcommand{\rkw}[1]{\ensuremath{\mbox{\sf{\small #1}}}}
\newcommand{\akw}[1]{\ensuremath{\mbox{\tt{\small #1}}}}
\newcommand{\qakw}[1]{{\QQ}\akw{#1}{\QQ}}    

\newcommand{\qnot}{\qkw{not}}
\newcommand{\xtrue}{\kw{true}}

\newcommand{\xfalse}{\kw{false}}

\newcommand{\qnull}{\qkw{null}}

\newcommand{\qone}{\qkw{oneOf}}

\newcommand{\qany}{\qkw{anyOf}}

\newcommand{\qall}{\qkw{allOf}}

\newcommand{\qreq}{\qkw{required}}

\newcommand{\qtype}{\qkw{type}}

\newcommand{\qprops}{\qkw{properties}}

\newcommand{\qpattProps}{\qkw{patternProperties}}

\newcommand{\qthen}{\qkw{then}}

\newcommand{\qif}{\qkw{if}}

\newcommand{\qelse}{\qkw{else}}

\newcommand{\qaddProps}{\qkw{additionalProperties}}

\newcommand{\qunProps}{\qkw{unevaluatedProperties}}

\newcommand{\xuniqIt}{\kw{uniqueItems}}

\newcommand{\qcont}{\qkw{contains}}

\newcommand{\qminC}{\qkw{minContains}}

\newcommand{\qmaxC}{\qkw{maxContains}}

\newcommand{\qits}{\qkw{items}}

\newcommand{\qprefIts}{\qkw{prefixItems}}

\newcommand{\qunIts}{\qkw{unevaluatedItems}}

\newcommand{\qdepS}{\qkw{dependentSchemas}}

\newcommand{\qconst}{\qkw{const}}

\newcommand{\qdref}{\qkw{\$ref}}

\newcommand{\qdrecA}{\qkw{\$recursiveAnchor}}

\newcommand{\qdrRef}{\qkw{\$recursiveRef}}

\newcommand{\qdefault}{\qkw{default}}

\newcommand{\qdefs}{\qkw{definitions}}

\newcommand{\qddefs}{\qkw{\$defs}}

\newcommand{\qtitle}{\qkw{title}}

\newcommand{\qobject}{\qkw{object}}

\newcommand{\qnumber}{\qkw{number}}

\newcommand{\qinteger}{\qkw{integer}}

\newcommand{\qstr}{\qkw{string}}

\newcommand{\qarray}{\qkw{array}}

\newcommand{\qboolean}{\qkw{boolean}}

\newcommand{\qdid}{\qkw{\$id}}

\newcommand{\qdschema}{\qkw{\$schema}}

\newcommand{\qda}{\qkw{\$anchor}}

\newcommand{\qddref}{\qkw{\$dynamicRef}}

\newcommand{\qdda}{\qkw{\$dynamicAnchor}}

\newcommand{\qexamples}{\qkw{examples}}

\newcommand{\eqdef}{\ensuremath{\stackrel{\vartriangle}{=}}}

\newcommand{\gcomment}[1]{}
\newcommand{\oldversion}[1]{}
\newcommand{\hideforspace}[1]{}
\newcommand{\hide}[1]{}
\newcommand{\save}[1]{}
\newcommand{\code}[1]{}

\newcommand{\Iff}{\Leftrightarrow}
\newcommand{\Implies}{\Rightarrow}

\newcommand{\Or}{\vee}

\newcommand{\TypeOf}{\key{TypeOf}}
\renewcommand{\comment}[1]{}

\renewcommand{\And}{\wedge}

\newcommand{\Not}{\neg}

\newcommand{\Num}{\akey{Num}}
\newcommand{\Str}{\akey{Str}}
\newcommand{\Int}{\akey{Int}}

\newcommand{\Min}{\akey{min}}

\newcommand{\keykey}[1]{\key{\underline{#1}}}

\newcommand{\DFour}{Draft-04}

\newcommand{\DSix}{Draft-06}

\newcommand{\DNineteen}{Draft 2019-09}
\newcommand{\DTwenty}{Draft 2020-12}
\newcommand{\JS}{JSON Sche\-ma}
\newcommand{\mJS}{Modern JSON Sche\-ma}
\newcommand{\MJS}{Modern JSON Sche\-ma}
\newcommand{\cJS}{Classical JSON Sche\-ma}

\newcommand{\custcom}[2]{\marginpar{\tiny #1: {#2}}}

\newcommand{\ST}[1]{\custcom{steffi}{#1}}

\newcommand{\M}{\ |\ }

\newlength{\NL}
\newlength{\SaveNL}

\newcommand{\EmptySet}{\emptyset}

\newcommand{\QQ}{\textnormal{\textquotedbl}}
\newcommand{\Set}[1]{\{\,{#1}\,\}}

\newcommand{\SetOpen}{\{\!|}
\newcommand{\SetClose}{|\!\}}
\renewcommand{\Set}[1]{\SetOpen{#1}\SetClose}

\newcommand{\SetST}[2]{\SetOpen{#1}\,\mid\,{#2}\SetClose}
\newcommand{\SetIIn}[3]{\Set{#1}^{{#2}\in{#3}}}
\newcommand{\SetTo}[1]{\Set{1\ldots{#1}}}

\newcommand{\SetFromTo}[2]{\Set{{#1}\ldots{#2}}}
\newcommand{\List}[1]{[\!|\,{#1}\,|\!]}
\newcommand{\pList}[1]{[\!|\,{#1}\,|\!]}

\newcommand{\str}{s}

\newcommand{\J}{{J}}

\newcommand{\semt}{\ensuremath{\mathit{JVal}}}

\newcommand{\rlan}[1]{L(#1)}

\newcommand{\jsonsch}{JSON Schema} 

\newcommand{\AnyOf}{\kw{AnyOf}}

\newcommand{\CCC}{\ensuremath{\mathcal{C}}}
\renewcommand{\SetOpen}{\{\!|\,}
\renewcommand{\SetClose}{\,|\!\}}
\renewcommand{\Set}[1]{\SetOpen{#1}\SetClose}
\renewcommand{\SetST}[2]{\SetOpen{#1}\,\mid\,{#2}\SetClose}

\newcommand{\Get}{\kw{get}}
\newcommand{\GetS}{\kw{gets}}
\newcommand{\GetK}{\kw{getk}}

\newcommand{\DGet}{\kw{dget}}
\newcommand{\DGetK}{\kw{dgetk}}
\newcommand{\DGetS}{\kw{dgets}}
\newcommand{\Load}{\kw{load}}
\newcommand{\StackDash}[1]{\stackrel{#1}{\vdash}}
\renewcommand{\StackDash}[1]{\mathrel{\vdash\raisebox{.90ex}{\tt\kern-0.5em {\tiny {#1}\kern0.5em}}}}

\newcommand{\Judg}[4]{{#2}\StackDash{#1}{#3}\ ? \ {#4}}
\renewcommand{\Judg}[4]{{#2}\StackDash{#1}{#3} : {#4}}
\newcommand{\JudgC}[3]{\Judg{#1}{C}{#2}{#3}}
\newcommand{\JudgCJ}[2]{\JudgC{#1}{J}{#2}}
\newcommand{\Ret}[1]{\,\,\stackrel{#1}{\rightarrow}\,\,}
\newcommand{\RetL}[1]{\,\,{\rightarrow}\,\,}
\renewcommand{\RetL}[1]{\,\,\stackrel{#1}{\rightarrow}\,\,}
\newcommand{\Kl}{\vec{K}}

\newcommand{\KJudg}[3]{\Judg{K}{#1}{#2}{#3}}

\newcommand{\KJudgCJ}[1]{\JudgCJ{K}{#1}}

\newcommand{\KLJudg}[3]{\Judg{L}{#1}{#2}{#3}}

\newcommand{\KLJudgCJ}[1]{\JudgCJ{L}{#1}}

\newcommand{\SJudg}[3]{\Judg{S}{#1}{#2}{#3}}
\newcommand{\SJudgC}[2]{\JudgC{S}{#1}{#2}}
\newcommand{\SJudgCJ}[1]{\JudgCJ{S}{#1}}

\newcommand{\SdJudg}[3]{\Judg{Sd}{#1}{#2}{#3}}
\newcommand{\KdJudg}[3]{\Judg{Kd}{#1}{#2}{#3}}
\newcommand{\KLdJudg}[3]{\Judg{Ld}{#1}{#2}{#3}}

\newcommand{\pk}{\ensuremath{\kappa}}
\newcommand{\ps}{\ensuremath{\sigma}}
\newcommand{\pkl}{\ensuremath{\vec{\pk}}}

\newcommand{\rl}{\vec{r}}

\newcommand{\anot}{\qakw{not}}
\newcommand{\atrue}{\akw{true}}
\newcommand{\afalse}{\akw{false}}
\newcommand{\anull}{\akw{null}}
\newcommand{\aone}{\qakw{oneOf}}
\newcommand{\aany}{\qakw{anyOf}}
\newcommand{\aall}{\qakw{allOf}}
\newcommand{\amin}{\qakw{minimum}}
\newcommand{\amax}{\qakw{maximum}}
\newcommand{\areq}{\qakw{required}}

\newcommand{\atype}{\qakw{type}}
\newcommand{\aexmin}{\qakw{exclusiveMinimum}}
\newcommand{\aexmax}{\qakw{exclusiveMaximum}}
\newcommand{\aprops}{\qakw{properties}}
\newcommand{\apropN}{\qakw{propertyNames}}
\newcommand{\apattProps}{\qakw{patternProperties}}
\newcommand{\aminP}{\qakw{minProperties}}
\newcommand{\amaxP}{\qakw{maxProperties}}
\newcommand{\athen}{\qakw{then}}
\newcommand{\aif}{\qakw{if}}
\newcommand{\aelse}{\qakw{else}}

\newcommand{\aaddProps}{\qakw{additionalProperties}}

\newcommand{\aunProps}{\qakw{unevaluatedProperties}}
\newcommand{\aunIts}{\qakw{unevaluatedItems}}

\newcommand{\amof}{\qakw{multipleOf}}

\newcommand{\amaxL}{\qakw{maxLength}}
\newcommand{\aminL}{\qakw{minLength}}
\newcommand{\apatt}{\qakw{pattern}}
\newcommand{\auniqIt}{\qakw{uniqueItems}}
\newcommand{\acont}{\qakw{contains}}

\newcommand{\aminC}{\qakw{minContains}}
\newcommand{\amaxC}{\qakw{maxContains}}
\newcommand{\aminIt}{\qakw{minItems}}
\newcommand{\amaxIt}{\qakw{maxItems}}
\newcommand{\ait}{\qakw{items}}

\newcommand{\aits}{\qakw{items}}
\newcommand{\aprefIts}{\qakw{prefixItems}}

\newcommand{\adepS}{\qakw{dependentSchemas}}

\newcommand{\adepR}{\qakw{dependentRequired}}

\newcommand{\aenum}{\qakw{enum}}
\newcommand{\aconst}{\qakw{const}}
\newcommand{\adref}{\qakw{\$ref}}

\newcommand{\adcomm}{\qakw{\$comment}}
\newcommand{\adescr}{\qakw{description}}
\newcommand{\adefault}{\qakw{default}}
\newcommand{\adefs}{\qakw{definitions}}
\newcommand{\addefs}{\qakw{\$defs}}
\newcommand{\atitle}{\qakw{title}}
\newcommand{\aname}{\qakw{name}}
\newcommand{\atags}{\qakw{tags}}
\newcommand{\acomment}{\qakw{comment}}
\newcommand{\aformat}{\qakw{format}}

\newcommand{\aid}{\qakw{id}}
\newcommand{\adid}{\qakw{\$id}}
\newcommand{\adschema}{\qakw{\$schema}}
\newcommand{\ada}{\qakw{\$anchor}}
\newcommand{\addref}{\qakw{\$dynamicRef}}
\newcommand{\adda}{\qakw{\$dynamicAnchor}}
\newcommand{\advoc}{\qakw{\$vocabulary}}

\newcommand{\adepr}{\qakw{deprecated}}
\newcommand{\areadOnly}{\qakw{readOnly}}
\newcommand{\awriteOnly}{\qakw{writeOnly}}
\newcommand{\aexamples}{\qakw{examples}}

\newcommand{\rnot}{\rkw{not}}
\newcommand{\rtrueS}{\rkw{true}}
\newcommand{\rfalseS}{\rkw{false}}

\newcommand{\rone}{\rkw{oneOf}}
\newcommand{\rany}{\rkw{anyOf}}
\newcommand{\rall}{\rkw{allOf}}
\newcommand{\rmin}{\rkw{minimum}}

\newcommand{\rtype}{\rkw{type}}

\newcommand{\rprops}{\rkw{properties}}
\newcommand{\rpropN}{\rkw{propertyNames}}
\newcommand{\rpattProps}{\rkw{patternProperties}}

\newcommand{\rthen}{\rkw{then}}
\newcommand{\rif}{\rkw{if}}
\newcommand{\relse}{\rkw{else}}

\newcommand{\raddProps}{\rkw{additionalProperties}}

\newcommand{\runProps}{\rkw{unevaluatedProperties}}
\newcommand{\runIts}{\rkw{unevaluatedItems}}

\newcommand{\rmof}{\rkw{multipleOf}}

\newcommand{\rcont}{\rkw{contains}}

\newcommand{\rits}{\rkw{items}}
\newcommand{\rprefIts}{\rkw{prefixItems}}
\newcommand{\rdepS}{\rkw{dependentSchemas}}

\newcommand{\rdepR}{\rkw{dependentRequired}}

\newcommand{\rdref}{\rkw{\$ref}}

\newcommand{\robject}{\rkw{object}}
\newcommand{\rnumber}{\rkw{number}}
\newcommand{\rstr}{\rkw{string}}
\newcommand{\rarray}{\rkw{array}}

\newcommand{\rddref}{\rkw{\$dynamicRef}}

\newcommand{\rschema}{\rkw{schema}}
\newcommand{\rklist}{\rkw{klist}}
\newcommand{\JObjOpen}{\{}
\newcommand{\JObjClose}{\}}
\newcommand{\JObj}[1]{\JObjOpen\,{#1}\,\JObjClose}
\newcommand{\JArrOpen}{[}
\newcommand{\JArrClose}{]}
\newcommand{\JArr}[1]{\JArrOpen\,{#1}\,\JArrClose}

\newcommand{\plus}{+}

\newcommand{\TTriv}{\rkw{Triv}}

\newcommand{\cat}{\!\cdot\!}
\newcommand{\lcat}{\cdot}
\newcommand{\catHash}{\cat\qkw{\#}\cat}
\newcommand{\lcatHash}{\lcat\qkw{\#}\lcat}

\newcommand{\RAsA}[1]{R\mbox{\ as\ }{A}}

\newcommand{\fstURI}{\kw{fstURI}}

\newcommand{\JE}{json-everything}
\newcommand{\JSD}{jschon.dev}
\newcommand{\BL}{Jim Blackler's JSON tools}
\newcommand{\HJ}{Hyperjump}

\newcommand{\bshort}{\noindent\begin{minipage}{0.5\textwidth}}
\newcommand{\eshort}{ \end{minipage}}

\newcommand{\shortrulefirst}[3]{\hspace*{-2.3em}\bshort \infrule[{#1}]{#2}{#3} \eshort}\newcommand{\shortrule}[3]{\hfill\bshort \infrule[{#1}]{#2}{#3} \eshort}

\newcommand{\shortaxfirst}[2]{\hspace*{-2.3em}\bshort \infax[{#1}]{#2} \eshort}
\newcommand{\shortax}[2]{\hfill\bshort \infax[{#1}]{#2} \eshort}

\newcommand{\kfailuretol}{failure-tolerating}
\newcommand{\successonly}{success-only}

\newcommand{\nullannotation}{null-annotation}

\newcommand{\evaluated}{\emph{evaluated}}

\newcommand{\evaluates}{\emph{evaluates}}

\newcommand{\webref}[2]{\href{#1}{#2}}
\newcommand{\webcite}[2]{\webref{#1}{\cite{#2}}}

\newcommand{\ES}{\EmptySet}

\newcommand{\setmax}{\key{max}}

\newcommand{\Eqd}[3]{{#1}\sim_{#3}{#2}}
\newcommand{\Eqdd}[2]{\Eqd{#1}{#2}{d}}
\newcommand{\Neqd}[3]{{#1}\not\sim_{#3}{#2}}

\newcommand{\UU}{\key{absURI}}
\newcommand{\sUU}{\key{sURI}}

\newcommand{\Bb}{(}
\newcommand{\Mm}{\ \ }
\newcommand{\Cc}{,)^?}

\renewcommand{\Bb}{}
\renewcommand{\Mm}{\ \M\ }
\renewcommand{\Cc}{}

\renewcommand{\ps}{\pk}
\renewcommand{\pkl}{\pk}
\renewcommand{\Ret}[1]{\rightarrow}

\newcommand{\CC}{C}

\newcommand{\btrue}{\mathcal{T}}
\newcommand{\bfalse}{\mathcal{F}}

\newcommand{\Alt}{\rkw{alt}}

\newcommand{\Index}{\key{Index}}
\newcommand{\AfterQ}{\key{AfterQ}}
\newcommand{\utfx}{\key{TOrFX}}
\newcommand{\Valid}{\key{Valid}}

\newcommand{\Encode}[1]{\underline{#1}}
\newcommand{\ECC}{\Encode{C}}
\newcommand{\TS}{\key{TS}}
\newcommand{\StS}{{\tt CI}}
\newcommand{\StG}{{\tt Static}}
\newcommand{\StK}{\StS}

\newcommand{\Class}[1]{\ensuremath{[{#1}]_{\sim}}}

\newcommand{\citeoneone}{\webcite{https://github.com/json-schema-org/json-schema-spec/issues/1172}{discussion1172}}
\newcommand{\citefiveseven}{\webcite{https://github.com/orgs/json-schema-org/discussions/57}{discussion57}}

\usepackage[skins]{tcolorbox}
\usepackage{listings}
\usepackage[framemethod=tikz]{mdframed}

\makeatletter
\newcommand\querysize{\@setfontsize\querysize\@vipt\@viipt}
\makeatother
\definecolor{mygray}{rgb}{0.643,0.643,0.643}
\lstdefinestyle{querynonumbers}{
    language=JSON,
    stepnumber=1,
    numbersep=1pt, 
    xleftmargin =-2pt,
    tabsize=4,
    showspaces=false,
    showstringspaces=false,
    basicstyle=\linespread{1}\fontfamily{lmtt}\selectfont\querysize,
    keywordstyle=\color{blue},
    stringstyle=\color{purple},
    upquote=true,
    breaklines=true,
    commentstyle=\color{CadetBlue},
}
\lstdefinestyle{oneliner}{
    language=JSON,
    aboveskip=0pt,
    belowskip=0pt,
    basicstyle=\linespread{1}\fontfamily{lmtt}\selectfont\scriptsize,
}
\lstdefinestyle{query}{
    language=JSON,
    numbers=left,
    stepnumber=1,
    numbersep=2pt, 
    numberstyle=\color{black!65},
    xleftmargin =-2pt,
    tabsize=4,
    showspaces=false,
    showstringspaces=false,
    basicstyle=\linespread{1}\fontfamily{lmtt}\selectfont\querysize,
    keywordstyle=\color{blue},
    stringstyle=\color{purple},
    upquote=true,
    breaklines=true,
    commentstyle=\color{CadetBlue}
}
\newtcolorbox{querybox}[2][]{%
    sidebyside align=top,
    enhanced,
    boxsep=0pt,
    arc=0pt,
    top=-3pt, bottom=-3pt,
    left=12pt, right=0pt,
    colback=background,
    colframe=gray!90,
    boxrule=0.5pt,
    leftrule=1pt,
    overlay unbroken and first ={%
    \node[rotate=90,
          minimum width=0.5cm,
          anchor=south,
          font=\scriptsize\rmfamily,
          yshift=-13.7pt,
          white]
    at (frame.west) {#2};
    }
}

\lstdefinelanguage{json}{
    literate=
      {:}{{{\color{punct}{:}}}}{1}
      {,}{{{\color{punct}{,}}}}{1}
      {\{}{{{\color{delim}{\{}}}}{1}
      {\}}{{{\color{delim}{\}}}}}{1}
      {[}{{{\color{delim}{[}}}}{1}
      {]}{{{\color{delim}{]}}}}{1},
}

\begin{document}

\title{Validation of Modern JSON Schema: Formalization and Complexity}

\author{Lyes Attouche}
\orcid{}
\affiliation{%
  \institution{Université Paris-Dauphine - PSL}
  \city{Paris}
  \country{France}
}
\email{lyes.attouche@dauphine.fr}

\author{Mohamed-Amine Baazizi}
\orcid{}
\affiliation{%
  \institution{Sorbonne University}
  \city{Paris}
  \country{France}
}
\email{baazizi@ia.lip6.fr}

\author{Dario Colazzo}
\orcid{}
\affiliation{%
  \institution{Université Paris-Dauphine - PSL}
  \city{Paris}
  \country{France}
}
\email{dario.colazzo@dauphine.fr}

\author{Giorgio Ghelli}
\orcid{}
\affiliation{%
  \institution{University of Pisa}
  \city{Pisa}
  \country{Italy}
}
\email{ghelli@di.unipi.it}

\author{Carlo Sartiani}
\orcid{}
\affiliation{%
  \institution{University of Basilicata}
  \city{Potenza}
  \country{Italy}
}
\email{carlo.sartiani@unibas.it}

\author{Stefanie Scherzinger}
\orcid{}
\affiliation{%
  \institution{University of Passau}
  \city{Passau}
  \country{Germany}
}
\email{stefanie.scherzinger@uni-passau.de}

\renewcommand{\shortauthors}{L. Attouche, M. Baazizi, D. Colazzo, G. Ghelli, C. Sartiani, S. Scherzinger}

\begin{abstract}
{\jsonsch}  is the de-facto standard schema language for {\json} data.
The language went through many minor revisions, but the most recent versions of the language, starting from {\DNineteen}, added two novel features,  \emph{dynamic references} and \emph{annotation-dependent validation}, that change the evaluation model. {\emph{\mJS}} is the name used to indicate all versions from {\DNineteen}, which are characterized by these new features, while {\emph{\cJS}} is used to indicate the previous versions.

These new ``modern'' features make the schema language quite difficult to understand and have generated many discussions about the correct interpretation of their official specifications; for this reason, we undertook the task of their formalization.
During this process, we also analyzed the complexity of data validation in {\mJS}, with the idea of confirming the polynomial complexity of {\cJS} validation, and we were surprised to discover a completely different truth: data validation, which is expected to be an extremely efficient process, acquires, with {\mJS} features, a PSPACE complexity.

In this paper, we give the first formal description of {\mJS}, which we have discussed with the community of {\jsonsch} tool developers,
and which we consider a central contribution of this work.
We then prove that its data validation problem is PSPACE-complete.
We prove that the origin of the problem lies in the {\DTwenty} version of dynamic references, and not in annotation-dependent validation. 
We study the schema and data complexities, showing that the problem is PSPACE-complete with respect to the schema size even with a fixed instance
but is in P when the schema is fixed and only the instance size is allowed to vary.
Finally, we run experiments that show that there are families of schemas where the difference in asymptotic complexity between dynamic
and static references is extremely visible, even with small schemas. 

\save{
{\jsonsch}  is the de-facto standard schema language for {\json} data. {\jsonsch} has been already studied and formalized, but the
most recent versions of the language, starting from {\DNineteen}, added two novel features, \emph{dynamic references} and \emph{annotation-dependent validation}, which drastically change the evaluation 
model. As a consequence, the theory previously developed for {\jsonsch} is no longer valid.
{\emph{\mJS}} is the name used to indicate all versions from {\DNineteen}, which are characterized by these new features, while {\emph{\cJS}} is used to indicate
the previous versions.

In this paper, we study the impact of the features that define {\mJS}.
We give the first formal description of the language, study the complexity of validation, and prove that, with the addition of dynamic references, this problem moves from P-complete to PSPACE-complete.
We also show that the problem is in P in the special case when there is a fixed bound on the number of distinct dynamic references in each schema. 
We study the schema and data complexities, showing that the problem is PSPACE-complete with respect to the schema size, but is in P when the schema is fixed and only the instance size is allowed to vary.
Finally, we run experiments that show that there are families of schemas where the difference in asymptotic complexity between dynamic
and static references is extremely visible, even with small schemas. 
}

\end{abstract}

\setcounter{tocdepth}{1} 

\ccsdesc[500]{Theory of computation~Type theory}

\keywords{JSON Schema, complexity of validation}

\maketitle

\section{Introduction}\label{sec:intro}


{\jsonsch}~\citep{jsonschema} is the de-facto standard schema language for {\json} data.
It is based on the combination of structural operators, which describe base values,
objects, and arrays, through logical operators such as disjunction, conjunction, negation, 
and recursive references.

JSON Schema passed through many versions, the most important being 
{\DFour} \citep{Draft04}, 
{\DSix}  \citep{Draft06}, which introduced extensions without changing the validation model,  
{\DNineteen}  \citep{Version09},  and {\DTwenty} \citep{specs2020}. 

The evaluation model of {\DFour} and {\DSix} is quite easy to understand and to formalize, 
and it has been studied in~\citep{DBLP:conf/www/PezoaRSUV16}, \citep{DBLP:conf/pods/BourhisRSV17,DBLP:journals/is/BourhisRV20},
\citep{DBLP:journals/pvldb/AttoucheBCGSS22}, yielding many interesting complexity results.
However, {\DNineteen} introduces two important novelties to the evaluation model: annotation-dependent validation, and dynamic recursive references, which have been generalized as general-purpose dynamic references in {\DTwenty}.
According to the terminology introduced by Henry Andrews in  
\webcite{https://modern-json-schema.com/}{modern}, because of these modifications to the evaluation model,
{\DNineteen} is the first 
Draft that defines \emph{\mJS}, while the previous Drafts define variations of \emph{{\cJS}}.
These novelties are motivated by application needs, but none of them is faithfully represented by the abstract models that had been developed and
studied for {\cJS}. Further, both novelties need formal clarification and specification, as documented by many online discussions, such as {\citeoneone} and {\citefiveseven}.
These discussions involve main actors behind the JSON Schema design and show that the informal JSON Schema specification leads to several, different, yet reasonable interpretations of the semantics of the new operators in {\mJS}.

\paragraph*{Our Contribution}\label{subsec:ourcontrib}

{\DTwenty} \citep{specs2020} is the version of {\mJS} that we study in this paper. We provide the following contributions.\footnote{Unless otherwise noticed, all proofs are in the full version \cite{attouche2023validation-arXiv}.}

\newcommand{\FV}{full version \cite{attouche2023validation-arXiv}}
\renewcommand{\FV}{full version}

\begin{description}

\item[i)]  We provide a formalization of {\mJS} through a set of rules that take into account both annotation-dependent validation and dynamic references (Sections \ref{sec:syntax} and  \ref{sec:validation});  to the best of our knowledge this is the first formal specification and study of {\mJS}.
In addition, we implemented, in Scala, a {\mJS} data validator by a direct translation of our formal system and used it to verify
the correctness of our formal system with respect to the JSON Schema standard test suite (Section~\ref{sec:experiments}).


\item[ii)] We analyze the complexity of validating a JSON instance against a schema and show that, surprisingly enough, when
the general-purpose dynamic references of {\DTwenty} come into play, the validation problem becomes PSPACE-hard (Section~\ref{sec:hardness}); validation was known to be P-complete for {\cJS} \citep{DBLP:conf/www/PezoaRSUV16,DBLP:conf/pods/BourhisRSV17}. We also prove that the bound is strict and, hence, the problem is PSPACE-complete (Section \ref{sec:upperbound}).

\item[iii)] We show that annotation-dependent validation alone, on the contrary, does not change the P complexity of validation, by providing an explicit algorithm for {\mJS} that runs in polynomial time on 
any family of schemas
   where the number of dynamic references is bounded by a constant (Section \ref{sec:ptime}). We also show that the validation for {\DNineteen} was still in P, because of the restrictions that it imposed on dynamic references.
      We study the \emph{data complexity} of validation and prove that, when fixing a schema $S$, validation remains polynomial even in the presence of dynamic references.  

\item[iv)] We provide a technique for substituting dynamic references
with static references, at the price of an exponential increase of the schema size (Section \ref{sec:elimination}).

\item[v)] We run experiments on a rich set of validators that show that there are families of schemas where the distinction between dynamic and static references is clearly visible in the experimental results; the experiments also show that many established validators exhibit an exponential behavior also on the P fragment of JSON Schema (Section~\ref{sec:experiments}).
\end{description}

This paper is about \emph{Formalization} and \emph{Complexity}.
We believe that the \emph{formalization} that we provide answers a need that has been clearly expressed in many public discussions, and we will use our connections
with the JSON Schema community in order to disseminate our results.

Our \emph{complexity} results, on the other hand, belong, for the moment, to the class of ``foundational'' results, that is, theoretical results that shed light on a phenomenon whose practical relevance \emph{may} manifest in the future.
Concretely, this is the first time that we encounter a construct of a data-validation language whose complexity is not in P (unless P=PSPACE).
As a consequence, from now on, we know that the addition of a dynamic re-binding mechanism to a data validation language may
have an unexpected impact on its computational complexity.

\section{{\mJS} through examples}\label{sec:mjsintro}

\subsection{An Example of {\mJS}}

A JSON Schema schema 
is a formal specification of a validation process that can be applied to a JSON value called ``the
instance''. A schema, such as the one in Figure \ref{fig1}, can contain nested schemas.
A schema is either $\atrue$, $\afalse$, or it is an object whose fields, such as $\qtype: \qarray$, are called \emph{keywords}.
Two keywords with the same parent object, such as {\qdid} : \qkw{https:.../schema} and {\qtype} : {\qobject} in Figure \ref{fig1}, are said to be \emph{adjacent}, so that keywords are \emph{adjacent} when they are siblings in the parse tree of the schema. 

\begin{figure}[htb!]
\begin{querybox}{}
\begin{lstlisting}[style=query,escapechar=Z]
{ "$schema": "https://json-schema.org/draft/2020-12/schema",
  "$id": "http://mjs.ex/st",
  "$anchor": "tree",
  "type": "object",
  "properties": {
     "data": true,
     "children": { "type": "array", "items": { "$ref": "http://mjs.ex/st#tree"}}
  },
  "examples": [
     { "data": 3, "children": [ { "data": null, "children": [] }, { "data": " ", "children": []}]},
     { "children": [ { "data": null }, { "children": [ { } ] } ] },
     { "daat": 3, "hcilreden": true }]
}
\end{lstlisting}
\end{querybox}
\caption{A schema representing trees.}
\label{fig1}
\end{figure}

Looking at Figure \ref{fig1}, the $\qdschema$ keyword specifies that this schema is based on {\DTwenty}.

Subschemas of a JSON schema can be identified through a URI with structure $\key{baseURI}\catHash\key{fragmentId}$
(we use $s_1\cat s_2$ for string concatenation);
the $\qdid$ keyword assigns a base URI to its parent schema, and the $\qda$ keyword assigns a fragment identifier to
its parent schema, so that, in this case, the fragment named \qkw{tree} by $\qda$ is the entire schema. Hence, this schema can be referred to as either
$\qkw{http://mjs.ex/st}$ or as $\qkw{http://mjs.ex/st\#tree}$.

The {\qtype} keyword specifies that this schema only validates objects, while it fails on 
arrays and base values.
The {\qprops} keyword specifies that, if the instance under validation contains a \qkw{data} property, then its value has no constraint 
($\qkw{data}: \atrue$),
and, if it contains a \qkw{children} property, then its value must satisfy the nested subschema of line 7: it must
be an array whose elements (if any) must all satisfy 
$\qdref: \qkw{http://mjs.ex/st\#tree}$.
{\qdref} is a reference operator that invokes a local or remote subschema, which, in this case, is the entire current 
schema.\footnote{%
By the standard rules of URI reference resolution \citep{RFC3986}, the URI \qkw{\footnotesize http://mjs.ex/st\#tree} 
could be substituted by the local reference $\qkw{\footnotesize \$ref}: \qkw{\footnotesize\#tree}$, since \qkw{\footnotesize http://mjs.ex/st} is the 
base URI of this schema.}

This schema is satisfied by all the instances that are listed in the $\qexamples$ array.\footnote{JSON Schema validation
will just ignore all non-validation keywords such as $\qkwfn{examples}$.} The second example shows that no field is mandatory
--- fields can be made mandatory by using the keyword {\qreq}.
The third example shows that fields that are different from  \qkw{data} and  \qkw{children} are just ignored. 
To sum up, the only JSON instances where this schema fails are those that associate \qkw{children} to a value that is not an array,
such as $\{ \qkw{data} : 3, \qkw{children} : \qkw{aaa} \}$.

\paragraph{Introducing dynamic references.}
Dynamic references have been added in {\mJS} as an extension mechanism, allowing one to first define a base form of a data structure
and then to refine it, very much in the spirit of ``self'' refinement in object-oriented languages.
To this aim, the basic data structure is named using $\qdda$ and is referred using the $\qddref$ keyword,
as in Figure \ref{fig2}, lines~3 and~7.
This combination  indicates that 
$\qddref : \qkw{\underline{http://mjs.ex/simple-tree}\#tree}$ is \emph{dynamic},
which means that, when this schema is accessed through a different context, as exemplified later,
this dynamic reference may refer to a fragment that is still named \qkw{tree}, but which is defined in a schema that is not 
\qkw{\underline{http://mjs.ex/simple-tree}}. We underline the absolute URI part of the dynamic reference to remind the
reader of the fact that this URI is only used if it is not redefined in the ``dynamic context'', as we will explain later.

\begin{figure}[htb!]
\begin{querybox}{}
\begin{lstlisting}[style=query,escapechar=Z]
{ "$schema": "https://json-schema.org/draft/2020-12/schema",
  "$id": "http://mjs.ex/simple-tree",
  "$dynamicAnchor": "tree",
  "type": "object",
  "properties": {
      "data": true,
      "children": {  "type": "array", "items": { "$dynamicRef": "Z\underline{http://mjs.ex/simple-tree}Z#tree" }}}
}
\end{lstlisting}
\end{querybox}
\caption{A schema representing extensible trees.}\label{fig2}
\end{figure}

The contextual redefinition mechanism is illustrated in the schema shown in Figure \ref{fig3}.
The schema \qkw{http://mjs.ex/strict-tree} redefines the dynamic anchor
\qkw{tree} (line 3), so that it now indicates a conjunction between  $\qdref: \qkw{http://mjs.ex/simple-tree\#tree}$ 
(line 4) and the keyword
$\qunProps: \afalse$ (line 5), which forbids the presence of any property whose name does not match those listed in \qkw{http://mjs.ex/simple-tree\#tree}.
If one applies this schema, it will invoke 
$\qdref: \qkw{http://mjs.ex/simple-tree\#tree}$ (Figure \ref{fig3} - line 4), which will execute the schema of
Figure \ref{fig2} in a ``dynamic scope'' 
where 
$\qkw{http://mjs.ex/strict-tree}$ has redefined the meaning of 
$\qddref: \qkw{\underline{http://mjs.ex/simple-tree}\#tree}$.
In detail, in {\DTwenty}, the ``outermost'' (or ``first'') schema that contains $\qdda: \key{fragmentName}$ is the one that 
fixes the meaning 
of that anchor for any other 
schema $S'$ that will invoke $\qddref : \key{\underline{absURI}}\catHash\key{fragmentName}$ later, independently from
the absolute URI \key{\underline{absURI}} used by $S'$, hence, in this case, 
the meaning of $\qkw{\underline{http://mjs.ex/simple-tree}\#tree}$ in Figure  \ref{fig2} is fixed to be
$\qkw{http://mjs.ex/strict-tree\#tree}$ 
because the schema of Figure \ref{fig3} is invoked before
the one in Figure  \ref{fig2}. 
Observe that the meaning of the static reference
$\qdref : \qkw{http://mjs.ex/simple-tree\#tree}$ to the dynamic anchor \qkw{\#tree} (Figure \ref{fig3} line 4) is fixed
and does not depend on the dynamic context.

\begin{figure}[htb!]
\begin{querybox}{}
\begin{lstlisting}[style=query,escapechar=Z]
{ "$schema": "https://json-schema.org/draft/2020-12/schema",
  "$id": "http://mjs.ex/strict-tree",
  "$dynamicAnchor": "tree",
  "$ref": "http://mjs.ex/simple-tree#tree",
  "unevaluatedProperties": false,
  "non-examples" : [
     { "dat": 3 },
     { "data": 3, "chldrn": [] },
     { "children": [ { "data": null }, { "chldrn": []}]}]
}
\end{lstlisting}
\end{querybox}
\caption{A schema that refines  trees.}\label{fig3}
\end{figure}

As another example, in Figure \ref{fig:schema}, we have a simplified version of the metaschema of {\JS}.

\begin{figure}[htb!]
\begin{querybox}{}
\begin{lstlisting}[style=query,escapechar=Z]
{  "$id": "https://json-schema.org/draft/2020-12/meta/applicator",
   "$dynamicAnchor": "meta",
   "properties": {
        "properties": {
            "type": "object",
            "patternProperties": {
                  ".*": { "$dynamicRef": "Z\underline{https://json-schema.org/draft/2020-12/meta/applicator}Z#meta" } 
            }
        }
    }
}

{   "$id": "https://json-schema.org/draft/2020-12/meta/uneval",
    "$dynamicAnchor": "meta",
    "properties": {
        "unevaluatedProperties": { "$dynamicRef": "Z\underline{https://json-schema.org/draft/2020-12/meta/uneval}Z#meta" }
    }
}

{   "$id": "https://json-schema.org/draft/2020-12/meta/core",
    "$dynamicAnchor": "meta",
    "type" : [ "boolean", "object" ],
    "properties": {
        "$id": { "type": "string"  },
        "$dynamicAnchor": { "type": "string"  },
        "$title": { "type": "string"  }
    }
}

{   "$id": "https://json-schema.org/draft/2020-12/schema",
    "$dynamicAnchor": "meta",
    "allOf": [ {"$ref": "https://json-schema.org/draft/2020-12/meta/applicator"},
               {"$ref": "https://json-schema.org/draft/2020-12/meta/uneval"},
               {"$ref": "https://json-schema.org/draft/2020-12/meta/core"} ]  
}

{   "$id": "https://json-schema.org/draft/2020-12/mini-schema",
    "$dynamicAnchor": "meta",
    "allOf": [ {"$ref": "https://json-schema.org/draft/2020-12/meta/applicator"},
               {"$ref": "https://json-schema.org/draft/2020-12/meta/core"} ]
}
\end{lstlisting}
\end{querybox}
\caption{{\JS} metaschema}\label{fig:schema}
\end{figure}

The metaschema \qkw{https://json-schema.org/draft/2020-12/meta/applicator} has a dynamic anchor (i.e., name) \qkw{meta}, and specifies that 
the value of a $\qprops$ keyword is an object $J_p$, where every field
of $J_p$ has an arbitrary name (like \qkw{data} or \qkw{children} in Figure \ref{fig2} line 6), and the value of every 
field of $J_p$ (such as $\atrue$ and $\JObj{\qtype: \qarray,\ldots}$ in Figure \ref{fig2} line 7) is itself a schema;
the use of {\qddref} : \qkw{\underline{https://json-schema.org/draft/2020-12/meta/applicator\#meta}} allows
the programmer to later redefine what constitutes a valid schema.

In the same way, the metaschema \qkw{https://json-schema.org/draft/2020-12/meta/uneval} has the same dynamic anchor (name) \qkw{meta}, and specifies that
the value of a $\qunProps$ keyword is itself a schema (we will describe $\qunProps$ later on).

The metaschema \qkw{https://json-schema.org/draft/2020-12/meta/core} just specifies that every schema is either a boolean or an object,
and that the value of keywords {\qdid}, {\qdda}, and {\qtitle}, 
is a string.

The metaschema \qkw{https://json-schema.org/draft/2020-12/schema} combines the three fragments above, and redefines the dynamic anchor \qkw{meta},
so that, when an instance is validated using \qkw{https://json-schema.org/draft/2020-12/schema}, any dynamic reference 
\qkw{\underline{...}\#meta} found inside any of the three fragments actually refers to \qkw{https://json-schema.org/draft/2020-12/schema},
that is, to their conjunction.

Some validators do not support the $\qunProps$ keyword; these validators would then use the
\qkw{https://json-schema.org/draft/2020-12/mini-schema} metaschema.
It combines only two of the fragments above, and redefines the dynamic anchor \qkw{meta},
so that, when an instance is validated using \qkw{.../mini-schema}, any$\qddref$: 
\qkw{\underline{...}\#meta} found inside any of the two fragments refers to \qkw{.../mini-schema}.

To sum up, the dynamic nature of \qkw{\underline{...}\#meta} allows one to define a ``customized'' metaschema by choosing the specific fragments to
combine. 
This example is very important, since it provided the initial motivation for the introduction of dynamic references:
allowing the definition of different versions of the metaschema by choosing which fragments to combine.

The semantics of dynamic references is not easy to understand, and we believe that it needs a formal definition.
We are going to provide that definition. 

\hide{
In order to test the above examples use this:

{
    "$schema": "https://json-schema.org/draft/2020-12/schema",
    "$id": "http://mjs.ex/root",
    "$ref": "http://mjs.ex/strict-tree",
    "$defs": {
       "ex2": 

{   "$schema": "https://json-schema.org/draft/2020-12/schema",
    "$id": "http://mjs.ex/simple-tree",
    "$dynamicAnchor": "tree",
    "type": "object",
    "properties": {
        "data": true,
        "children": {
            "type": "array",
            "items": { "$dynamicRef": "Z\underline{http://mjs.ex/root}Z#tree" }
        }
    }
}

       ,"ex3":

{
    "$schema": "https://json-schema.org/draft/2020-12/schema",
    "$id": "http://mjs.ex/strict-tree",
    "$dynamicAnchor": "tree",
    "$ref": "http://mjs.ex/simple-tree#tree",
    "unevaluatedProperties": false,
    "non-examples" : [
       { "dat": 3 },
       { "data": 3, "chldrn": [ ] },
       { "children": [ { "data": null }, { "chldrn": [ ] } ] }
    ]
}

   }
}

That is, put this scaffolding around the examples:

{
    "$schema": "https://json-schema.org/draft/2020-12/schema",
    "$ref": "http://mjs.ex/strict-tree",   <--- this is the Id of the example where I want to start
    "$defs": {
       "ex2": 

HERE GOES EX2

       ,"ex3":

HERE GOES EX3

   }
}

}

\hide{
Consider this schema:

{
    "$schema": "https://json-schema.org/draft/2020-12/schema",
    "$id": "http://mjs.ex/root",
    "$ref": "http://mjs.ex/strict-tree-of-trees",
    "$defs": {
       "http://mjs.ex/tree": {
          "$schema": "https://json-schema.org/draft/2020-12/schema",
          "$id": "http://mjs.ex/tree",
          "$dynamicAnchor": "node",
          "type": "object",
          "properties": {
              "data": true,
              "children": {
                  "type": "array",
                  "items": { "$dynamicRef": "#node" }
              }
          }
       },
       "http://mjs.ex/strict-tree-of-trees": {
          "$schema": "https://json-schema.org/draft/2020-12/schema",
          "$id": "http://mjs.ex/strict-tree-of-trees",
          "$dynamicAnchor": "node",
          "$ref": "http://mjs.ex/tree",
          "properties": {
              "data": {  "$ref": "http://mjs.ex/tree" }
          },
          "unevaluatedProperties": false
       }
   }
}

It validates 
{ "data" : { "daa" : {}, "children": [ {"data" : {} } ] } 
}

but it does not validate
{ "data" : { "data" : {}, "children": [ {"daa" : {} } ] } 
}

}

\subsection{Annotation Dependency}\label{sec:annotations}

In {\mJS}, the application of a keyword to an instance produces \emph{annotations}, and the validation result of a keyword may
depend on the annotations
produced by adjacent keywords.
These annotations carry a lot of information, but the information that is relevant for validation is which
children of the current instance (that is, which properties, if it is an object, or which items, if it is an array)
have already been {\evaluated}.
This information is then used by the operators $\qunProps$ and $\qunIts$,
since they are only applied to children that have \emph{not}
been {\evaluated}. 

For example, the assertion $\qunProps: \afalse$ in the schema of Figure \ref{fig3} depends on the annotations returned
by the adjacent keyword $\qdref : \qkw{http://mjs.ex/simple-tree\#tree}$
(which refers to the schema in Figure \ref{fig2}).
In this case, $\qdref$ {\evaluates} all and only
the fields whose name is either $\qkw{data}$ or $\qkw{children}$. Hence, $\qunProps: \afalse$  is applied to any other field,
and it fails if, and only if, fields with a different name exist.

The order in which the keywords appear in the schema is irrelevant for this mechanism;
as formalized later, the result is the same as if the
$\qunProps$ keyword was always evaluated last among the keywords inside its schema (i.e., its \emph{adjacent}
keywords).


The definition of {\evaluated} in the specifications of {\DTwenty} (\citep{specs2020}) presents many ambiguities,
as testified by online discussions such as {\citeoneone} and {\citefiveseven}.
These ambiguities do not affect the final result of the evaluation, but only the error messages that are
generated; these aspects are further discussed in the {\FV}, and we will formalize here the interpretation that is more widely accepted.
We believe that an important contribution of this work is that it provides a precise and succinct language
in which these ambiguities can be discussed and settled.

\section{Formalizing JSON Schema syntax}\label{sec:syntax}


A key contribution of this work is a formalization of the entire {\mJS} language, but, 
for reasons of space, we only report here a crucial subset that illustrates the approach and is sufficient to carry out
the complexity analysis; the remaining part is in the {\FV}.

In this section, we formalize the syntax, which does not present any technical difficulty; the validation behavior is
defined in the next section.

Schemas are structured as \emph{resources}, which are collected into \emph{documents} and refer to each other using URIs.
We deal with these aspects through a ``normalization process'' (Section \ref{sec:URI}), where we eliminate the issues related
to URI resolution and to the collection of many resources in just one document; we then formalize the syntax of normalized JSON Schema in Section \ref{sec:normalizedsyntax}.


\subsection{URI Resolution, Resource Flattening, Schema Closing}\label{sec:URI}

The keywords $\qdid$, $\qdref$, and $\qddref$ accept any URI reference as value, they apply the \emph{resolution} process defined in 
\citep{RFC3986}, and then interpret the resulting resolved URI according to the JSON Schema rules. Since resolution is already specified in \citep{RFC3986}, we will not formalize it here, and we will assume that, in every schema that is interpreted through our rules,
the values of these three keywords have already been resolved, and that the result of that resolution has the shape
$\UU\catHash\key{fragmentId}$ for $\qdref$, and $\qddref$, where \key{fragmentId} may be empty
and has the shape $\key{absURI}$, with no fragment,
for $\qdid$.

\begin{figure}[htb!]
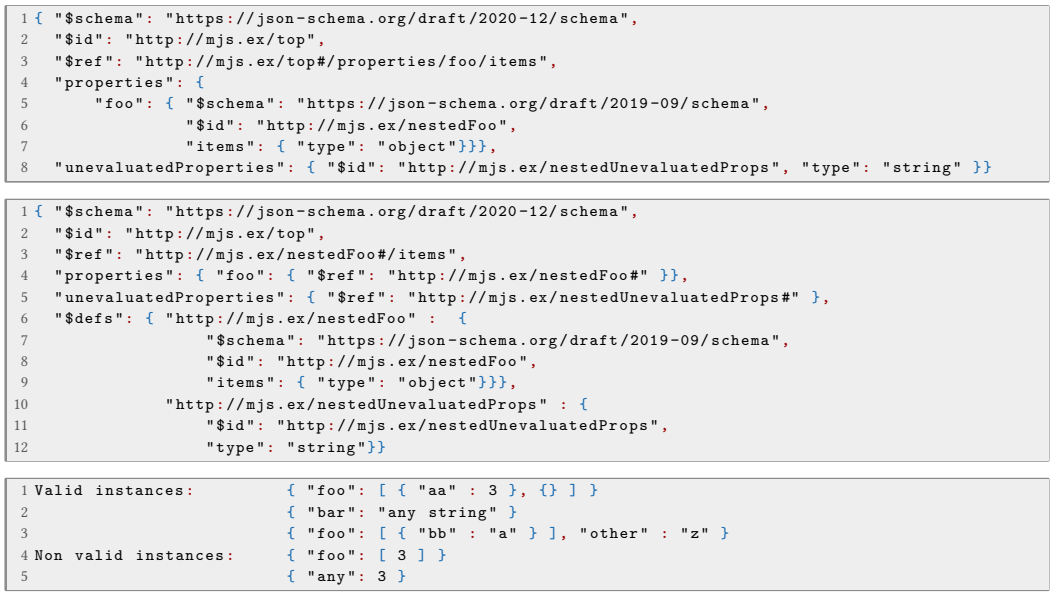

\begin{querybox}{}
\begin{lstlisting}[style=query,escapechar=Z]
{ "$schema": "https://json-schema.org/draft/2020-12/schema",
  "$id": "http://mjs.ex/top",
  "$ref": "http://mjs.ex/top#/properties/foo/items",
  "properties": {
      "foo": { "$schema": "https://json-schema.org/draft/2019-09/schema",
               "$id": "http://mjs.ex/nestedFoo",
               "items": { "type": "object"}}},
  "unevaluatedProperties": { "$id": "http://mjs.ex/nestedUnevaluatedProps", "type": "string" }}
\end{lstlisting}
\end{querybox}

\begin{querybox}{}
\begin{lstlisting}[style=query,escapechar=Z]
{ "$schema": "https://json-schema.org/draft/2020-12/schema",
  "$id": "http://mjs.ex/top",
  "$ref": "http://mjs.ex/nestedFoo#/items",
  "properties": { "foo": { "$ref": "http://mjs.ex/nestedFoo#" }},
  "unevaluatedProperties": { "$ref": "http://mjs.ex/nestedUnevaluatedProps#" },
  "$defs": { "http://mjs.ex/nestedFoo" :  {
                 "$schema": "https://json-schema.org/draft/2019-09/schema",
                 "$id": "http://mjs.ex/nestedFoo",
                 "items": { "type": "object"}}},
             "http://mjs.ex/nestedUnevaluatedProps" : {
                 "$id": "http://mjs.ex/nestedUnevaluatedProps",
                 "type": "string"}}
\end{lstlisting}
\end{querybox}

\begin{querybox}{}
\begin{lstlisting}[style=query,escapechar=Z]
Valid instances:         { "foo": [ { "aa" : 3 }, {} ] }
                         { "bar": "any string" }
                         { "foo": [ { "bb" : "a" } ], "other" : "z" }
Non valid instances:     { "foo": [ 3 ] }
                         { "any": 3 }


\end{lstlisting}
\end{querybox}
\caption{A schema with embedded resources, its flattened version, and some examples.}\label{fig:embedded}
\end{figure}

A $\qdid : \key{absURI}$ keyword at the top-level of a schema object $S_{\ensuremath{id}}$ that is nested inside a {\JS} document 
$S$ indicates that the schema
object $S_{\ensuremath{id}}$ is a separate resource, identified by $\key{absURI}$, that is embedded inside the document $S$ but is otherwise 
independent.\hide{\footnote{%
When the  $\qdid : \key{absURI}$ keyword is found in an object that is not interpreted as a schema, for example 
inside an unknown keyword or a non validation keyword such as $\qdefault$, then it has no special meaning.
The set of positions that are interpreted as schemas is defined by the grammar presented in Section \ref{sec:normalizedsyntax}.}}
Embedded resources are an important feature, since they allow the distribution of different resources with just one file,
but present some problems when they are nested inside arbitrary keywords,
and when a reference crosses the boundaries between resources, as does $\qdref: \qkw{http://mjs.ex/top\#/properties/foo/items}$
in Figure \ref{fig:embedded}, line~3 
(see also Section 9.2.1 of the specifications \citep{specs2020})
(a fragment identifier following \qkw{\#} may be either an anchor, as in the previous examples, or a JSON path, as in this case). 


To avoid this kind of problem, we assume that every
JSON Schema document is \emph{resource-flattened} before validation.
Resource flattening consists in moving every embedded resource identified by
$\qdid: \key{absURI}$ into the value of a field named \key{absURI} of a $\qddefs$ keyword at the top level of the
document, and replacing the moved resource with an equivalent schema $\{ \qdref: \key{absURI}\cat\qkw{\#} \}$ that invokes 
that resource;\footnote{The empty fragment identifier after $\key{absURI}\cat\qkw{\#}$ refers to the root of the resource $\key{absURI}$.}
{\qddefs} is a placeholder keyword that is not evaluated  but is a place to collect schemas
that can be invoked using $\qdref$ or 
$\qddref$.
\footnote{The $\qddefs$ keyword is the ``Modern'' version of the $\qdefs$ keyword of \cJS, and the act of collecting all embedded resources
in the $\qddefs$ section is described as ``bundling'' in {\DTwenty}, 
\webref{https://json-schema.org/draft/2020-12/json-schema-core\#name-compound-documents}{Section 9.3}.} 
During this phase, we also replace any reference that crosses resource boundaries 
with an equivalent 
reference whose URI is the base URI of the target, as suggested in Section 8.2 of the specifications \citep{specs2020} 
(e.g., line 3 in the first schema of Figure \ref{fig:embedded},
which crosses the boundary of the internal resource \qkw{http://mjs.ex/nestedFoo},
is substituted with the equivalent reference of line 3 in the second schema);
the two steps are exemplified in Figure  \ref{fig:embedded}.

\paragraph{Closed schemas}

The input of a validation problem includes a schema $S$ and all schemas that are recursively reachable from
$S$ by following the URIs used in the $\qdref$ and $\qddref$ operators. For complexity evaluation, we will only consider \emph{closed} schemas, that is, schemas that include all the different
resources that can be recursively reached from the top-level schema. 
There is no loss of generality, since external schemas can be embedded in a top-level one by copying
them in the $\qddefs$ section, using the $\qdid$ operator to preserve their base URI.


\subsection{JSON Schema Normalized Grammar}\label{sec:ordering}\label{sec:normalizedsyntax}

JSON Schema syntax is a subset of JSON syntax. 
We present in Figure \ref{fig:grammar} the grammar for a subset of the keywords, which is rich enough to present our
results. 
In this grammar, the meta-symbols are $(\key{X})^*$, which is Kleene star of $X$,
and $(\key{X})^?$, which is an optional $X$.
Non-terminals are italic words,
and everything else --- including \{ [ , : ] \} --- are terminal symbols.


JSON Schema allows the keywords to appear in any order and evaluates them in an order that
respects the dependencies among keywords.
We formalize this behavior by assuming that,
before validation, each schema is
reordered to respect the grammar in Figure \ref{fig:grammar}.
The grammar specifies that a schema $S$ is either a boolean schema, that matches any value ($\xtrue$) or no value at all ($\xfalse$), or it begins with a 
possibly empty sequence of Independent Keywords or triples (IK), followed by
a possibly empty sequence of First-Level Dependent keywords (FLD), followed by
a possibly empty sequence of Second-Level Dependent keywords (SLD).
\save{
The grammar also groups the keywords $\qif$-$\qthen$-$\qelse$,
and specifies that the presence of any keyword among
$\aif$-$\athen$-$\aelse$ implies the presence of $\athen$ and $\aelse$,
which is enforced by adding a trivial $\qthen : \{\}$, $\qelse : \{\}$ when it is missing;
this presentation reduces the number of rules needed to formalize $\qif$-$\qthen$-$\qelse$.
}
%
Specifically, the two keywords in  \key{FLD}, {\qaddProps} and {\qits},  
depend on some keywords in
\key{IK} (such as $\qprops$ and $\qpattProps$),
and the two keywords in  \key{SLD} depend on the keywords in \key{FLD}, 
and on many keywords in \key{IK}, such as $\qprops$, $\qpattProps$,
$\qany$, $\qall$, 
$\qdref$, and others.


\begin{figure}[h!]
$$
\begin{array}{lllllll}
\multicolumn{3}{l}{
q\in \Num, i \in \Int, k \in \Str, \key{absURI} \in \Str, f \in \Str, \key{format} \in \Str, p \in \Str, J \in \semt}\\[\NL]
\key{Tp} &::=& \qobject \Mm \qnumber \Mm \qinteger \Mm \qstr  \Mm \qarray \Mm \qboolean \Mm \qnull  \\[\NL]
\key{S} &::=
&  \xtrue \ \M\ \xfalse  \ \M\ 
        \JObjOpen\   \key{IK}
             \ (, \key{IK})^*
             \ (, \key{FLD})^*
             \ (, \key{SLD})^*  
            \ \JObjClose \ \\
 & &    \M\ \JObjOpen\  \key{FLD}
             \ (, \key{FLD})^*
             \ (, \key{SLD})^*  
            \ \JObjClose
            \ \M\ \JObjOpen\  \key{SLD}
              \ (, \key{SLD})^*  
            \ \JObjClose
            \ \M\  \JObjOpen\   \JObjClose 
            \\[\NL]
\key{IK} & ::= &
\Bb \amin: q   \Cc \Mm
\Bb \amax: q   \Cc \Mm
\Bb \apatt: p   \Cc  \Mm
\Bb  \areq: \JArr{k_1,\ldots,k_n}\Cc \\ &&  \Mm 
\Bb  \atype: \key{Tp} \Cc \Mm
\Bb  \atype: [ \key{Tp_1},\ldots,  \key{Tp_n}] \Cc \Mm
\Bb \addefs :  \JObj{k_1: S_1,\ldots,k_n:S_n} \Cc  \\
& & \Mm
\Bb \adid: \key{absURI}  \Cc  \Mm   
\Bb  \adref: \key{absURI}\catHash\key{f}  \Cc  \Mm
\Bb  \addref: \key{absURI}\catHash\key{f}\Cc \\ && \Mm
\Bb \ada: \emph{plain-name}  \Cc  \Mm
\Bb \adda: \emph{plain-name}  \Cc \\ && \Mm
\Bb  \aany: \JArr{S_1,\ldots,S_n} \Cc  \Mm
\Bb  \aall: \JArr{S_1,\ldots,S_n} \Cc  \Mm
\Bb  \anot: S \Cc \\
& & \Mm
\Bb \apattProps: \JObj{ p_1 : S_1,\ldots,p_m : S_m }  \Cc 
\\ &&  \Mm
\Bb \aprops: \JObj{ k_1 : S_1,\ldots,k_m : S_m }  \Cc \\ && \Mm
\Bb k: J \Cc \text{\ \ (with $k$ not previously cited)}
 \\[\NL]
$\key{FLD}\!\!$ & ::= &   
\Bb \aaddProps: S \Cc \Mm
\Bb \ait: S \Cc \\
$\key{SLD}\!\!$  & ::= & 
\Bb \aunProps: S  \Cc  \Mm
\Bb \aunIts: S \Cc
\end{array}
$$
\caption{Minimal grammar of normalized JSON Schema Draft 2020-12.}
\label{fig:grammar}
\end{figure}

This grammar specifies the predefined keywords, the type of the associated value (here $\semt$ is the set of
all JSON values, and \emph{plain-name} denotes any alphanumeric string starting with a letter), and their order.
We do not formalize here further restrictions on patterns $p$, absolute URIs \key{absURI}, and fragment identifiers~$f$.
A valid schema must also satisfy two more constraints: (1) every URI that is the argument of
$\qdref$ or $\qddref$ must reference a schema, and (2) any two adjacent keywords must have different names. 

\section{JSON Schema Validation}\label{sec:validation}

\renewcommand{\ps}{\pk}
\renewcommand{\pkl}{\pk}
\renewcommand{\Ret}[1]{\rightarrow}


{\json} values $\J$ are either base values, or nested arrays and objects; the order of object fields is irrelevant.
\[
\begin{array}{lllllllllll}
\multicolumn{2}{l}{ s\in\Str, d\in\Num, n\in \Int, n \geq 0, l_{i} \in \Str } \\[\NL]
\J ::=  & \anull \mid \atrue \mid \afalse \mid d \mid \str         \mid  \JArr{\J_1, \ldots, \J_n}  \mid   \JObj{l_1:\J_1,\ldots,l_n:\J_n } &  i\neq j \Rightarrow l_i\neq l_j
\end{array}
\]


In this paper, we reserve the notation $\JObj{ \ldots }$ to JSON objects, hence we
use $\Set{a_1,\ldots,a_n}$ and $\SetIIn{a_i}{i}{I}$ to indicate a set.
When the order of the elements is relevant, we use 
the list notation $\List{a_1,\ldots,a_n}$; 
we also use $\vec{a}$ to indicate a list. 

Given a pattern $p$, we will use $\rlan{p}$ to denote the language generated by $p$, i.e., the set of all strings that match that pattern.

\subsection{Introduction to the Proof System}

We are going to define a judgment that describes the result and the annotations that are returned
when a keyword  $K=k\!:\!P$ is applied 
to an instance $\J$ in a context $C$, where $C$ provides the information needed to interpret dynamic references.
Hence, we read the judgment $\KJudg{C}{J}{K}\Ret{r}{(r,\pk)}$
as: the application of the keyword $K$ to the instance $J$, in the context $C$, 
returns the boolean~$r$ and the annotations~$\pk$.
The annotations, as defined in \citep{specs2020}, are a complex data structure,
but we only represent here the small subset that is relevant for validation, that is, the set of {\evaluated} 
children, of the instance $J$.
The {\evaluated} children of an object are represented by their names, and the {\evaluated} children of an array by their position, so that:
$$
\begin{array}{llllllll}
\KJudg{C}{J}{K}\Ret{r}{(r,\pk)}\ &\And\ &J = \JObj{k_1:J_1,\ldots,k_n:J_n} &\Implies&  \pk \subseteq\Set{k_1,\ldots,k_n} \\[\NL]
\KJudg{C}{J}{K}\Ret{r}{(r,\pk)} &\And& J = \JArr{J_1,\ldots,J_n}              &\Implies&  \pk \subseteq\Set{1,\ldots,n} \\[\NL]
\KJudg{C}{J}{K}\Ret{r}{(r,\pk)} &\And&  J \mbox{ a base value} &\Implies&  \pk = \ES \\[\NL]
\end{array}
$$

Hence, the set of annotations $\pk$ can contain member names (strings) or array positions (integers).

We define a similar \emph{schema judgment} $\SJudgCJ{S}\Ret{r}{(r,\ps)}$
in order to describe the result of applying a schema $S$ to an instance $J$,
and we define a 
\emph{list evaluation} judgment
$\KLJudgCJ{\List{K_1,\ldots,K_n}}\Ret{}{(r,\pkl)}$ in order to apply a list of keywords to $J$, 
passing the annotations produced by a sublist $\List{K_1,\ldots,K_i}$ to the following keyword $K_{i+1}$.
Observe that the letters {\tt \small{K}}, {\tt \small{S}}, and {\tt \small{L}} that appear on top of $\,\StackDash{}{}\!$ are not
metavariables but just symbols used to differentiate the three judgments.

\hide{
\begin{example}
For example, in our system we can prove the following judgment.

\newcommand{\CC}{C}

$$
\SJudg{\CC}{9}{\JObj{\aany: [\JObj{\amof:3},\JObj{\amof:2}]}}\Ret{\btrue}{\ps}
$$
where the proof-term $\ps$ says: the schema $\JObj{\aany: [\JObj{\amof:3},\JObj{\amof:2}]}$
applied to 9 in the context $\CC$
returns $\btrue$ since its only keyword $\aany: [\JObj{\amof:3},\JObj{\amof:2}]$ returns $\btrue$,
which happens since the first item $\JObj{\amof:3}$ returns $\btrue$ over 9 by definition of 
$\amof$;
the second item $\JObj{\amof:2}$ returns $\bfalse$ over 9 by definition of 
$\amof$.

\begin{tabbing}
aaa\=aaa\=aaa\=aaa\=aaa\=aaa\=aaa\kill
$\ps$ \>= \> $\rschema(\CC,9,\JObj{\aany: [\JObj{\amof:3},\JObj{\amof:2}]},\btrue,\pk))$ \+ \\
$\pk$ \>= \> $\rany(\CC,9,\aany: [\JObj{\amof:3},\JObj{\amof:2}],\btrue,\pList{\ps_1,\ps_2}) $ \+\\
$\ps_1$ \>= \> $\rschema(\CC,9,\JObj{\amof:3},\btrue,\pk_1))$ \+ \\
$\pk_1$ \>= \> $\rmof(\CC,9,\amof:3,\btrue)$  \-\\
$\ps_2$ \>= \> $\rschema(\CC,9,\JObj{\amof:2},\bfalse,\pk_2))$ \+ \\
$\pk_2$ \>= \> $\rmof(\CC,9,\amof:2,\bfalse)$  \-\\

\end{tabbing}
\end{example}
}

In the next sections, we define the rules for keywords and for schemas.
Keywords are called \emph{assertions} when they assert properties of the analyzed instance,
so that $\qdid: \key{absURI}$ is not an assertion, while $\qtype:\key{Tp}$ is.
Assertions are called \emph{applicators} when they have schema parameters,
such as $S_1$ and $S_2$ in $\qany : [ S_1, S_2 ]$, that they apply either to the instance,
in which case they are \emph{in-place applicators} (e.g, $\qany : [ S_1, S_2 ]$), 
or to elements or items of the instance, in which
case they are \emph{object applicators} or \emph{array applicators} (e.g., $\qprops: \JObj{k_1:S_1, k_2:S_2}$).

In the following, we present the rules for ``terminal'' assertions, Boolean in-place applicators, and for object and
array applicators. We also illustrate the rules for sequential evaluation and for the schema judgments and, finally, for static and dynamic references.  
Rules are shown in Figure~\ref{fig:valrules}.




\begin{figure*}
\tiny
\begin{multicols}{2}
\infrule[\rmin\TTriv]
{
\TypeOf(J) \neq \rnumber
}
{\KJudgCJ{\amin: q}
 \Ret{T}
 (\btrue,\ES)
} 

 \infrule[\rtype]
{
r = (\,\TypeOf(J) =\rkw{Tp}\,)
}
{\KJudgCJ{\atype: \rkw{Tp}}
 \Ret{r}
  (r,\ES)
}

\columnbreak

\infrule[\rmin]
{
\TypeOf(J) = \rnumber \andalso r = (J \geq q)
}
{\KJudgCJ{\amin: q}
 \Ret{r}
 (r,\ES)
}

\end{multicols}

\begin{multicols}{2}

\infrule[\rnot]
{
\SJudg{C}{J}{S}\Ret{r}{(r,\ps)} 
}
{\KJudgCJ{\anot:S}
 \Ret{\Not r}
 (\Not r,\ps)
}

\end{multicols}

\begin{multicols}{2}

\infrule[\rany]
{
\forall i\in \SetTo{n}.\ \ \SJudg{C}{J}{S_i}\Ret{r_i}{(r_i,\ps_i)} \andalso
r=\Or(\SetIIn{r_i}{i}{\SetTo{n}}) 
}
{\KJudg{C}{J}{\aany:[S_1,...,S_n]}
 \Ret{r}
 (r,{\bigcup_{i\in\SetTo{n}}\ps_{i}} )
}
\columnbreak

\infrule[\rall]
{
\forall i\in \SetTo{n}.\ \SJudg{C}{J}{S_i}\Ret{r_i}{(r_i,\ps_i)} \andalso
r=\And(\SetIIn{r_i}{i}{\SetTo{n}}) 
}
{\KJudgCJ{\aall:[S_1,...,S_n]}
 \Ret{r}
 (r,{\bigcup_{i\in\SetTo{n}}}\ps_{i} )
}
\end{multicols}

\infrule[\rpattProps\TTriv]
{
\TypeOf(J) \neq \robject
}
{\KJudgCJ{\apattProps: \JObj{ p_1 : S_1,\ldots,p_m : S_m }}
 \Ret{T}
 (\btrue,\ES)
}

\infrule[\rpattProps]
{
J = \JObj{k'_1:J_1,\ldots,k'_n:J_n} \andalso
\Set{(i_1,j_1),\ldots,(i_l,j_l)} = \SetST{(i,j)}{k'_i \in \rlan{p_j}} \\[\NL]
\forall q\in \SetTo{l}.\ 
\SJudgC{J_{i_q}}{S_{j_q}}\Ret{r_q}{(r_q,\ps_q)} \andalso
r=\And(\SetIIn{r_q}{q}{\SetTo{l}}) 
}
{
\KJudgCJ{\apattProps: \JObj{ p_1 : S_1,\ldots,p_m : S_m }} 
 \Ret{r}
 {(r,\Set{k'_{i_1},\ldots,k'_{i_l}})}
}

\infrule[\rprops]
{
J = \JObj{k'_1:J_1,\ldots,k'_n:J_n} \andalso
\Set{(i_1,j_1),\ldots,(i_l,j_l)} = \SetST{(i,j)}{k'_i=k_j} \\[\NL]
\forall q\in \SetTo{l}.\ 
\SJudgC{J_{i_q}}{S_{j_q}}\Ret{r_q}{(r_q,\ps_q)} \andalso
r=\And(\SetIIn{r_q}{q}{\SetTo{l}})
}
{
\KJudgCJ{\aprops: \JObj{ k_1 : S_1,\ldots,k_m : S_m }} 
 \Ret{r}
 {(r,\Set{k'_{i_1},\ldots,k'_{i_l}})}
}

\begin{multicols}{2}
\infax[\rtrueS]
{\SJudgCJ{\atrue}
 \Ret{T}
 {(\btrue,\ES)}
}

\infrule[\rschema\rkw{-true}]
{
\KLJudgCJ{\List{K_1,\ldots,K_n}}\Ret{\List{r_1,\ldots,r_n}}{(\btrue,\pk)} 
}
{\SJudgCJ{\JObj{K_1,\ldots,K_n}}
 \Ret{r}
 {(\btrue,\pk)}
}

\columnbreak

\infax[\rfalseS]
{\SJudgCJ{\afalse}
 \Ret{F}
 {(\bfalse,\ES)}
}

\infrule[\rschema\rkw{-false}]
{
\KLJudgCJ{\List{K_1,\ldots,K_n}}\Ret{\List{r_1,\ldots,r_n}}{(\bfalse,\pk)} 
}
{\SJudgCJ{\JObj{K_1,\ldots,K_n}}
 \Ret{r}
 {(\bfalse,\ES)}
}

\end{multicols}

\infrule[\runProps]
{
J=\JObj{k_1 : J_1,\ldots,k_n:J_n} \andalso
\KLJudgCJ{\Kl}\Ret{\rl}{(r,\pk)} \\[\NL]
\Set{{i_1},\ldots,{i_l}} = \SetST{i}{1\leq i \leq n \And\ k_i \not\in \pk} \\[\NL]
\forall q\in \SetTo{l}.\ 
\SJudgC{J_{i_q}}{S}\Ret{r_q}{(r_q,\ps_q)} \andalso
r'=\And(\SetIIn{r_q}{q}{\SetTo{l}}) 
}
{
\KLJudgCJ{(\Kl\plus\aunProps : S)}
 \Ret{\rl\plus r}{(r\And r',\Set{k_{1}\ldots,k_{n}})}
}

\infrule[\raddProps]
{
J=\JObj{k_1 : J_1,\ldots,k_n:J_n} \andalso
\KLJudgCJ{\Kl}\Ret{}{(r,\pk)} \\[\NL]
\Set{{i_1},\ldots,{i_l}} = \SetST{i}{1\leq i \leq n \And\ k_i \not\in \rlan{\akw{propsOf}(\Kl)}} \\[\NL]
\forall q\in \SetTo{l}.\ 
\SJudgC{J_{i_q}}{S}\Ret{r_q}{(r_q,\ps_q)} \andalso
r'= \And(\SetIIn{r_q}{q}{\SetTo{l}}) 
}
{
\KLJudgCJ{(\Kl\plus\aaddProps : S)}
\Ret{\rl\plus r}{(r\And r',\Set{k_{1}\ldots,k_{n}})}
}

\begin{multicols}{2}
\infrule[\rklist-(n+1)]
{
K\in\key{IK} \andalso
\KLJudgCJ{\Kl}\Ret{\rl}{(r_l,\pk_l)} \andalso
\KJudg{C}{J}{K}\Ret{r}{(r,\pk)} \\[\NL]
}
{\KLJudgCJ{(\Kl\plus K)}
\Ret{\rl\plus r}{(r_l \And r,\pk_l \cup \pk)}
}
\columnbreak 

\infax[\rklist-0]
{\KLJudgCJ{\List{}}
 \Ret{\List{}}{(\btrue,\ES)}
}
\end{multicols}

\begin{multicols}{2}

\infrule[\rdref]
{
S' = \Get(\Load(\key{absURI}),\key{f}) \andalso
\SJudg{C  + \key{absURI}\ }{J}{S'}\Ret{r}{(r,\ps)}
}
{
 \KJudg{C}{J}{\adref: \key{absURI}\catHash\key{f}}
 \Ret{r}
 (r,\ps)
}

\infrule[\rddref\rkw{AsRef}]
{
\DGet(\Load(\key{absURI}),\key{f}) = \bot  \\[\NL]
S' = \Get(\Load(\key{absURI}),\key{f}) \andalso
\SJudg{C  + \key{absURI}\ }{J}{S'}\Ret{r}{(r,\ps)}
}
{
 \KJudg{C}{J}{\addref: \key{absURI}\catHash\key{f}}
  \Ret{r}
 (r,\ps)
}

\columnbreak

\infrule[\rddref]
{
\DGet(\Load(\key{absURI}),\key{f}) \neq \bot  \andalso
\key{fURI} = \fstURI(C+\key{absURI},f) 
 \\[\NL]
S' = \DGet(\Load(\key{fURI}),\key{f}) \andalso
\SJudg{C  + \key{fURI}\ }{J}{S'}\Ret{r}{(r,\ps)}
}
{\KJudg{C}{J}{\addref: \key{absURI}\catHash\key{f}}
  \Ret{r}
 (r,\ps)
}


\end{multicols}

\caption{Validation rules.}
\label{fig:valrules}
\end{figure*}

\subsection{Terminal Assertions}\label{sec:terminal}

Terminal assertions are those that do not contain any subschema to reapply.
The great majority of them are conditional on a type $T$: they are trivially satisfied when the instance $J$ does
not belong to $T$, and they otherwise verify a specific condition on $J$. 
Hence, these keywords are defined by a couple of rules,
as exemplified here for the keyword $\amin: q$.
Rule (\rmin\TTriv) (Figure \ref{fig:valrules}) always returns $\btrue$ (true) when $\J$ is not a number, while rule (\rmin),
applied to numbers, returns the same boolean $r\in\Set{\btrue,\bfalse}$ as checking whether $J \geq q$.
The set of {\evaluated} children is $\ES$: in our model, these two rules do not generate an annotation.



These typed terminal assertions are completely defined by a type and a condition;
a complete list of these keywords, with the associated type and condition, 
is in the {\FV}.

We also have four \emph{type-uniform} terminal assertions,  that do not single out any specific type for a special treatment;
they are {\aenum}, {\aconst}, $\atype: [\rkw{Tp}_1,\ldots,\rkw{Tp}_n]$, and ${\atype} : \rkw{Tp}$.
The rule for a type-uniform terminal assertion is completely defined by a condition, as
reported in Table \ref{tab:terms}.

In Figure~\ref{fig:valrules} we show the rule for ${\atype} : \rkw{Tp}$,
where $\TypeOf(J)$ extracts the type of the instance $J$.


\begin{table}[htb]
\centering
\caption{Boolean conditions for type-uniform terminal assertions.}
\label{tab:terms}

\begin{tabular}{| l | l |}
\hline
\textbf{assertion: \key{kw}\! :\! J'}  & \textbf{condition: \rkw{cond}(J,\ \key{kw}\! :\! J')} \\ 
\hline 
\hline
$\aenum: [J_1,\ldots,J_n]$   &   $J \in \Set{J_1,\ldots,J_n}$ \\ 
$\aconst: J_c $  &  $J = J_c$ \\ 
$\atype: \rkw{Tp}$ &   $\TypeOf(J)= \rkw{Tp}$ \\ 
$\atype: [\rkw{Tp}_1,\ldots,\rkw{Tp}_n]$  &   $\TypeOf(J) \in \Set{\rkw{Tp}_1,\ldots,\rkw{Tp}_n}$ \\ 
\hline
\end{tabular}

\end{table}

\subsection{Boolean Applicators}\label{sec:inplace}

%


JSON Schema boolean applicators, 
such as $\aany:[S_1,...,S_n]$,
apply a list of schemas to the instance, obtain a list of intermediate boolean results,
and combine the intermediate results using a boolean operator.
For the annotations, all assertions always return a union of the annotations
produced by their subschemas, even when the assertion fails;
this should be contrasted with the behavior of schemas, where a failing schema never
returns any annotation (Section \ref{sec:sequential}).\footnote{Other interpretations of
the specifications of {\DTwenty} are possible; see the {\FV} for a discussion.
}

The rule for the disjunctive applicator {\aany} combines the intermediate results using the
$\Or$ operator, and a child of $J$ is {\evaluated} if, and only if, it has been {\evaluated} by any subschema
$S_i$.


The rules for $\aall$ and for $\anot$ are analogous:
$\aall$ is successful if all premises are successful,
and negation is successful if its premise fails.


\hide{
\infrule[\rkw{if-true-then}]
{
\KLJudg{C}{J}{\Kl}\Ret{\rl}{\pkl} \andalso
\KJudg{C}{J}{S_i}\Ret{\btrue}{\ps_i} \andalso
\KJudg{C}{J}{S_t}\Ret{r_t}{\ps_t}
}
{
\KLJudgCJ{(\ \Kl\plus 
  \aif: S_i \plus
  \athen: S_t \ (\plus
  \aelse: S_e)^?\ )}
  \Ret{\rl \plus \btrue \plus r_t}{\pkl\plus\rif(\_,\ps_i)\plus\rthen(\_,\ps_t)}
}

\infrule[\rkw{if-true-no-then}]
{
\KLJudg{C}{J}{\Kl}\Ret{\rl}{\pkl} \andalso
\KJudg{C}{J}{S_i}\Ret{\btrue}{\ps_i}
}
{
\KLJudgCJ{(\ \Kl\plus 
  \aif: S_i
  \ (\plus
  \aelse: S_e)^?\ )}
  \Ret{\rl \plus \btrue}{\pkl\plus\rif(\_,\ps_i)}
}

\infrule[\rkw{if-false-else}]
{
\KLJudg{C}{J}{\Kl}\Ret{\rl}{\pkl} \andalso
\KJudg{C}{J}{S_i}\Ret{\bfalse}{\ps_i} \andalso
\KJudg{C}{J}{S_e}\Ret{r_e}{\ps_e}
}
{
\KLJudgCJ{(\ \Kl\plus 
  \aif: S_i  
  \ (\plus \athen: S_t \ )^?
  \ \plus \aelse: S_e)}
  \Ret{\rl \plus \bfalse \plus r_e}{\pkl\plus\rif(\_,\ps_i)\plus\relse(\_,\ps_e)}
}

\infrule[\rkw{if-false-no-else}]
{
\KLJudg{C}{J}{\Kl}\Ret{\rl}{\pkl} \andalso
\KJudg{C}{J}{S_i}\Ret{\bfalse}{\ps_i}
}
{
\KLJudgCJ{(\ \Kl\plus 
  \aif: S_i 
  \ (\plus \athen: S_t \ )^?\ )
  }
  \Ret{\rl \plus \bfalse}{\pkl\plus\rif(\_,\ps_i)}
}
}

\subsection{Independent Object and Array Applicators (Independent Structural Applicators)}\label{sec:structural}


Independent structural applicators are those that reapply a subschema to some children of the instance
(\emph{structural})
and whose behavior does not depend on adjacent keywords (\emph{independent}).

We start with the rules for the $\apattProps$ applicator that asserts that
if $J$ is an object, then every property of $J$ whose name matches a pattern $p_j$ has a value that satisfies $S_j$.
This rule constrains all instance fields whose name matches any pattern $p_j$ in the applicator, but it does not
force any of the $p_j$'s to be matched by any property name, nor any property name to match any $p_j$;
if there is no match, the keyword is satisfied. Patterns $p_j$ may have a non-empty intersection of their languages,
so that a single instance field may match two or more patterns.

We first have the trivial
rule (\rpattProps\TTriv) for the case where $J$ is not an object: non-object instances trivially satisfy the operator.


In the non-trivial case (rule (\rpattProps)), where $J = \JObj{k'_1:J_1,\ldots,k'_n:J_n}$, we first collect the set $\Set{(i_1,j_1),\ldots,(i_l,j_l)}$ of all pairs
${(i,j)}$ such that $k'_i \in \rlan{p_j}$, where $\rlan{p_j}$ is the language of the pattern $p_j$.
For each such pair $(i_q,j_q)$, we collect the boolean $r_q$ that specifies whether $\SJudgC{J_{i_q}}{S_{j_q}}$
holds or not, and the 
entire keyword is successful over $J$ if the conjunction $r=\And(\SetIIn{r_q}{q}{\SetTo{l}})$ is $\btrue$ (\emph{true}).
According to the standard convention, the empty conjunction $\And(\Set{})$ evaluates to $\btrue$, hence this
rule does not force any matching.


%

The {\evaluated} properties are all the properties $k_{i_q}$ for which a corresponding pattern
$p_{i_q}$ exists, independently of the result $r_q$ of the corresponding validation, and independently of the overall result
$r$ of the keyword.
Observe that the sets $\ps_q$ of the children that are {\evaluated} in the subproofs are discarded; this
happens because elements of $\ps_q$ are children of a child $J_{i_q}$ of $J$; we collect information
about the evaluation of the children of $J$, and we are not interested in children of children.



The rule (\rprops) for $\aprops: \JObj{ k_1 : S_1,\ldots,k_m : S_m }$ is essentially the same, with
equality $k'_i=k_j$ taking the place of matching $k'_i \in \rlan{p_j}$, hence $l$ is the number of pairs $(i,j)$ such that $k'_i=k_j$; likewise, no name match is required, but in case of a match, the corresponding
child of $J$ must satisfy the subschema with the same name. 

Of course, we also have the trivial rule $(\rprops\TTriv)$,  analogous to rule
$(\rpattProps\TTriv)$:  when $J$ is not an object, {\qprops} is trivially satisfied.
The rules for the other independent object 
and array applicators  
can be found in the {\FV}.


The independent keywords presented in this section (and in the previous one) produce (respectively, collect and transmit)
annotations that influence the behavior of the
dependent keywords, which are $\aaddProps$, $\aits$, $\aunProps$, and $\aunIts$. 
All these dependencies are formalized in the following sections.

\subsection{The Semantics of Schemas: Sequential Evaluation of Keywords}\label{sec:sequential}


We have defined the semantics of the independent keywords.
We now introduce the rules for schemas and for sequential executions of keywords.

The rules (\rtrueS) and (\rfalseS) for the $\atrue$ and $\afalse$ schemas are trivial.

%

The rules ({\rschema\rkw{-true}}) and ({\rschema\rkw{-false}}) 
for an object schema $\JObj{\Kl}$ are based on the keyword-list 
judgment $\KLJudgCJ{\Kl}\RetL{\rl}{\pkl}$, which
applies the keywords in the ordered list $\Kl$, passing the annotations from left to right. 

Rule ({\rschema\rkw{-true}}) just reuses the result of the keyword-list 
judgment, but ({\rschema\rkw{-false}}) specifies that, as dictated by \citep{specs2020}, when schema validation
fails, all annotations are removed, and hence no instance child is regarded as {\evaluated}.
This is a crucial difference with keyword-lists, since the $\KJudgCJ{K}\Ret{}{(r,\pk)} $
judgment may return non-empty annotations even when $r=\bfalse$.\footnote{See the discussion in the {\FV}.}


\hide{
\infrule[\rschema\rkw{\$id}]
{
\KLJudg{C+\key{absURI}}{J}{\List{K_1,\ldots,K_n}}\RetL{\List{r_1,\ldots,r_n}}{\List{\pk_1,\ldots,\pk_n}} \andalso
r=\And(\SetIIn{r(\pk_i)}{i}{\SetTo{n}}) 
}
{\SJudgCJ{\JObj{\qdid: \key{absURI}, K_1,\ldots,K_n}}
 \Ret{r}
 \rschema\rkw{\$id}(\_,\Set{\pk_1,\ldots,\pk_n})
}
}

We now describe the rules for the sequential evaluation judgment $\KLJudgCJ{\Kl}\Ret{\rl}{(r,\pk)}$.
The rules are  specified for each list 
$\Kl\plus K$
by induction on $|\Kl|+|K|$ and by cases on $K$.

We start with the crucial rule (\runProps),  for $\Kl\plus\aunProps: S$, which forces all unevaluated properties to conform to $S$.
To evaluate $\Kl\plus\aunProps: S$ we first evaluate $\Kl$, which yields  a set of {\evaluated} children
$\pk$, we then evaluate $S$ on the other children, and we combine the results by conjunction.
We return every property of the instance as {\evaluated}.


The rule for $\Kl\plus\aaddProps: S$ (\raddProps) is identical, apart from the fact that we only eliminate the
properties that have been {\evaluated} by \emph{adjacent} keywords.
The specifications \citep{specs2020} indicate that this information should be passed as annotation, but
that a static analysis is acceptable if it gives the same result. We formalize this second approach since it is slightly simpler.
We define a function $\akw{propsOf}(\Kl)$ that extracts all the patterns and all the names that appear
in any $\qprops$ and $\qpattProps$ keywords that appear in $\Kl$ and combines them into a pattern;
a property is directly evaluated by a keyword in $\Kl$ if, and only if, it belongs to $\rlan{\akw{propsOf}(\Kl)}$.
The notation $\keykey{k_i}$ used in the first line indicates a pattern whose language is $\Set{k_i}$;
$\ES$ in the third line is a pattern such that $\rlan{\ES} = \ES$.

$$
\!\!\begin{array}{llll}
\akw{propsOf}(\qprops: \JObj{ k_1 : S_1,\ldots,k_m : S_m }) &=& 
 \keykey{k_1} \cat \qkw{|}\cat\ldots \cat \qkw{|}\cat \keykey{k_n} \\[\NL]
\akw{propsOf}(\qpattProps: \JObj{ p_1 : S_1,\ldots,p_m : S_m }) &=& 
 p_1 \cat \qkw{|}\cat\ldots \cat \qkw{|}\cat p_m \\[\NL]
\akw{propsOf}(K) &=& 
 \ES \qquad\qquad\qquad\qquad \mbox{otherwise} \\[\NL]
\akw{propsOf}(\List{K_1,\ldots,K_n}) &=& 
 \akw{propsOf}(K_1) \cat \qkw{|}\cat\ldots \cat \qkw{|}\cat \akw{propsOf}(K_n)
\end{array}
$$

The keyword $\qaddProps$ was already present in {\cJS}, and, as shown by our formalization,
it does not need to access the annotations passed
by the previous keywords, but can be implemented on the basis of information that can be statically extracted from
$\qprops$ and $\qpattProps$; critically, it is not influenced
by what is {\evaluated} by an adjacent $\qdref$, as happens to $\qunProps$ in the example of Figure \ref{fig3}.
{\MJS}  introduced the new keyword $\qunProps$
in order to overcome this limitation.


The rules for $\aunIts : S$ and $\aits : S$ are similar and can be found in the {\FV}.


Having exhausted the rules for the dependent keywords, we have a catch-all rule (\rklist-(n+1)) for all other keywords, that
says that, when $K$ is an independent keyword, we combine the results
of $\KLJudgCJ{\Kl}\Ret{\rl}{(r_l,\pk_l)}$ and $\KJudg{C}{J}{K}\Ret{r}{(r,\pk)}$, but no information
is passed between the two judgments.
Rule (\rklist-0) is just the base case for induction.


%

\subsection{Static and Dynamic References}\label{sec:refs}

Annotation-dependent validation and dynamic references are the two additions that characterize {\mJS}.
Dynamic references are those that had the greatest need for formalization.

The reference operators ${\qdref} : {\UU}\lcatHash\key{fragmentId}$ and
${\qddref} : {\UU}\lcatHash\key{fragmentId}$ are in-place applicators that allow a URI-identified subschema to be applied
to the current instance,
but the two applicators interpret the URI in a very different way. 

$\qdref: \key{absURI}\catHash\key{fragmentId}$ retrieves the resource $S$ identified by {\UU},
which may be the current schema or a different one, retrieves the subschema $S'$ of $S$ identified by
$\key{fragmentId}$, and applies $S'$ to the current instance $J$ (rule $(\adref)$).
The $\qddref$ keyword, instead, interprets the reference in a way that depends on the 
\emph{dynamic scope}, which is, informally, the ordered list of all resources that have been visited in the current branch of the
proof tree, which we represent in the rules by listing their URIs in the context $C$.

Specifically, as shown in rule $(\adref)$, the evaluation of $\qdref: \key{absURI}\catHash\key{fragmentId}$
changes the dynamic scope, by extending the context $C$ in the premise  with ${\UU}$
--- in the rule, $L + e$  denotes the operation of adding an element $e$ at the end of 
a list $L$.

In rule (\rdref), $\Load(\key{absURI})$ returns the schema $S$ identified by $\key{absURI}$, an operation that
we cannot formalize
since the standards leave it undefined \citep{specs2020,RFC3986}; we can regard it as an access to an immutable
store that associated URIs to schemas.
$\Get(S,\key{f})$ returns the subschema identified by $f$ inside $S$; the fragment $f$ may either be empty,
hence identifying the entire $S$, or a plain-name, which is matched by a corresponding $\qda$ operator inside
$S$,\footnote{Actually, it can also be matched by a $\qkw{\footnotesize \$dynamicAnchor}$ operator, which, in this case, is interpreted as
exactly as $\qkw{\footnotesize \$anchor}$.
}
or a JSON Pointer, that begins with ``/'' and is interpreted by navigation. The $\Get$ function
is formally defined in the {\FV}.

\hide{ DO NOT DELETE
\webref{https://json-schema.org/draft/2020-12/json-schema-core.html\#name-lexical-scope-and-dynamic-s}{7.1}
Lexical and dynamic scopes align until a reference keyword is encountered. While following the reference keyword moves processing from one lexical scope into a different one, from the perspective of dynamic scope, following a reference is no different from descending into a subschema present as a value. A keyword on the far side of that reference that resolves information through the dynamic scope will consider the originating side of the reference to be their dynamic parent, rather than examining the local lexically enclosing parent.

If the initially resolved starting point URI includes a fragment that was created by the "$dynamicAnchor" keyword, the initial URI MUST be replaced by the URI (including the fragment) for the outermost schema resource in the dynamic scope (Section 7.1) that defines an identically named fragment with "$dynamicAnchor".
}



For simplicity, we assume that the schema has already been analyzed to ensure the following properties; it would not be difficult
to formalize these conditions in the rules:
\begin{enumerate}
\item the $\Load(\key{absURI})$ invocation will not fail, that is, \key{absURI} is a valid URI;
\item the $\Get(\Load(\key{absURI}),\key{f})$ invocations will not fail, that is, 
     $\Load(\key{absURI})$ contains a subschema identified by $f$ (which is either a JSON Pointer
      or an anchor name);
\item the $\DGet(\Load(\key{fURI}),\key{f})$ invocation will not fail, that is, 
     $\Load(\key{fURI})$ contains a subschema identified by the dynamic anchor $f$;
\item every $\qdid$ operator assigns to its schema a URI that is different from the
   URI of any other resource recursively reachable from its schema;
\item there exist no two $\qda$ and $\qdda$ keywords that assign the same name 
to two different schemas inside one specific resource.
\end{enumerate}

If any of these conditions does not hold, the validator should raise a failure.


The behavior of {\qddref} is very different from that of {\qdref}, and is defined as follows (see \citep{specs2020} \webref{https://json-schema.org/draft/2020-12/json-schema-core.html\#name-dynamic-references-with-dyn}{Section 8.2.3.2}):
{\begin{quote}{If the initially resolved starting point URI includes a fragment that was created by the "\$dynamicAnchor" keyword, the initial URI MUST be replaced by the URI (including the fragment) for the outermost schema resource in the dynamic scope (Section 7.1) that defines an identically named fragment with \qdda. Otherwise, its behavior is identical to {\qdref}, and no runtime resolution is needed.}\end{quote}}

This sentence is not easy to decode, but it means that, given an assertion
$\qddref: \key{absURI}\catHash\key{f}$, 
one first verifies whether the resource referenced by the ``starting point URI'' \key{absURI} contains a
dynamic anchor $\qdda: f'$ with $f'=f$.
If this is the case, $\qddref: \key{absURI}\catHash\key{f}$ will
be interpreted according to the dynamic interpretation specified in the second part of the sentence,
otherwise it will be interpreted as if it were a static reference ${\qdref}$; this verification is formalized
by the premises $\DGet(\Load(\key{absURI}),\key{f}) \neq \bot$ and
$\DGet(\Load(\key{absURI}),\key{f}) = \bot$ of the two rules that we present above for 
$\qddref$. The function $\DGet(S,f)$  looks inside $S$ for a subschema that contains $\qdda: f$,
but it returns $\bot$ if there is no such subschema.\footnote{%
Observe that $\akwfn{dget}(\akwfn{load}(\key{absURI}),\key{f}) \neq \bot$ is a static check that may be performed 
once for all when the schema is loaded.
This check is called ``the bookending requirement'', and it may be dropped 
in future Drafts (see
\webref{https://github.com/json-schema-org/json-schema-spec/issues/1064}{Remove \$dynamicRef bookending requirement}),
yet this decision would not affect our results.}
After this check is passed, the dynamic interpretation focuses on the fragment $f$, and it looks 
for the first (the ``outermost'') resource in $C^+$ that contains a subschema identified by $\qdda: f$,
where $C^+$ is the dynamic context $C$ extended with the initial URI \key{absURI}.

We formalize this specification using two functions: $\DGet(S,\key{f})$ and
$\fstURI(C,\key{f})$.
The function $\DGet(S,\key{f})$ returns the subschema $S'$ that is identified in $S$ by a plain-name $\key{f}$
that has been defined by $\qdda: \qkw{f}$, and returns
$\bot$ when no such subschema is found in $S$, and its definition is given in the {\FV}.
The function $\fstURI(L,\key{f})$ returns the first URI in the list $L$ that defines $f$, that is, such that
$\DGet(\Load(\key{absURI}),\key{f}) \neq \bot$.

We can finally formalize the dynamic reference rule (\rddref). 
It first checks that the initial URI refers to a dynamic anchor, but after this check, the result of
$\Load(\key{absURI})$ is forgotten. Instead, we look for the first URI \key{fURI} in $C+\key{absURI}$ where the dynamic anchor $f$
is defined, and extract the corresponding subschema $S'$ by executing $\DGet(\Load(\key{fURI}),f)$.


\begin{example}
The dynamic references mechanism allows one to define multiple successive refinements of a same schema.
One may first refine the trees of Figure \ref{fig2} to the strict trees of Figure \ref{fig3}, and then to the
integer strict trees of Figure \ref{strict-int-trees} below.

\begin{figure}[htb!]
\begin{querybox}{}
\begin{lstlisting}[style=query,escapechar=Z]
{ "$id": "http://mjs.ex/strict-int-tree",
  "$dynamicAnchor": "tree",
  "$ref": "http://mjs.ex/strict-tree#tree",
  "properties": { "data" : { "type" : "integer"  } } 
}
\end{lstlisting}
\end{querybox}
\caption{Refining a refined type.}\label{strict-int-trees}
\end{figure}

In an empty context, a reference $\qdref: \qkw{http://mjs.ex/strict-int-tree\#tree}$ would invoke this schema, which
would then invoke the strict-tree of Figure \ref{fig3}, which would invoke the simple-tree of Figure \ref{fig2}.
At this point we find the only dynamic reference, the following line in Figure \ref{fig2}, which says that the
children of a tree are trees:
\begin{querybox}{}
\begin{lstlisting}[style=query,escapechar=Z]
    "children": {  "type": "array", "items": { "$dynamicRef": "Z\underline{http://mjs.ex/simple-tree}Z#tree" }}}}
\end{lstlisting}
\end{querybox}

Here, the dynamic context $C$ is:
{\small
\[
\List{\qkw{http://mjs.ex/strict-int-tree},\qkw{http://mjs.ex/strict-tree},\qkw{http://mjs.ex/simple-tree}}
\]}
The dynamic reference is resolved as \qkw{http://mjs.ex/strict-int-tree\#tree}, since\\
\qkw{http://mjs.ex/strict-int-tree} is the first resource in $C$ that defines
the dynamic anchor \qkw{tree}.
\end{example}

\begin{remark}\label{rem:nodynref}
Observe that $\fstURI(C+\key{absURI},f)$ searches $\key{fstURI}$ into a list that contains the dynamic context
extended with the URI $\key{absURI}$. We have the impression that the specifications (as copied above) would rather
require $\key{fURI} = \fstURI(C,f)$, but we contacted the authors and checked some online 
verifiers that are widely adopted. There seems to be a general agreement that $\key{fURI} = \fstURI(C+\key{absURI},f)$
is the correct formula (see the {\FV} for a concrete example).

This is a typical example of the problems generated by natural language specifications, where different readers interpret
the same document in different ways, and one needs to discover the current consensus by social interaction and
experiments.
Formal specifications would be extremely useful to address this kind of problem.
%
\end{remark}

The second rule for $\qddref$ (rule (\rddref\rkw{AsRef})) applies when the initially resolved starting point URI does not include a fragment that was created by the 
{\qdda} keyword ($\DGet(\Load(\key{absURI}),\key{f}) = \bot$), 
in which case {\qddref} behaves as {\qdref}.


\hide{
What happens when you use a JSON path in order to enter in a non-canonical way inside a subschema? 
Nothing, it is ok.

{
  "$id": "https://json-schema.hyperjump.io/schema",
  "$schema": "https://json-schema.org/draft/2020-12/schema",
  "properties" : {
      "a" : {
          "$id": "https://json-schema.hyperjump.io/schema/inside",
          "multipleOf" : 7,
          "$anchor" : "seven"
     }
  },
  "$ref" : "https://json-schema.hyperjump.io/schema#/properties/a"
}

}

\subsection{Compressing the Context by \emph{Saturation}}

The only rule that depends on the context $C$ is rule $(\rddref)$ that uses 
$\fstURI(C+\key{absURI},f)$ to retrieve, in $C+\key{absURI}$, the first $URI$  that identifies a
schema that contains $\key{f}$ as a dynamic anchor. 
When $URI$ is already present in $C$,
its addition at the end of $C$ does not affect the result of $\fstURI$, hence, for each $URI$, we could just retain 
its first occurrence in $C$. Let us define $C+? URI$, that we read as $C$ \emph{saturated with} $URI$, as $C+? URI=C$ when 
$URI\in C$ and $C+?URI=C+URI$ when $URI\not\in C$. By the observation above, we can substitute
$C+URI$ with $C+?URI$ in the premises of rules $(\rdref)$, $(\rddref)$, and $(\rddref\rkw{AsRef})$,
obtaining, for example the following  rule.

\infrule[$\rddref_c$]
{
\DGet(\Load(\key{absURI}),\key{f}) \neq \bot  \andalso
\key{fURI} = \fstURI(C+?\key{absURI},f) 
\\[\NL]
S' = \DGet(\Load(\key{fURI}),\key{f}) \andalso
\SJudg{C  +? \key{fURI}\ }{J}{S'}\Ret{r}{(r,\ps)}
}
{\KJudg{C}{J}{\addref: \key{absURI}\catHash\key{f}}
  \Ret{r}
 (r,\ps)
}

This observation will be crucial in our complexity evaluations. From now on, we adopt
this version of the reference rules and assume that contexts are URI lists with no repetition.

\subsection{Ruling Out Infinite Proof Trees}\label{sec:termination}

A \emph{proof tree} is a tree whose nodes are judgments and such that, for every node $N$ of the tree, 
there is a deduction rule that allows
$N$ to be deduced from its children. 
A judgment $N$ is \emph{proved} when there is a \emph{finite} proof tree whose root is $N$.

The \emph{naive application algorithm}, given a triple $C$, $J$, $S$, builds the proof tree
rooted in $\SJudg{C}{J}{S}\Ret{r}(r,\ps)$ by
finding the deduction rule whose conclusion matches $C$, $J$, $S$, 
and by recurring on all the judgments in its premises.

Consider now a schema that reapplies itself to the current instance, such as:
$$S_{\key{loop}}\ =\ \JObj{\qdid: \qkw{here}, \qall : \JArr { \JObj { \qdref : \qkw{here}} } }.$$
The naive algorithm would produce an infinite loop when applied to a triple $(C,J,S_{\key{loop}})$, for any $C$ and
$J$, which reflects the fact that any proof tree whose root is $\SJudg{C}{J}{S_{\key{loop}}}\Ret{r}(r,\ps)$ is infinite.

The JSON Schema specifications forbid any schema which may generate infinite proof trees.
Pezoa et al. \citep{DBLP:conf/www/PezoaRSUV16} formalized this constraint for {\cJS} as follows
(we use the terminology of \citep{DBLP:journals/pvldb/AttoucheBCGSS22}). 


\begin{definition}[Unguardedly references in \cJS; well-formed schema]\label{def:urc}
Given a closed {\cJS} schema $S$, 
a subschema~$S_i$ of~$S$ ``unguardedly references'' a subschema~$S_j$ of~$S$ if
the following three conditions hold:
\begin{enumerate}
\item $\qdref :  \key{absURI}\catHash\key{f}$ is a keyword of a subschema $S'_i$ of $S_i$
  (that is,  $\qdref :  \key{absURI}\catHash\key{f}$ is one of the fields of the object $S'_i$);
\item  every keyword (if any) in the path from $S_i$ to $S'_i$ is a boolean applicator ($S'_i$ is \emph{unguarded});
\item  $\qdref :  \key{absURI}\catHash\key{f}$ refers to $S_j$, that is: $\Get(\Load(\key{absURI}),\key{f}) = S_j$,
\end{enumerate}
A closed schema $S$ is well-formed if the graph of the ``unguardedly references'' relation 
is acyclic.
\end{definition}

For example, the schema $S_{\key{loop}}$ above unguardedly references itself
(all operators in the path 
to the subschema $\JObj { \qdref : \qkw{here}}$ are boolean)
 hence it is not well-formed;
instead, the reference in the following schema
is guarded by a $\qprops$ keyword, hence the schema is well-formed.

\begin{querybox}{}
\begin{lstlisting}[style=query,escapechar=Z]
{  "$id": "http://mjs.ex/gd", "properties": { "data": { "$ref": "http://mjs.ex/gd" }}}
\end{lstlisting}
\end{querybox}

Observe that this is an over-conservative criterion: a schema containing
an unguarded cycle that is unreachable from the root would be judged ill-formed,
but would never generate
infinite proofs.

We can extend this definition to {\mJS} as follows. Observe that the new form of (3) means that,
while a static keyword $\qdref :  \key{absURI}\catHash\key{f}$ ``unguardedly references'' exactly one schema, a dynamic
keyword $\qddref :  \key{absURI}\catHash\key{f}$ ``unguardedly references'' all the schemas that define $f$
as a dynamic anchor, regardless of their $\key{URI}$.

\begin{definition}[Unguardedly references for {\mJS}, well-formed]
Given a closed {\mJS} schema $S$,
a subschema $S_i$ of $S$ ``unguardedly references'' a subschema~$S_j$ of~$S$ if either
$S_i$ ``unguardedly references'' subschema~$S_j$ accordingly to Definition~\ref{def:urc}, or if:
\begin{enumerate}
\item $\qddref :  \key{absURI}\catHash\key{f}$ is a keyword of a subschema $S'_i$ of $S_i$;
\item every keyword (if any) in the path from $S_i$ to $S'_i$ is a boolean applicator;
\item $\qdda : \key{f}$ is a keyword of $S_j$.
\end{enumerate}
A closed schema $S$ is well-formed if the graph of the ``unguardedly references'' relation is acyclic.
\end{definition}

For example, consider a closed schema that embeds the two schemas in Figures~\ref{fig2} and~\ref{fig3}.
Let us use $S_{dr}$ to indicate the subschema that immediately
encloses the dynamic reference $\qddref : \qkw{\underline{http://mjs.ex/simple-tree}\#tree}$ in Figure \ref{fig2}.
The subschema $S_{dr}$ unguardedly references
the entire schemas of Figure~\ref{fig2} and of~\ref{fig3},
since both contain a dynamic anchor $\key{tree}$. However, the schema
of Figure~\ref{fig2}
does not unguardedly reference itself, nor the schema of Figure~\ref{fig3},
since its subschema~$S_{dr}$ is \emph{guarded} by the intermediate $\qprops$ keyword.
Moreover, no other subschema references $S_{dr}$, hence the graph of the \emph{unguardedly
references} relation is acyclic.

When a closed schema $S$ is well formed, then every proof about any subschema of $S$ is finite.

\begin{theoremrep}[Termination]\label{the:termination}
If a closed schema $S$ is well formed, then, for any $C$ that is formed using URIs of $S$, 
for any $J$, $r$, and $\ps$, there exists one and only one proof tree whose
root is $\SJudg{C}{J}{S}\Ret{r}(r,\ps)$, and that proof tree is finite.
\end{theoremrep}

\begin{appendixproof}
We prove a more general property:
assume that a closed schema $S_0$ is well-formed, and consider any subschema $S$ of $S_0$, consider
any keyword $K$ that appears in such a subschema $S$, and consider any context $C$ that is formed using URIs of the schema $S$.
Then:
\begin{itemize}
\item there exists one unique finite proof tree whose root is 
$\SJudg{C}{J}{S}\Ret{r}(r,\ps)$ (1)
\item and there exists one unique finite proof tree whose root is 
$\KJudg{C}{J}{K}\Ret{r}(r,\ps)$ (2).
\end{itemize}

In order to prove (1) and (2), we first consider the graph $G$ of the ``unguardedly references'' relation for $S_0$,
and we define the degree $d(S)$ of any subschema $S$ of $S_0$ as the length of the longest
path in $G$ that starts from $S$; $d(S)$ is well-defined since $G$ is acyclic by hypothesis;
the degree of any keyword $d(K)$ is defined as the
degree of the schema that immediately encloses the keyword.

Hence, we can prove our result by simultaneous lexicographic induction on $(J, d(O),|O|)$,
where $O = S$ for property (1) and $O = K$ for property (2). 

We analyze all operators and we observe that, for every judgment $\SJudg{C}{J}{S}\Ret{r}(r,\ps)$
or $\KJudg{C}{J}{K}\Ret{r}(r,\ps)$, there is always exactly one rule that can be applied.

For the structural applicators, we conclude by induction on $J$.

For the referencing applicators $\qdref$ and $\qddref$, the premises are applied to the same $J$, and we conclude by induction of $d(O)$.

For a boolean applicator $K$, the premises are applied to the same $J$, and all the schemas $S'$ that are arguments of the applicator
are unguarded, hence their degree is either smaller than $d(K)$, or equal. When it is strictly smaller, we conclude by induction
on $d(S')<d(K)$. When $d(S')=d(K)$, we conclude because $|S'| < |K|$.
\end{appendixproof}

From now on, we will assume that our rules are only applied to well-formed schemas, so that every proof-tree is guaranteed to be finite.
Of course, all schemas that we will use in our examples will be well-formed.

\section{PSPACE hardness: using dynamic references to encode a QBF sentence}\label{sec:hardness}

Validation is usually regarded as a low-cost test to be embedded in efficient processes such as distributed function invocation: The server
declares the expected schema of its parameters and, for every invocation, each parameter is validated by the declared schema. 
Hence, data-definition languages are usually designed in order to get a high expressive power with a low validation cost.
In practice, one would like the validation problem to run in polynomial time in the worst case.
This is the case for {\cJS}, whose validation is P-complete \cite{DBLP:conf/www/PezoaRSUV16}.
The input of the validation problem is a couple $(J,S)$, for which we ask whether $\SJudg{\List{\key{baseURI}}}{J}{S}\Ret{\btrue}(\btrue,\ps)$,
(where $\key{baseURI}$ is the base URI of $S$),
for some $\ps$;
hence, for {\cJS}, the time bound is a polynomial function $f$ whose argument
is the total input size, $|J|+|S|$.

Unfortunately, this is not the case for {\mJS}.
Dynamic references add a seemingly minor twist to the validation rules, but this twist has a dramatic effect on the computational
complexity of validation:
we prove here that dynamic references make validation PSPACE-hard.
Following \cite{sipser}, we recall here that PSPACE is the class of decision problems that can be solved in polynomial \emph{space} by a deterministic Turing machine; a decision problem B is PSPACE-hard if every problem in PSPACE is polynomial time reducible to B.
It is well known that $P \subseteq NP \subseteq PSPACE$. While there is still no proof that the inclusions are proper,
no reduction from $PSPACE$ to $NP$, or to $P$, is known, hence $PSPACE$-hard problems are currently regarded as ``intractable''.

We prove that validation is PSPACE-hard by reducing quantified Boolean formulas (QBF) validity, a well-known PSPACE-complete problem \citep{DBLP:conf/stoc/StockmeyerM73},
to JSON Schema validation.  In detail, we encode an arbitrary closed QBF formula~$\psi$ as a schema 
$S_{\psi}$ whose size is linear in $\psi$ and with the property that, given any {\json} instance $J$, the assertion 
$\SJudg{\List{\key{baseURI}}}{J}{S_{\psi}}$ returns $(\btrue,\ES)$
if, and only if,  the formula $\psi$ is valid. 

Observe that the actual value of $J$ is irrelevant: in our encoding, the schema $S_{\psi}$ is either satisfied by
any instance, or by none at all: $S_{\psi}$ is a \emph{trivial} schema, where \emph{trivial} indicates a schema
that returns the same result when applied to any instance $J$, as happens for the schemas
$\atrue$ and $\afalse$.
Hence, we actually prove that the validation problem is PSPACE-hard even when restricted to trivial schemas only.

We start with an example.
Consider the following QBF formula: $\forall x1.\ \exists x2.\ (x1 \And x2) \Or (\Not x1 \And \Not x2)$;
Figure \ref{fig:encoding} shows how it can be encoded as a {\JS} schema (we use here URIs based on \qkw{urn:} rather than on \qkw{https:}, for space reasons).


\begin{figure}
\begin{querybox}{}
\begin{lstlisting}[style=query,escapechar=Z]
{ "id": "urn:psi",
  "$schema": "https://json-schema.org/draft/2020-12/schema",
  "allOf": [ { "$ref": "urn:truex1#afterq1" }, { "$ref": "urn:falsex1#afterq1" }],   Z\label{l:allof}Z
  "$defs": {
    "urn:truex1": {      Z\label{l:btruex1}Z
      "$id": "urn:truex1",
      "$defs": {
        "x1": Z\label{l:bvars1}Z { "$dynamicAnchor": "x1", "anyOf": [true] },
        "not.x1":  { "$dynamicAnchor": "not.x1", "anyOf": [false] },Z\label{l:evars1}Z
        "afterq1":
           { "$anchor": "afterq1",
              "anyOf": [ { "$ref": "urn:truex2#afterq2" },  { "$ref": "urn:falsex2#afterq2" }]
    }}},    Z\label{l:etruex1}Z
    "urn:falsex1": {              Z\label{l:bfalsex1}Z
      "$id": "urn:falsex1",
      "$defs": {
        "x1":  { "$dynamicAnchor": "x1", "anyOf": [false] },
        "not.x1":  { "$dynamicAnchor": "not.x1", "anyOf": [true] },
        "afterq1":
           { "$anchor": "afterq1",
              "anyOf": [ { "$ref": "urn:truex2#afterq2" },  { "$ref": "urn:falsex2#afterq2" }]
    }}},    Z\label{l:efalsex1}Z
    "urn:truex2": {      Z\label{l:btruex2}Z
      "$id": "urn:truex2",
      "$defs": {
        "x2":  Z\label{l:bvars2}Z { "$dynamicAnchor": "x2", "anyOf": [true] },
        "not.x2":  { "$dynamicAnchor": "not.x2", "anyOf": [false] },Z\label{l:evars2}Z
        "afterq2":  { "$anchor": "afterq2", "$ref": "urn:phi#phi" }
    }},      Z\label{l:etruex2}Z
    "urn:falsex2": {       Z\label{l:bfalsex2}Z
      "$id": "urn:falsex2",
      "$defs": {
        "x2":  { "$dynamicAnchor": "x2", "anyOf": [false] },
        "not.x2":  { "$dynamicAnchor": "not.x2", "anyOf": [true] },
        "afterq2":   { "$anchor": "afterq2", "$ref": "urn:phi#phi" }
    }},      Z\label{l:efalsex2}Z
    "urn:phi": {   
      "$id": "urn:phi",
      "$defs": {
        "phi": {    Z\label{l:bphi}Z
           "$anchor": "phi", 
           "anyOf": [
             { "allOf":[{ "$dynamicRef": "Z\underline{urn:truex1}Z#x1" }, { "$dynamicRef": "Z\underline{urn:truex2}Z#x2" }]},
             { "allOf":[{ "$dynamicRef": "Z\underline{urn:truex1}Z#not.x1"}, { "$dynamicRef": "Z\underline{urn:truex2}Z#not.x2"}]}] 
}}}}       Z\label{l:ephi}Z
\end{lstlisting}
\end{querybox}
\caption{Encoding  $\forall x1.\ \exists x2.\ (x1 \And x2) \Or (\Not x1 \And \Not x2)$.} 
\label{fig:encoding}
\end{figure}

For each variable $xi$ we define a resource $\qkw{urn:truex}\cat i$ (lines \ref{l:btruex1}-\ref{l:etruex1} 
and \ref{l:btruex2}-\ref{l:etruex2}  of Figure \ref{fig:encoding}), 
which defines two dynamic schemas,
one with plain-name $\qkw{x}\cat i$ and value $\atrue$, and the other one with plain-name $\qkw{not.x}\cat i$ and value 
$\afalse$\footnote{More precisely, it is $\qkw{\footnotesize anyOf}: [\akw{\footnotesize false}]$,
since we cannot add an anchor to a schema that is just $\akw{\footnotesize false}$; 
$\qkw{\footnotesize anyOf}: [\akw{\footnotesize true}]$ in the body of  $\qkw{\footnotesize x}\cat i$ is clearly redundant, and is there only for readability.}
(lines \ref{l:bvars1}-\ref{l:evars1}  and \ref{l:bvars2}-\ref{l:evars2}).
For each variable $xi$ we also define a resource $\qkw{urn:falsex}\cat i$ 
(lines \ref{l:bfalsex1}-\ref{l:efalsex1}  and \ref{l:bfalsex2}-\ref{l:efalsex2}),
where, on the contrary, $\qkw{x}\cat i$ has value $\afalse$, and $\qkw{not.x}\cat i$ has value $\atrue$.

The formula $\phi$ is encoded in the schema $\qkw{urn:phi\#phi}$  (lines \ref{l:bphi}-\ref{l:ephi}).
All variables in $\qkw{urn:phi\#phi}$ 
are encoded as dynamic references, so that their value depends on the resources that are in-scope when 
$\qkw{urn:phi\#phi}$ is evaluated.
Consider, for example, $\qddref: \qkw{\underline{urn:truex1}\#x1}$ inside $\qkw{urn:phi\#phi}$.
A dynamic anchor \qkw{x1} is defined in 
 $\qkw{urn:truex1}$ and in $\qkw{urn:false1}$,
hence, if the context of the evaluation of $\qkw{urn:phi\#phi}$ contains the URI $\qkw{urn:truex1}$
before $\qkw{urn:false1}$, or without
$\qkw{urn:false1}$,
then $\qddref: \qkw{\underline{urn:truex1}\#x1}$ resolves to the $\atrue$ subschema defined in $\qkw{urn:true1}$. 
If the context of the evaluation contains $\qkw{urn:false1}$
before, or without, $\qkw{urn:true1}$, then $\qddref: \qkw{\underline{urn:truex1}\#x1}$ resolves to the $\afalse$ subschema
defined in $\qkw{urn:falsex1\#x1}$.
Observe that a dynamic reference $\qddref: \qkw{\underline{urn:falsex1}\#x1}$ would behave exactly as
$\qddref: \qkw{\underline{urn:truex1}\#x1}$ --- a fundamental feature of dynamic references is that the 
absolute URI before the \qkw{\#} is substantially irrelevant, a fact that we indicate by underlining it.

Now we describe how we encode the quantifiers.
The first quantifier is encoded in the root schema; if the quantifier is $\forall$, as in this case,
then we apply \qall\ to two references (line \ref{l:allof}),
one that checks whether the rest of the formula holds when $x1$ is true, by invoking \qkw{urn:truex1\#afterq1},
which sets $x1$ to true by bringing \qkw{urn:truex1} in scope,
and the second one that checks whether the rest of the formula holds when $x1$ is false, by invoking \qkw{urn:falsex1\#afterq1},
which sets $x1$ to false by bringing \qkw{urn:falsex1} in scope.

The formulas \qkw{urn:truex1\#afterq1} and  \qkw{urn:falsex1\#afterq1},
which are identical, encode the evaluation,
in two different contexts, of the rest of the formula $\exists x2.\ (x1 \And x2) \Or (\Not x1 \And \Not x2)$.
They encode the existential quantifier (lines 12 and 21) in the same way as the universal one in line \ref{l:allof}, with the only difference that 
\qany\ substitutes \qall, so that
\qkw{urn:truex1\#afterq1} holds if the rest of the formula holds for at least one boolean value of $x2$, when $x1$ is true, 
and similarly for \qkw{urn:falsex1\#afterq1} when $x1$ is false.
This technique allows one to encode any QBF formula with a schema whose size is linear in the size of the formula
$Q_1 x_1 \ldots Q_n x_n. \ \phi$:
the size of \qkw{urn:phi\#phi}  is linear in $|\phi|$, and the rest of the schema is linear in $|Q_1 x_1 \ldots Q_n x_n|$.

Observe that the schema is well-formed:
every maximal path in the unguardedly-references graph has shape
\emph{r-aq1-aq2-phi-var},
where 
\begin{enumerate}
\item \emph{r} is the root schema;

\item$\kw{aq1}$ matches \qkw{urn:(true|false)x1\#afterq1};

\item$\kw{aq2}$ matches \qkw{urn:(true|false)x2\#afterq2};

\item \emph{phi} is \qkw{urn:phi\#phi};

\item  \emph{var} matches \qkw{urn:(true|false)xi\#[not.]xi}.

\end{enumerate}

We now formalize this encoding.

\begin{definition}[$S_{\psi}$]
Consider a generic closed QBF formula: $ \psi = Q_1 x_1 \ldots Q_n x_n. \ \phi$,
where $Q_i\in\Set{\forall,\exists}$ and $\phi$ is generated by:
$$
\phi ::= x_i \M \Not x_i \M \phi \Or \phi \M \phi \And \phi.
$$

The schema $S_{\psi}$ contains $2n+2$ resources: $\qkw{urn:psi}$ (the root), 
 $\qkw{urn:truex}\cat i$ and  $\qkw{urn:falsex}\cat i$, for $i$ in $\SetTo{n}$,
 and
$\qkw{urn:phi}$.

The root encodes the outermost quantifiers  as follows:
\[
\begin{array}{llll}
\key{boolOp}: [\ & \{ \qdref: \qkw{urn:true1\#afterq1} \}\ ,\ 
          &  \{ \qdref: \qkw{urn:false1\#afterq1}  \}
\ ]
\end{array}
\]   
where $\key{boolOp}=\qall$ when $Q_{1}=\forall$,
and $\key{boolOp}=\qany$ when $Q_{1}=\exists$.

Every other resource defines a set of named subschemas,
each containing a $\qda$ or a $\qdda$ keyword that assigns it a name, and one more keyword
that we call its ``body''.

For each
$x_i$, the resource  $\qkw{urn:truex}\cat i$ 
contains $3$ named subschemas:
$\qkw{x}\cat i$, 
$\qkw{not.x}\cat i$, and
$\qkw{afterq}\cat i $.
The body of $\qkw{x}\cat i$ is $\qany: [ \atrue ]$,
and the body of $\qkw{not.x}\cat i$ is $\qany: [ \afalse ]$.

When $i<n$, the body of $\qkw{afterq}\cat i $ encodes
 $Q_{i+1} x_{i}. \ \phi$ as follows:
\[
\begin{array}{llll}
\key{boolOp}: [\ & \{ \qdref: \qkw{urn:truex} \cat i  \cat \qkw{\#afterq} \cat i \}\ ,\ 
          &  \{ \qdref: \qkw{urn:falsex} \cat i  \cat \qkw{\#afterq} \cat i \}
\ ]
\end{array}
\]   
where $\key{boolOp}=\qall$ when $Q_{i+1}=\forall$,
and $\key{boolOp}=\qany$ when $Q_{i+1}=\exists$.
When $i=n$, the body of $\qkw{afterq}\cat i $
is just $\qdref: \qkw{urn:phi\#phi}$.

Finally, the $\qkw{urn:phi}$ resource only contains a schema $\qkw{phi}$,
whose body is $S_{\phi}$, which is recursively defined as follows:
$$
\begin{array}{llllllll}
S_{x_i} &=& \{ \qddref : \qkw{\underline{urn:truex}}\cat \underline{i}\#\qkw{x} \cat i \}  & S_{\phi_1 \Or \phi_2} &=& \qany : [\ S_{\phi_1} , S_{\phi_2} \ ] \\[\NL]
S_{\Not x_i} &=& \{ \qddref : \qkw{\underline{urn:truex}}\cat \underline{i}\#\qkw{not.x} \cat i \} & S_{\phi_1 \And \phi_2} &=& \qall : [\ S_{\phi_1} , S_{\phi_2} \ ]. \\[\NL]
\end{array}
$$
\end{definition}

The following theorem states the correctness of the translation.

\begin{theoremrep}
Given a QBF closed formula $\psi = Q_1 x_1\ldots Q_n x_n.\ \phi$ and the corresponding schema $S_{\psi}$,
$\psi$ is valid if, and only if, for every $J$, $\SJudg{\List{\qkw{urn:psi}}}{J}{S_{\psi}}\Ret{} (\btrue,\ES)$.
\end{theoremrep}

\begin{appendixproof}


Consider a QBF formula $\psi = Q_1 x_1\ldots Q_n x_n.\ \phi$ and the corresponding schema $S_{\psi}$.
We say that a context $C$ is well-formed for $(S_{\psi},i)$ with $i\leq n$ if 
(1)~all of its URIs denote a resource of $S_{\psi}$
and (2)~for every $j\leq i$ either $\qkw{urn:truex}\cat i\in C$ or $\qkw{urn:falsex}\cat i\in C$
holds, but not both
and (3)~for every $j> i$ neither  $\qkw{urn:truex}\cat i\in C$ nor $\qkw{urn:falsex}\cat i\in C$.
We say that a context $C$ is well-formed for $S_{\psi}$ if there exists $i$ such that $C$ is well formed
for $(S_{\psi},i)$. The index of a well-formed context $C$, $\Index(C)$, is the only $i$ such that
$C$ is well formed for  $(S_{\psi},i)$

We associate every well-formed context $C$  to an assignment $A_C$ defined over $x_1,\ldots,x_{\Index(C)}$ as
follows:
$$
\begin{array}{llll}
A_C(x_i)=\btrue \ \Iff\ \qkw{urn:truex}\cat i \in C \\[\NL]
 A_C(x_i)=\bfalse \ \Iff\ \qkw{urn:falsex}\cat i \in C
\end{array}
$$
Given a QBF formula $\psi'$ that may contain open variables and an assignment $A$ that is defined for every open variable of
$\psi'$, we use  $ \Valid(\psi',A)$ to indicate the fact that $\psi'$ is valid when every open variable is substituted with its
value in $A$.


The evaluation of every subschema of $S_{\psi}$ either returns $(\btrue,\ES)$ or $(\bfalse,\ES)$, since no annotation
is generated. We will prove, by induction, the following properties that describe how the boolean returned by some
crucial subschemas are related to the validity of the subformulas of $\psi$, where  we define 
$\AfterQ_i(\psi)=Q_{i+1} x_{i+1}\ldots Q_n x_n.\ \phi$,
so that 
$\AfterQ_0(\psi) = Q_{1} x_{1}\ldots Q_n x_n.\ \phi$
and 
$\AfterQ_{n}(\psi)=\phi$.

$$
\begin{array}{llllllllllllll}
(1a) & \multicolumn{3}{l}{\Index(C) \geq i, \utfx \in \Set{ \qkw{urn:truex}, \qkw{urn:falsex}} \ \Implies}\\ 
     &\ \ \   \SJudg{C}{J}{\{\ \qddref :  \underline{\utfx\cat i}\#\qkw{x}\cat i\ \} }\Ret{} (\btrue,\ES) &\Iff&  A_C(x_i) = \btrue \\[\NL]
(1b) & \multicolumn{3}{l}{\Index(C) \geq i, \utfx \in \Set{ \qkw{urn:truex}, \qkw{urn:falsex}} \ \Implies}\\ 
     & \multicolumn{3}{l}{\ \ \   
                         \SJudg{C}{J}{\{\ \qddref :  \underline{\utfx\cat i}\#\qkw{not.x}\cat i\ \} }\Ret{} (\btrue,\ES) \ \Iff\   A_C(x_i) = \bfalse
                } 
             \\[\NL]
(2) & \Index(C) =n  \ \Implies\\ 
     &\ \ \    \SJudg{C}{J}{\{\ \qdref: \qkw{urn:phi\#phi}\ \}}\Ret{} (\btrue,\ES)
      \ \ &\Iff&\ \  \Valid( \phi,A_{C})  \\[\NL]
(3a) & \multicolumn{3}{l}{\Index(C) = i-1\Implies} \\
     &\ \ \   \SJudg{C}{J}{\{\ \qdref: \qkw{urn:truex}\cat i\#\qkw{afterq}\cat i\ \}}\Ret{} (\btrue,\ES)
      \ \ &\Iff&\ \  \Valid( \AfterQ_{i}(\psi),A_{C}[x_{i}\leftarrow \btrue]) \\[\NL]
(3b) & \multicolumn{3}{l}{\Index(C) = i-1\Implies} \\
     &\ \ \   \SJudg{C}{J}{\{\ \qdref: \qkw{urn:falsex}\cat i\#\qkw{afterq}\cat i\ \}}\Ret{} (\btrue,\ES)
      \ \ &\Iff&\ \  \Valid( \AfterQ_{i}(\psi),A_{C}[x_{i}\leftarrow \bfalse]) \\[\NL]
(4) & \Index(C) =0\Implies \\
     &\ \ \   \SJudg{C}{J}{\{\ \qdref: \qkw{urn:psi}\ \}}\Ret{} (\btrue,\ES)
      \ \ &\Iff&\ \  \Valid( \psi,A_C) \\[\NL]
\end{array}
$$

The theorem follows from case (4), since $\List{\qkw{urn:psi}}$ is well-formed and its index is 0.


Property (1a):
 it holds since either $\qkw{urn:truex} \cat\ i\in C$ or $\qkw{urn:falsex} \cat\ i\in C$ but not both.
In the first case, $ \qddref : \underline{\utfx\cat i}\#\qkw{x}\cat i$ evaluates to $(\btrue,\ES)$ and $A_C(x_i) = \btrue$,
in the second case $ \qddref : \underline{\utfx\cat i}\#\qkw{x}\cat i$ evaluates to $(\bfalse,\ES)$ and $A_C(x_i) = \bfalse$.
Observe that the underlined absolute URI $ \underline{\utfx\cat i}$ is irrelevant when the dynamic reference is evaluated.

Property (1b):
either $\qkw{urn:truex} \cat\ i\in C$ or $\qkw{urn:falsex} \cat\ i\in C$ but not both.
In the first case, $ \qddref : \underline{\utfx\cat i}\#\qkw{not.x}\cat i$ evaluates to $(\bfalse,\ES)$ and $A_C(x_i) = \btrue$,
in the second case $ \qddref :  \underline{\utfx\cat i}\#\qkw{not.x}\cat i$ evaluates to $(\btrue,\ES)$ and $A_C(x_i) = \bfalse$.

Property (2): by induction on ${\phi}$, using property (1) for $x_i$ and $\Not x_i$, and induction for
$S_1 \And S_2$ and $S_1 \Or S_2$ .

Property (3a) and (3b): is proved by induction on $n-i$, so that the base case has $n-i=0$, that is, $i=n$, and,
when applying induction, we can assume that the property holds for every $j > i$.

(3a), base case $i=n$, : we want to prove that 
$$
\begin{array}{lll}
\Index(C) = n-1\Implies   \\
\SJudg{C}{J}{\{\ \qdref: \qkw{urn:truex}\cat n\#\qkw{afterq}\cat n\ \}}\Ret{} (\btrue,\ES)
      \ \ \Iff\ \  \Valid( \AfterQ_{n}(\psi),A_{C}[x_{n}\leftarrow \btrue])
\end{array}
$$
The $\qdref$ keyword evaluates the body of $ \qkw{urn:truex}\lcat n\#\qkw{afterq}\lcat n$ in a context 
$C^{n:\btrue}=C +? \qkw{urn:truex}\cat n $,
whose index is $n$, and which has the property that 
$A_{C^{n:\btrue}} = A_{C}[x_{n}\leftarrow \btrue]$.
The body of $ \qkw{urn:truex}\cat n\#\qkw{afterq}\cat n$
just invokes
$\qkw{urn:phi\#phi}$ in this context, whose index in $n$, and we have that $ \AfterQ_n(\psi) =\phi$. 
Hence, by property (2), we have the following list of equivalences:
$$
\begin{array}{lll}
\SJudg{C}{J}{\{\ \qdref: \qkw{urn:truex}\cat n\#\qkw{afterq}\cat n\ \}}\Ret{} (\btrue,\ES) \\[\NL]
  \ \ \Iff\ \ \SJudg{C^{n:\btrue}}{J}{\{\ \qkw{urn:phi\#phi} \}}\Ret{} (\btrue,\ES)\\[\NL]
  \ \ \Iff\ \ \Valid( \phi,A_{C^{n:\btrue}})\\[\NL]
      \ \ \Iff\ \  \Valid( \AfterQ_{n}(\psi),A_{C}[x_{n}\leftarrow \btrue])
\end{array}
$$

(3b), base case $i=n$, is analogous.

(3a), case $n-i>0$, i.e., $i < n$: we want to prove the following property, 
where $A^{i:\btrue}_C=A_C[x_{i}\leftarrow \btrue]$
$$
\begin{array}{lll}
\Index(C) = i-1 \Implies \\[\NL]
\ \ \SJudg{C}{J}{\{\ \qdref:  \qkw{urn:truex}\cat i\#\qkw{afterq}\cat i\ \}}\Ret{} (\btrue,\ES)
    \ \   \Iff\ \  \Valid( \AfterQ_{i}(\psi),A^{i:\btrue}_C) 
\end{array}
$$
By definition, 
$\AfterQ_{i}(\psi) = Q_{i+1}x_{i+1}.\AfterQ_{i+1}(\psi)$.
We assume that $Q_{i+1}=\forall$; the case for $Q_{i+1}=\exists$ is analogous.

Since  $Q_{i+1}=\forall$, the body of
$\qkw{urn:truex} \cat i\#\qkw{afterq}\cat i$ is  
$$
\begin{array}{lll}
\qall: [\ &\{ \qdref: \qkw{urn:truex} \lcat {(i+1)}  \cat \qkw{\#afterq} \lcat {(i+1)}  \}\ ,\\[\NL]
            &\{ \qdref: \qkw{urn:falsex} \lcat {(i+1)}  \cat \qkw{\#afterq} \lcat {(i+1)}  \}
&\ ]
\end{array}
$$
The body of $\qkw{urn:truex}\cat i\#\qkw{afterq}\cat i$ is evaluated in a context
$C^{i:\btrue} = C+?\qkw{urn:truex} \cat i$, whose index is $i$ and which has the property that
$A_{C^{i:\btrue}} = A^{i:\btrue}_{C}$.

From $Q_{i+1}=\forall$ we get:
$$\begin{array}{lll}
\Valid( \AfterQ_i.(\psi),A^{i:\btrue}_C) \Iff \\[\NL]
\ \  (\Valid(\AfterQ_{i+1}(\psi),A^{i:\btrue}_C[x_{i+1}\leftarrow \btrue]) \And \Valid(\AfterQ_{i+1}(\psi),A^{i:\btrue}_C[x_{i+1}\leftarrow \bfalse]))
\end{array}
$$
The thesis follows since, by induction,  $\qkw{urn:truex} \lcat {(i+1)} \cat \qkw{\#afterq} \lcat {(i+1)}$ evaluated in $C^{i:\btrue}$
returns $(\btrue,\ES)$ if and only if 
$\Valid( \AfterQ_{i+1}(\psi),A_{C^{i:\btrue}}[x_{i+1}\leftarrow \btrue])$,
and  $\qkw{urn:falsex} \lcat {(i+1)} \cat \qkw{\#afterq} \lcat {(i+1)}$ evaluated in $C^{i:\btrue}$
returns $(\btrue,\ES)$ if and only if 
$\Valid( \AfterQ_{i+1}(\psi),A_{C^{i:\btrue}}[x_{i+1}\leftarrow \bfalse])$.

(3b), case $n-i>0$, i.e., $i < n$: same as (3a).

(4): this is analogous to case (3).
We want to prove the following property:
$$
\begin{array}{llllllll}
 \Index(C) =0 &\Implies &\   (\ \SJudg{C}{J}{\{\ \qdref: \qkw{urn:psi}\ \}}\Ret{} (\btrue,\ES)
      \ \ &\Iff&\ \  \Valid( \psi,A_C) \ )
\end{array}
$$
By definition, 
$\psi = Q_{1}x_{1}.\AfterQ_{1}(\psi)$.
We assume that $Q_{1}=\forall$; the case for $Q_{1}=\exists$ is analogous.

Since $Q_{1}=\forall$, the body of
$\qkw{urn:psi}$ is  
$$
\begin{array}{lll}
\qall: [\ &\{ \qdref: \qkw{urn:truex} \lcat {1}  \cat \qkw{\#afterq} \lcat {1}  \}\ ,\\[\NL]
            &\{ \qdref: \qkw{urn:falsex} \lcat {1}  \cat \qkw{\#afterq} \lcat {1}  \}
&\ ]
\end{array}
$$

From $Q_{1}=\forall$ we get:
$$\begin{array}{lll}
\Valid(\psi,A_C) & \Iff &
\ \ (\Valid(\AfterQ_{1}(\psi),A_C[x_{1}\leftarrow \btrue]) \And \Valid(\AfterQ_{1}(\psi),A_C[x_{1}\leftarrow \bfalse]))
\end{array}$$
The thesis follows since, by induction,  $\qkw{urn:truex} \lcat {1} \cat \qkw{\#afterq} \lcat {1}$ evaluated in $C$
returns $(\btrue,\ES)$ if and only if 
$\Valid( \AfterQ_{1}(\psi),A_{C}[x_{1}\leftarrow \btrue])$,
and  $\qkw{urn:falsex} \lcat {1} \cat \qkw{\#afterq} \lcat {1}$ evaluated in $C$
returns $(\btrue,\ES)$ if and only if 
$\Valid( \AfterQ_{1}(\psi),A_{C}[x_{1}\leftarrow \bfalse])$.

\end{appendixproof}

Since the encoding has linear size, PSPACE-hardness is an immediate corollary.
In our encoding we use the eight operators $\qdref$, $\qda$, $\qddref$, 
$\qdda$,
$\qany$, $\qall$, $\atrue$, and $\afalse$, but $\atrue$, $\qdref$, and $\qda$ are only used to improve readability:
every $\qany: [ \atrue ]$ could just be removed, while $\qdref$ and $\qda$ could be substituted by
$\qddref$ and $\qdda$. Hence, only the operators we list below are really needed for our results.

\begin{corollary}[PSPACE-hardness]
Validation in any fragment of {\mJS} that includes $\qddref$, $\qdda$,
$\qany$, $\qall$, and $\afalse$, is PSPACE-hard.
\end{corollary}

\section{Validation is in PSPACE}\label{sec:upperbound}

We present here a polynomial-space validation algorithm, hence proving that the PSPACE bound is tight. 
To this aim, we consider the algorithm that applies the typing rules through recursive calls,
using a list of already-met subproblems in order to cut infinite loops.
{This list could be replaced by a static check of well-formedness,
but we prefer to employ this dynamic approach, since the list is useful for the complexity evaluation.}

For each schema, Algorithm \ref{alg:valid} evaluates its keywords, passing the current value of the boolean result and of the
{\evaluated} children from one keyword to the next.
Independent
keywords (such as $\qany$ and $\qpattProps$) execute their own rule
and update the current result and the current {\evaluated}
items using conjunction and union, as dictated by rule (\rklist-(n+1)),
while each dependent keyword (such as $\qunProps$),
updates these two values as specified by its own rule.

In Algorithm~\ref{alg:valid} we exemplify in-place independent applicators ($\qany$),
 in-place applicators that update the context ($\qddref$),
 structural independent applicators ($\qpattProps$),
and  dependent applicators ($\qunProps$).

\RestyleAlgo{ruled}
\begin{algorithm}
\footnotesize
\caption{Validation\label{alg:valid}
}
\SetKwProg{Fn}{}{}{end}
\SetKw{kwWhere}{where}
\SetKw{Raise}{raise}

\SetKwFunction{SchemaValidate}{SchemaValidate}
\SetKwFunction{KeywordValidate}{KeywordValidate}
\SetKwFunction{Properties}{Properties}
\SetKwFunction{UnProperties}{UnevaluatedProperties}
\SetKwFunction{PattProperties}{PatternProperties}
\SetKwFunction{DynRef}{DynamicRef}
\SetKwFunction{StaticRef}{StaticRef}
\SetKwFunction{AnyOf}{AnyOf}
\SetKwFunction{AndK}{And}
\SetKwFunction{UnionK}{Union}
\SetKwFunction{OrK}{Or}
\SetKwFunction{NamesOf}{NamesOf}
\SetKwFunction{Keywords}{Keywords}
\SetKwFunction{Insert}{Insert}
\SetKwFunction{Present}{Present}
\SetKwFunction{TrueK}{True}
\SetKwFunction{FalseK}{False}
\SetKwFunction{EmptySetK}{EmptySet}
\SetKwFunction{Saturate}{Saturate}
\SetKwFunction{Singleton}{Singleton}
\SetKwFunction{AddK}{Add}

\Fn{\SchemaValidate{Cont,Inst,Schema,StopList}}{ 
\lIf{(Schema == True))}{\Return (\TrueK,\EmptySetK) }
\lIf{(Schema == False))}{\Return (\FalseK,\EmptySetK) }
\tcc{``Input'' represents a triple}
Input := (Cont,Inst,Schema)\;
\lIf{(\Present(Input,StopList))}{
            \Raise (``error: infinite loop'')
}
  \tcc{Res and Eval are initialized, and then updated by each call to \KeywordValidate }
 Res := \TrueK;
 Eval := \EmptySetK\;
  \lFor*{Kw in \Keywords(Schema)}{ (Res,Eval) := \KeywordValidate(Cont, Inst, Kw, Res, Eval, StopList+Input) \;}
\leIf{Result == \TrueK}
  {  \Return (Res,Eval)\;  }
  {  \Return (Res,\EmptySetK)  }
} \mbox{\ \ }\\

\Fn{\KeywordValidate{Cont,Inst, Kw, PrevResult, PrevEval, StopList}}{ 
   \Switch{Kw}{
      \Case{``anyOf'': List}{
          \Return (\AnyOf(Cont,Inst,List,PrevResult,PrevEval,StopList))\;
      }
   \Case{``dynamicRef'': absURI ``\#'' fragmentId}{
          \Return (\DynRef(Cont,Inst,absURI,fragment,PrevResult,PrevEval,StopList))\;
      }
  \ldots
   }
}
\mbox{\ \ }\\

\Fn{\AnyOf{Cont, Inst, List, PrevResult, PrevEval, StopList}}{ 
 Res := \TrueK;
 Eval := \EmptySetK\;
 \For{Schema in List}{
  (SchRes,SchEval) = \SchemaValidate(Cont, Inst, Kw, StopList) \;
     Res := \OrK(Res,SchRes);
     Eval := \UnionK(Eval,SchEval);
  }
  {  \Return (\AndK(PrevResult,Res), \UnionK(PrevEval,Eval))\;  }
} \mbox{\ \ }\\

\Fn{\DynRef{Cont, Inst, AbsURI, fragment, PrevResult, PrevEval, StopList}}{ 
\lIf*{(dget(load(AbsURI),fragment) = bottom))}
                {\Return(\StaticRef(...))\;}
 \lFor*{URI in Cont+AbsURI}{
   \lIf*{(dget(load(URI),fragment) != bottom)}{
          \{ fstURI := URI; break; \}
       }
  } \\
  fstSch ::= get(load(fstURI),fragment)\;
  { (SchRes,SchEval) = \SchemaValidate{\Saturate(Cont,fstURI), Inst,fstSch,StopList}\;  
    \Return (\AndK(PrevResult,SchRes), \UnionK(PrevEval,SchEval))\;  
  } 
}\mbox{\ \ }\\

\Fn{\PattProperties{Cont, Inst, Schema, PrevResult, PrevEval, StopList}}{ 
\lIf{(Inst is not Object)}{
            \Return(\TrueK,\EmptySetK)
}
 Res := \TrueK;
 Eval := \EmptySetK\;
 \For{(name,J) in Inst}{
   \For{(patt,Schema) in Schema}{
     \If{(name matches patt)}{
            (SchRes,Ign) = \SchemaValidate(Cont, J, Schema, StopList) \;
             Res := \AndK(Result,SchRes);
             Eval := \UnionK(\Singleton(name),Eval)\;
         }
     }
  }
  {  \Return (\AndK(PrevResult,Res), \UnionK(PrevEval,Eval))\;  }
}  \mbox{\ \ }\\

\Fn{\UnProperties{Cont, Inst, Schema, PrevResult, PrevEval, StopList}}{ 
\lIf{(Inst is not Object)}{
            \Return(\TrueK,\EmptySetK)
}
 Res := \TrueK\;
 \For{(a,J) in Inst}{
   \If{(a not in PrevEval)}{
          (SchRes,Ign) = \SchemaValidate(Cont, J, Schema, StopList) \;
           Res := \AndK(Result,SchRes)\;
       }
  }
  {  \Return (\AndK(PrevResult,Res), \NamesOf(Inst))\;  }
}


\end{algorithm}

Function \SchemaValidate(\emph{Cont, Inst, Schema, StopList}) applies \emph{Schema} in the context
\emph{Cont}, that is, a list of absolute URIs without repetitions, to \emph{Inst}, and uses \emph{StopList}
in order to avoid infinite recursion. 
The \emph{Cont} list is extended by the evaluation of dynamic and static references using the
function {\Saturate(\emph{Cont, URI})} (line 30),
which adds \emph{URI} to \emph{Cont} only if it is not already there.
The \emph{StopList} records the
(Cont, Inst, Schema) triples that have been met in the current call stack.
It stops the algorithm when the same triple is met twice in the same evaluation branch,
which prevents infinite loops, since any infinite branch
must find the same triple infinitely many times, because every instance and schema that is met
is a subterm of the input, and only finitely many different contexts can be generated.

Now we prove that this algorithm runs in polynomial space. To this aim, the key observation is the fact that we
have a polynomial bound of the length of the call stack.
The call stack is a sequence of alternating tuples 
\SchemaValidate(Cont, Inst, Schema, StopList) - 
\KeywordValidate(\ldots) -
k(...) -
\SchemaValidate(Cont$'$, Inst$'$, Schema$'$, StopList$'$), 
where $k$(\dots) is the keyword-specific function invoked by \KeywordValidate.
We focus on the sequence of \SchemaValidate(Cont, Inst, Schema, StopList) tuples,
ignoring the intermediate calls. This sequence
can be divided in at most~$n$ subsequences, if $n$ is the input size,
the first one with a context that contains only one
URI, the second one with contexts with two URIs, and the last one having a number of URIs that is bound by the input size,
since no URI is repeated twice in a context. In each subsequence all the (Inst, Schema) pairs are different,
since the stoplist test would otherwise raise a failure. Since every instance in a call stack tuple
is a subinstance of the initial one, and every schema is a subschema of the initial one, we have at
most $n^2$ elements in each subsequence, and hence the entire call stack never exceeds $n^3$.
We finally observe that every single function invocation can be executed in polynomial space plus the space used by the functions it invokes, directly and indirectly; the result follows, since these functions are never more
than $O(n^3)$ at the same time.
This is the basic idea behind the following Theorem, whose full proof can be found in the {\FV}.
Since our algorithm runs in polynomial space, the problem of validation
for {\mJS} is PSPACE-complete.

\begin{theoremrep}
For any closed schema $S$ and instance $J$ whose total size is less than $n$, Algorithm~\ref{alg:valid}
applied to $J$ and $S$ requires an amount of space that is polynomial in $n$.
\end{theoremrep}

\begin{appendixproof}
We first prove that the call stack has a polynomial depth, thanks to the following properties,
where we use $n$ for the size of the input instance and $m$ for the size of the input schema.
\begin{enumerate}
\item The depth of the call stack is always less than $|\key{StopList}|\times 3$.
\item Each StopList generated during validation can be divided into $M$ distinct sublists $sl_1,\ldots,sl_M$, such that all elements in the same sublist have the same context, and the context of the sublist $sl_{i+1}$, with $i \geq 1$, contains one more URI than the context of the sublist $sl_{i}$.
\item The size of each sublist $sl_{i}$ is at most $n\times m$.
\end{enumerate}

Property (1) holds since \emph{SchemaValidate(\ldots,StopList)} invokes  \emph{KeywordValidate} with a stoplist
of length $|\key{StopList}|+1$, and \emph{KeywordValidate} invokes a rule-specific function
that may invoke \emph{SchemaValidate} with the same stoplist,
so that the StopList grows by 1 every time the call stack grows by 3, and vice versa.

Property (2) holds by induction: the function  \emph{SchemaValidate} passes to \emph{KeywordValidate} the same context it receives, and 
the function  \emph{KeywordValidate} passes to \emph{SchemaValidate} either the same context it receives or a context that contains one more URI, which happens when the keyword analyzed is either $\qdref$ or $\qddref$ and the target URI was not yet in the context.
 
 Property (3) holds since no two elements of the stoplist can be equal, since a failure is raised when the input is already in the StopList parameter. All elements in a sublist have the same context, and hence they must differ either in the instance or in the schema.
 The instance must be a sub-instance of the input instance, hence we have at most $n$ choices. 
 The schema must be a sub-schema of the input schema, hence we have at most $m$ choices. 
 Hence, the size of each sublist $sl_{i}$ is at most $n\times m$.
 
 The combination of these three properties implies that the StopList parameter contains at most $(n\times m) \times (m+1)$ elements,
 since it can be decomposed into at most $m+1$ sublists, implying that this parameter has a polynomial size and that the call stack, by property~(1), has a polynomial depth.
 
 We now provide the rest of the proof, which is quite straightforward.

%
%
We observe that
\emph{\SchemaValidate(Cont,Inst,Schema,StopList)} scans all keywords inside \emph{Schema} in sequence,
and it only needs enough space to keep the pair \emph{(Result,Eval)} between one call and the other,
and the size of this pair is in $O(n)$, since the list \emph{Eval} of {\evaluated} properties or items cannot be larger than the instance.
Since we reuse the same space for all keywords, we only need to prove that each single keyword can be
recursively evaluated in polynomial space, including the space needed for the call stack and for the parameters.

Every keyword has its own specific algorithm, some of which are exemplified in Algorithm \ref{alg:valid}.

We first analyze \qpattProps. All four parameters can be stored in polynomial space; \emph{Cont} since it is
a list of URIs that belong to the schema and because it does not contain repetition, and the other parameters have already been discussed.
Each matching pair can be analyzed in polynomial space, apart from the recursive call.
For the recursive call, the call stack has a polynomial length, and every called function employs polynomial space.

We need to repeat the same analysis for all rules, but none of them is more complex than \qpattProps.
For example, the dependent keywords such as \qunProps\ have one extra parameter that contains the already {\evaluated} 
properties and items, but it only takes polynomial space, as discussed.

This completes the proof of the fact that the algorithm runs in polynomial space, hence the problem of validation
for JSON Schema 2020-12 is PSPACE-complete.
\end{appendixproof}

\section{Polynomial time validation for static references  and polynomial time data complexity}\label{sec:ptime}

While dynamic references make validation PSPACE-hard, annotation-dependent validation alone does not
change the P complexity of  {\cJS} validation. We prove this fact by restricting our attention to the set of all schemas that contain at most $k$ references, where $k$ is a fixed integer; for this set of schemas we define an optimized variant of Algorithm \ref{alg:valid} that runs in polynomial time in situations where
there is a fixed bound on the maximum number of dynamic references, hence, a fortiori, for schemas where no dynamic
reference is present.

Our optimized algorithm exploits a memoization technique: When, during the computation of $\SJudg{\List{b}}{J}{S} \Ret{} (r,\pk)$, we
complete the evaluation of an intermediate judgment $\SJudg{C'}{J'}{S'} \Ret{} (r',\pk')$, we store this intermediate result.
However, while there is only a polynomial number of $S'$ and $J'$ that may be generated while proving $\SJudg{\List{b}}{J}{S} \Ret{} (r,\pk)$,
there is an exponential number of different $C'$, corresponding to different subsets of URIs that appear in~$S$ and to different reordering
of these subsets; this phenomenon occurs, for example, in our leading example (Figure~\ref{fig:encoding}). 
We solve this problem in the case of a fixed bound on the number of dynamic references by observing that two different contexts~$C_1$ and~$C_2$ are equivalent,
with respect to a specific validation problem $\SJudg{C'}{J'}{S'} \Ret{} (r',\pk')$ , when the two resolve in the same way any dynamic reference that is actually
expanded during the analysis of that specific problem. In the bounded case, this equivalence relation on contexts has a polynomial
number of equivalence classes, which allows us to recompute the result of $\SJudg{C'}{J'}{S'}$, for a fixed pair $J'$, $S'$, only for a polynomial number of different contexts $C'$.

In greater detail, our algorithm returns, for each evaluation of $S$ over $J$ in a context $C$, not only the boolean result
and the {\evaluated} children, but also a \emph{DFragSet}, that returns the set of fragment ids
$f$ such that $\fstURI(\_,f)$ has been computed during that evaluation.
For each evaluated judgment $\SJudgCJ{S} \Ret{} (r,\pk)$,
we add the tuple $(C,J,S,r,\pk,\key{DFragSet})$ to an updatable store.
When, during the same validation, we evaluate again $J$ and $S$ in an arbitrary context $C'$, we retrieve any previous evaluation with the same pair
$(J,S)$, and we verify whether the new $C'$ is equivalent to the context $C$ used for that evaluation, with respect to the
set of fragments that have been actually evaluated, reported in the \emph{DFragSet}; here \emph{equivalent}
means that, for each fragment $f$ in \emph{DFragSet}, $\fstURI(C,f)$ and $\fstURI(C',f)$ coincide.
If the two contexts are equivalent, then we do not recompute the result, but we just return the previous $(r,\pk,\key{DFragSet})$ triple.
It is easy to prove that,
when the number of different dynamic references is bounded, this equivalence relation has a number of equivalence
classes that are polynomial in size of~$S$, hence that memoization limits the total number of
function calls below a polynomial bound.

For simplicity, in our algorithm, we keep the UpdatableStore and the StopList separated; it would not be difficult to
merge them in a single data structure that can be used for the purposes of both.
We show here how $\SchemaValidate$ changes from Algorithm \ref{alg:valid}.
In the {\FV}, we also report how $\KeywordValidate$ is modified.

\RestyleAlgo{ruled}
\begin{algorithm}
\footnotesize
\caption{Polynomial Time Validation\label{alg:ptime}
}
\SetKwProg{Fn}{}{}{end}
\SetKw{kwWhere}{where}
\SetKw{Raise}{raise}

\SetKwFunction{SchemaValidateAS}{SchemaValidateAndStore}
\SetKwFunction{SVAS}{SVAS}
\SetKwFunction{KeywordValidate}{KeywordValidate}
\SetKwFunction{Properties}{Properties}
\SetKwFunction{UnProperties}{UnevaluatedProperties}
\SetKwFunction{PattProperties}{PatternProperties}
\SetKwFunction{DynRef}{DynamicRef}
\SetKwFunction{StaticRef}{StaticRef}
\SetKwFunction{AnyOf}{AnyOf}
\SetKwFunction{AndK}{And}
\SetKwFunction{DTUnionK}{DTUnion}
\SetKwFunction{OrK}{Or}
\SetKwFunction{NamesOf}{NamesOf}
\SetKwFunction{Keywords}{Keywords}
\SetKwFunction{Insert}{Insert}
\SetKwFunction{Present}{Present}
\SetKwFunction{Multiple}{Multiple}
\SetKwFunction{MayBe}{Maybe}
\SetKwFunction{Bottom}{Bottom}
\SetKwFunction{Equivalent}{Equivalent}
\SetKwFunction{FirstURI}{FirstURI}

\tcc{The UpdatableStore maps each evaluated Instance-Schema pair to the list, maybe empty, of all contexts
   where it has been evaluated, each context paired to the associated result}
\Fn{\SchemaValidateAS{Cont,Inst,Schema,StopList,UpdatableStore}}{ 
\lIf{(Schema == True)}{\Return (\TrueK,\EmptySetK,\EmptySetK) }
\lIf{(Schema == False)}{\Return (\FalseK,\EmptySetK,\EmptySetK) }
Input := (Cont,Inst,Schema)\;
\lIf{(\Present(Input,StopList))}{
            \Raise (``error: infinite loop'')
}
 PrevResultSameSchemaInst := UpdatableStore.get(Inst,Schema)\;
  \For{(OldCont,OldRes,OldEval,OldDFragSet) in PrevResultSameSchemaInst}{
        \lIf*{(\Equivalent(Cont,OldCont,OldDFragSet))}{
        \tcc{If the current Cont is equivalent to the OldCont we reuse the old output}
            \Return(OldOutput)\;
       }
  }
 Output := (\TrueK,\EmptySetK,\EmptySetK)\;
 \For{Keyword in \Keywords(Schema)}{
  Output := \KeywordValidate(Cont, Inst, Keyword, Output, StopList+Input,UpdatableStore) \;
  }
  ()
  (Res,Eval,DFragSet) := Output\;
  UpdatableStore.addToList((Inst,Schema),(Cont,Res,Eval,DFragSet))\;
 \lIf{Res == \TrueK}
  {  \Return (Res, Eval, DFragSet) }
  \lElse{   
       \Return (Res, \EmptySetK, DFragSet)}
} \mbox{\ \ }\\

\Fn{\Equivalent(Cont,OldCont,DFragSet)}{ 
 Res := \TrueK\;
 \lFor*{f in DFragSet}{
      Res := \AndK(Res, (\FirstURI(Cont,f) ==\FirstURI(OldCont,f)))\;
  }
  \Return(Result)\;
}

\end{algorithm}

This optimized algorithm returns the same result as the base algorithm and runs in polynomial time if the number of different dynamic fragments is limited by a fixed bound.

\begin{theoremrep}\label{the:ptimecorrect}
Algorithm \ref{alg:ptime} applied to $(C,J,S,\ES,\ES)$ returns $(r,\pk,d)$, for some $d$, if, and only if,
$\SJudgCJ{S} \Ret{} (r,\pk)$.
\end{theoremrep}

\begin{appendixproof}
In the following proof, we use the following metavariables conventions:

\begin{itemize}
\item $sl$ (stoplist) denotes a list of triples $(C,J,S)$;
\item $d$ (DFragSet) is a list of fragments, i.e.\ of dynamic anchors, $\key{f}$;
\item $O$ (Output) is a triple $(r,\pk,d)$;
\item $us$ (UpdatableStore) is a list of quadruples $(C,J,S,O)$.
\end{itemize}
This means that when we say ``for any $sl$\dots'' we actually mean ``for any list $sl$ of tuples $(C,J,S)$\dots''.

We first define an enriched version of the typing rules that defines a judgment
$\SdJudg{C}{J}{S}\Ret{\rl}{(r,\pk,d)}$ that returns in $d$ the fragment names of the dynamic references that are
resolved. The most important rule is $(\rddref_d)$, which adds the resolved $\key{f}$ to the list $d$.

\infrule[$\rddref_d$]
{
\DGet(\Load(\key{absURI}),\key{f}) \neq \bot  \andalso
\key{fURI} = \fstURI(C+?\key{absURI},f) \\[\NL]
S' = \DGet(\Load(\key{fURI}),\key{f}) \andalso
\SJudg{C  +? \key{fURI}\ }{J}{S'}\Ret{r}{(r,\ps,d)}
}
{
 \KJudg{C}{J}{\addref: \key{absURI}\cat\qkw{\#}\cat\key{f}}
 \Ret{r}
 (r,\ps,d \cup \Set{\key{f}} )
}

All the applicators pass the evaluated dynamic references to their result, 
and a schema passes all the evaluated dynamic references even when it fails.
We present here the failing schema rule and the rule for $(\rpattProps_d)$,
$(\rschema\rkw{-false}_d)$, and $(\rklist$-$(n+1)_d)$, the other rules being analogous.

\infrule[$\rpattProps_d$]
{
J = \JObj{k'_1:J_1,\ldots,k'_n:J_n} \andalso
\Set{(i_1,j_1),\ldots,(i_l,j_l)} = \SetST{(i,j)}{k'_i \in \rlan{p_j}} \\[\NL]
\forall q\in \SetTo{l}.\ 
\SJudg{C}{J_{i_q}}{S_{j_q}}\Ret{r_q}{(r_q,\ps_q,d_q)} \andalso
r=\And(\SetIIn{r_q}{q}{\SetTo{l}}) 
}
{
\KdJudg{C}{J}{\apattProps: \JObj{ p_1 : S_1,\ldots,p_m : S_m }} 
 \Ret{r}
 {(r,\Set{k'_{i_1},\ldots,k'_{i_l}},\bigcup_{q\in\SetTo{l}}d_q)}
}

\infrule[$\rschema\rkw{-false}_d$]
{
\KLdJudg{C}{J}{\List{K_1,\ldots,K_n}}\Ret{\List{r_1,\ldots,r_n}}{(\bfalse,\pk,d)} 
}
{\SdJudg{C}{J}{\JObj{K_1,\ldots,K_n}}
 \Ret{r}
 {(\bfalse,\ES,d)}
}

\infrule[$\rklist$-$(n+1)_d$]
{
k:J'\in\key{IKOrT} \andalso
\KLdJudg{C}{J}{\Kl}\Ret{\rl}{(r_l,\pk_l,d_l)} \andalso
\KdJudg{C}{J}{K}\Ret{r}{(r,\pk,d)} \\[\NL]
}
{\KLdJudg{C}{J}{(\Kl\plus K)}
\Ret{\rl\plus r}{(r_l \And r,\pk_l \cup \pk,d_l \cup d)}
}

These new judgments just return some extra information with respect to the basic judgment, and
hence the following facts are immediate.
$$
\begin{array}{llllll}
\KJudg{C}{J}{K}\Ret{\rl}{(r,\pk)}  &\Iff& \exists d.\ \KdJudg{C}{J}{K}\Ret{\rl}{(r,\pk,d)} \\[\NL]
\SJudg{C}{J}{S}\Ret{\rl}{(r,\pk)}   &\Iff& \exists d.\ \SdJudg{C}{J}{S}\Ret{\rl}{(r,\pk,d)}\\[\NL]
\KLJudg{C}{J}{\Kl}\Ret{\rl}{(r,\pk)} &\Iff& \exists d.\ \KLdJudg{C}{J}{\Kl}\Ret{\rl}{(r,\pk,d)}  \\[\NL]
\end{array}
$$

We define an equivalence relation $\Eqdd{C}{C'}$ defined as:
$$
\Eqdd{C}{C'} \ \Iff\ \forall \key{f} \in d.\ 
    \fstURI(C,f) = \fstURI(C',f) 
$$

We prove the following property:
$$
\begin{array}{llllll}
\Eqdd{C}{C'}\ \And\ \KdJudg{C}{J}{K}\Ret{\rl}{(r,\pk,d')}
                     \And d'\subseteq d  &\Implies
                         & \KdJudg{C'}{J}{K}\Ret{\rl}{(r,\pk,d')}\\[\NL]
\Eqdd{C}{C'}\ \And\ \KLdJudg{C}{J}{\Kl}\Ret{\rl}{(r,\pk,d')}
                     \And d'\subseteq d  &\Implies
                         & \KLdJudg{C'}{J}{\Kl}\Ret{\rl}{(r,\pk,d')}\\[\NL]
\Eqdd{C}{C'}\ \And\ \SdJudg{C}{J}{S}\Ret{\rl}{(r,\pk,d')}
                     \And d'\subseteq d  &\Implies
                         & \SdJudg{C'}{J}{S}\Ret{\rl}{(r,\pk,d')}
\end{array}
$$
We prove it by induction on the rules. The proof is immediate for the terminal rules.
For the applicators, consider the rule $(\rpattProps_d)$.

\infrule[$\rpattProps_d$]
{
J = \JObj{k'_1:J_1,\ldots,k'_n:J_n} \andalso
\Set{(i_1,j_1),\ldots,(i_l,j_l)} = \SetST{(i,j)}{k'_i \in \rlan{p_j}} \\[\NL]
\forall q\in \SetTo{l}.\ 
\SJudg{C}{J_{i_q}}{S_{j_q}}\Ret{r_q}{(r_q,\ps_q,d_q)} \andalso
r=\And(\SetIIn{r_q}{q}{\SetTo{l}}) 
}
{
\KdJudg{C}{J}{\apattProps: \JObj{ p_1 : S_1,\ldots,p_m : S_m }} 
 \Ret{r}
 {(r,\Set{k'_{i_1},\ldots,k'_{i_l}},\bigcup_{q\in\SetTo{l}}d_q)}
}

From the hypothesis that $\bigcup_{q\in\SetTo{l}}d_q \subseteq d$
we immediately deduce that for $q\in\SetTo{l}$ we have $d_q \subseteq d$,
hence we can apply the inductive hypothesis to each
$\SJudg{C}{J_{i_q}}{S_{j_q}}\Ret{r_q}{(r_q,\ps_q,d_q)}$, and conclude.
All the rules that do not modify the context are proved in the same way.
We are left with the rules that modify the context, which are $(\rdref_d)$ and $(\rddref_d)$.
This is the rule $(\rdref_d)$.

\infrule[$\rdref_d$]
{
S' = \Get(\Load(\key{absURI}),\key{f}) \andalso
\SdJudg{C  +? \key{absURI}\ }{J}{S'}\Ret{r}{(r,\ps,d)}
}
{
 \KdJudg{C}{J}{\adref: \key{absURI}\catHash\key{f}}
 \Ret{r}
 (r,\ps,d)
}

We now prove that $\Eqdd{C}{C'}$ implies $\Eqdd{C+?\key{absURI}}{C'+?\key{absURI}}$.
Consider any $\key{f}\in d$;
if $\fstURI(C,f)$ are both different from $\bot$ we have:
$$\fstURI(C+?\key{absURI},f) = 
\fstURI(C,f) = 
\fstURI(C',f) = 
\fstURI(C'+?\key{absURI},f) 
$$
If $\fstURI(C,f)$ are both equal to $\bot$, we have:
$$\fstURI(C+?\key{absURI},f) = 
\key{absURI} =
\fstURI(C'+?\key{absURI},f).
$$
Hence, $\Eqdd{C+?\key{absURI}}{C'+?\key{absURI}}$, and we can apply induction to the $(\rdref)$ rule.

Consider finally rule $(\rddref_d)$, applied to $C$ and to $C'$.

\infrule[$\rddref_d$]
{
\DGet(\Load(\key{absURI}),\key{f}) \neq \bot  \andalso
\key{fURI} = \fstURI(C+?\key{absURI},f) \\[\NL]
S_0 = \DGet(\Load(\key{fURI}),\key{f}) \andalso
\SJudg{C  +? \key{fURI}\ }{J}{S_0}\Ret{r}{(r,\ps,d_0)}
}
{
 \KJudg{C}{J}{\addref: \key{absURI}\cat\qkw{\#}\cat\key{f}}
 \Ret{r}
 (r,\ps,d_0 \cup \Set{\key{f}} )
}

\infrule[$\rddref_d$]
{
\DGet(\Load(\key{absURI}),\key{f}) \neq \bot  \andalso
\key{fURI'} = \fstURI(C'+?\key{absURI},f) \\[\NL]
S'_0 = \DGet(\Load(\key{fURI'}),\key{f}) \andalso
\SJudg{C'  +? \key{fURI'}\ }{J}{S'_0}\Ret{r}{(r',\ps',d'_0)}
}
{
 \KJudg{C'}{J}{\addref: \key{absURI}\cat\qkw{\#}\cat\key{f}}
 \Ret{r}
 (r',\ps',d'_0 \cup \Set{\key{f}} )
}

By hypothesis, $(d_0 \cup \Set{\key{f}})\subseteq d$, hence
$\key{f} \in d$. By reasoning as for rule $(\rdref)$,
we deduce that
$ \fstURI(C+?\key{absURI},f)= \fstURI(C'+?\key{absURI},f)$,
hence $\key{fURI}=\key{fURI'}$,
hence $S'_0=S_0$.
By reasoning as for rule $(\rdref)$, $\key{absURI}=\key{absURI'}$
implies that $\Eqdd{C  +? \key{fURI}}{C'  +? \key{fURI'}}$.
Hence, since $d_0 \subseteq (d_0 \cup \Set{\key{f}} ) \subseteq d$, we can apply the induction hypothesis
and deduce that $(r',\ps',d'_0)=(r,\ps,d_0)$, from which we deduce
that 
$$
(r',\ps',d'_0 \cup \Set{\key{f}} ) =  (r,\ps,d_0 \cup \Set{\key{f}} )
$$

Hence, we have proved that the following deduction rule is sound, that is, that its application does not allow
any new judgment to be deduced:

\infrule[$\rkw{reuse}_d$]
{
 \SdJudg{C}{J}{S} \Ret{r} (r,\ps,d) \andalso 
 \Eqd{C}{C'}{d} 
}
{
 \SdJudg{C'}{J}{S}
 \Ret{r}
 (r,\ps,d)
}

Our algorithm applies the rules of $\SdJudg{C}{J}{S}$ plus $(\rkw{reuse}_d)$ that is sound, hence it can only
prove correct statements. 
On the other direction, completeness of the algorithm follows from the fact that it is correct and that it always terminates.
\end{appendixproof}

%

\begin{theoremrep}\label{the:ptime}
Consider a family of closed schemas $S$ and judgments $\J$ such that
$(|S|+|\J|) \leq n$, and let~$D$ be the set of different fragments~$f$ that appear in the argument 
of a $\qddref: \key{initURI}\catHash\key{f}$ in $S$.
Then, Algorithm \ref{alg:ptime} runs on $S$ and $J$ in time $O(n^{k+|D|})$ for some constant $k$.
\end{theoremrep}

\begin{appendixproof}
Consider $S$, $J$, $n$, and $D$, as in the statement of the Theorem.
Consider the call tree of a run of Algorithm \ref{alg:ptime} applied to the tuple $(C,J,S,\ES,\ES)$, with $S$.
The top node of this tree is a call to $\SchemaValidateAS$, from now on abbreviates as
$\SVAS$, applied to $(C,J,S,\ES,\ES)$,
which in turn invokes from 0 to $n$ times
$\KeywordValidate(C,J,K_i,\ldots)$, each applying one specific rule, 
each rule invoking a certain number of times  $\SVAS(C',J',S',sl',up')$.

We first observe that, for any invocation $\SVAS(C',J',S',sl',up')$ 
in the call tree rooted in $\SVAS(C,J,S,\ES,\ES)$, with $S$ closed,
$J'$ is a subtree of $J$, $S'$ is a subschema of $S$, and $C'$ is a list of URIs of resources in
$S$ without repetitions.
We also observe that the call tree is finite, since any infinite call tree would have
an infinite branch, every infinite branch would contain two different invocations of
$\SVAS$ with the same triple $(C,J,S)$, and this possibility is prevented by the
test executed on $sl$.

We observe that every two nodes labeled with $\SVAS(C',J',S',sl',up')$
and $\SVAS(C'',J'',S'',sl'',up'')$ in that call tree either differ in $J'$,
or they differ in $S'$, or they enjoy the property that $\Neqd{C}{C'}{D}$, since the check performed
on $us$ prevents a second call to $\SVAS$ with the same $J$ and $S$ when
$\Eqd{C}{C'}{d'}$, where $d'$ is the \emph{DFragSet} returned for $(C,J,S)$, hence, for any two calls
with the same $(J,S)$, the corresponding $C$ and $C'$ enjoy $\Neqd{C}{C'}{d'}$ for some $d' \subseteq D$,
hence they enjoy $\Neqd{C}{C'}{D}$. 
Each equivalence class of the relation $\Eqd{C}{C'}{D}$ is characterized by a different function that maps each $\key{f}\in D$
to a URI or to $\bot$; if we have $n$ different URIs in $S$, the equivalence relation has at most $(n+1)^{|D|}$
distinct classes, since we have at most $n+1$ possible different choices for each fragment identifier.
We have at most $n$ subschemas of $S$ and subterms of $\J$, and $(n+1)^{|D|}$ different equivalence classes
for $C$, hence the call tree for $\SVAS(C,J,S,\ES,\ES)$ has at most $(n+1)^{|D|}\times n \times n$
nodes labeled with $\SVAS$. The time needed to invoke the at most $n$ instances of
$\SVAS$ that are immediately invoked by a call to $\SVAS$ is polynomial --- 
the most complicated case is $(\rpattProps)$ where each of the $n$ label in $J$ must be matched against
each of the $n$ regular expressions of the patterns of $K$, hence $\SVAS(C,J,S,\ES,\ES)$ runs in time 
$O(n^{k+|D|})$ for some constant $k$.
\end{appendixproof}

\begin{corollary}\label{cor:ptime}
Validation is in P if we fix a constant bound on the maximum number of different  fragments~$f$ that appear in the argument
of a $\qddref: \key{initURI}\catHash\key{f}$ in $S$.
\end{corollary}

\paragraph{P data complexity} 


There are situations where the schema is fixed and has a very small size by comparison to the instance size, hence
it is important to understand how the cost of evaluating $\SJudg{\List{b}}{J}{S}$ depends on the size of $J$,
when $S$ is fixed; this is analogous to the notion of 
\emph{data complexity} that is standard in the database field \citep{DBLP:conf/stoc/Vardi82}.
When the schema is fixed, then, a fortiori, also the number of
different fragments that are argument of $\qddref$ is fixed; hence,
by Corollary \ref{cor:ptime}, the problem of validating
arbitrary instances using any fixed schema is in P.

\begin{corollary}[Fixed-schema complexity]\label{the:datacomplexity}
When $S$ is fixed, the validation problem $\SJudg{\List{b}}{J}{S}\Ret{r}{(r,\ps)}$ 
is in P with respect to $|J|$.
\end{corollary}

Fixed-schema complexity is similar to data complexity in query evaluation, but the parallelism is not
precise: while queries are, in practical cases, almost invariably much smaller than data, there are many situations
where JSON Schema documents are bigger than the checked instances, for example, when
complex schemas are used in order to validate function parameters.

\section{Elimination of dynamic references} 
\label{sec:elimination}


As we have seen, dynamic references change the computational and algebraic properties of JSON Schema.
We define here a process to eliminate dynamic references, by substituting them with static references; 
this allows us to reuse results and algorithms that have been defined for {\cJS}.
Specifically, we will prove here that dynamic references can be 
substituted with static references, at the price of a potentially exponential increase in the 
size of the schema.





A dynamic reference $\qddref: \key{initURI}\catHash\key{f}$ is resolved, during validation, to a
URI reference $\fstURI(C+?\key{initURI},f)\catHash\key{f}$ that depends on the context $C$ of the
validation (Section \ref{sec:refs}), 
so that the same schema $S$ behaves in different ways when applied in different contexts.
This context-dependency extends to static references: A static reference
$\qdref: \key{absURI}\catHash\key{f}$ is always resolved to the same subschema; however, when
this subschema invokes some dynamic reference, directly or through a chain of static references,
then the validation behavior of this subschema depends on the context, as happens with
$\qdref: \qkw{urn:phi\#phi}$ in our example, which is a static reference, but the behavior of the
schema it refers to depends on the context.

To obtain the same effect without dynamic references, we observe that, if the context $C$ is fixed, 
then every dynamic reference has a fixed behavior, and it can be encoded
using a static reference $\qdref:  \fstURI(C+?\key{initURI},f)\catHash\key{f}$.
Every dynamic reference can be eliminated if we iterate this process by defining, for each subschema $S'$ and 
for each context $C$, a context-injected version $\StS(C,S')$, which describes how $S'$ behaves when the
context is $C$.
The context-injected $\StS(C,S')$ is obtained by (1) substituting in $S'$ every dynamic reference
$\qddref: \key{initURI}\catHash\key{f}$ 
with a static reference to the context-injected version of the schema identified by 
$\fstURI(C+?\key{initURI},f)\catHash\key{f}$, and  (2) substituting every static reference
$\qdref: \key{absURI}\catHash\key{f}$ 
with a static reference to the context-injected version of the schema identified by
$\key{absURI}\catHash\key{f}$. Step (2) is crucial, since a static reference may recursively invoke a
dynamic one, hence the context must be propagated through the static references.

Before giving a formal definition of the process that we outlined, we start with an example.
Consider the context 
 $C=\List{\qkw{urn:psi}, \qkw{urn:truex1}, \qkw{urn:falsex2}, \qkw{urn:phi}}$ and a
reference $\qdref: \qkw{urn:phi\#phi}$, which refers to the following schema $S'$,
which contains four dynamic references, and which can be found inside the resource $\qkw{urn:phi}$.
\begin{querybox}{}
\begin{lstlisting}[style=query,escapechar=Z]
{ "$anchor": "phi", 
  "anyOf": [ {"allOf": [{"$dynamicRef": "Z\underline{urn:truex1}Z#x1"}, {"$dynamicRef": "Z\underline{urn:truex2}Z#x2"}]},
             {"allOf": [{"$dynamicRef": "Z\underline{urn:truex1}Z#not.x1"},  {"$dynamicRef": "Z\underline{urn:truex2}Z#not.x2"}]}]}
\end{lstlisting}
\end{querybox}

The corresponding context-injected schema $\StS(C,S')$ is the following.
When a schema is identified by $\key{absURI}\catHash\key{f}$, we identify its 
context-injected version $\StS(C,S')$ using $\key{absURI}\catHash\ECC\cat\key{f}$, 
where $\ECC$ is an invertible encoding of 
$C$ into a plain-name.\footnote{In the example, we encode a sequence of absolute URIs such as 
$\List{\qkwfn{urn:psi}, \qkwfn{urn:truex1}, \qkwfn{urn:falsex2}, \qkwfn{urn:phi}}$ as ${\qkwfn{urn:psi\_urn:truex1\_urn:falsex2\_urn:phi\_}}$,
that is, we escape any underscore inside the URIs (not exemplified here), and we terminate each 
URI with an underscore.} 

\begin{querybox}{}
\begin{lstlisting}[style=query,escapechar=Z]
{ "$anchor": "urn:psi_urn:truex1_urn:falsex2_urn:phi_phi", 
  "anyOf": [
    {"allOf":  [{"$ref":  "urn:truex1#urn:psi_urn:truex1_urn:falsex2_urn:phi_x1" },
                {"$ref":  "urn:falsex2#urn:psi_urn:truex1_urn:falsex2_urn:phi_x2" }]},
    {"allOf":  [{"$ref":  "urn:truex1#urn:psi_urn:truex1_urn:falsex2_urn:phi_not.x1"},
                {"$ref":  "urn:falsex2#urn:psi_urn:truex1_urn:falsex2_urn:phi_not.x2"}
    ]}]},
\end{lstlisting}
\end{querybox}

In this example, it is interesting to observe how the underlined absolute URI $\qkw{\underline{urn:truex2}}$
of $\qddref: \qkw{\underline{urn:truex2}\#x2}$ and
$\qddref: \qkw{\underline{urn:truex2}\#not.x2}$ (lines 3 and 4)
has been substituted with $\qkw{urn:falsex2}$ (lines 4 and 6), which reflects the fact that the context $C$ contains
$\qkw{urn:falsex2}$ but does not contain $\qkw{urn:truex2}$. 
On the other hand,  the underlined absolute URI $\qkw{\underline{urn:truex1}}$ has been substituted with
a static reference to $\qkw{urn:truex1}$, reflecting the presence of $\qkw{urn:truex1}$ in $C$;
the complete unfolding is in the {\FV}.

\hide{ do not delete Observe also that the injected context of the $\qdref$'s in lines 4-7 extends 
${\qkw{urn:psi\_urn:truex1\_}}$ with
$\qkw{urn:phi\_}$. This corresponds to what happens in the rule $(\rddref)$:

\infrule[$\rddref$]
{
\DGet(\Load(\key{absURI}),\key{f}) \neq \bot  \andalso
\key{fURI} = \fstURI(C+?\key{absURI},f) 
\\[\NL]
S' = \DGet(\Load(\key{fURI}),\key{f}) \andalso
\SJudg{C  +? \key{fURI}\ }{J}{S'}\Ret{r}{(r,\ps)}
}
{\KJudg{C}{J}{\addref: \key{absURI}\catHash\key{f}}
  \Ret{r}
 (r,\ps)
}

The rule dictates that the schema $S'$ must be analyzed in the extended context $C^+$, not in the
original context.
Hence, in $\StS(C,S')$, each dynamic reference to $\qkw{urn:phi\#}\cat f'_i$  has been substituted with a
static reference to the context-injected schema
$\StS(C^+,\fstURI(C^+,f'_i)\catHash\key{f'_i})$, whose URI-reference is
$\fstURI(C^+,f'_i)\catHash \Encode{C^+}\cat\key{f'_i}$.
}
%

\hide{
\infrule[\rdref]
{
S'_1 = \Get(\Load(\key{fURI}),\key{\ECC\cat\key{f}}) \andalso
\SJudg{C  +? \key{fURI}\ }{J}{S'_1}\Ret{r}{(r_1,\ps_1)}
}
{
\KJudg{C}{J}{\adref: \key{fURI}\catHash\ECC\cat\key{f}}
  \Ret{r}
 (r_1,\ps_1)
}
}

We can now give a formal definition of the translation process.
For simplicity, we assume that all fragment identifiers referred to by $\qdref$ are plain-names
defined using $\qda$, without loss of generality, since
JSON Pointers can be easily translated using the anchor mechanism.

Given a judgment $S_0$ with base URI $b$,
we first define a ``local'' translation function $\StS$ that maps
every pair $(C,S)$, where $C$ is a list of URIs from $S_0$ without repetitions
and $S$ is a subschema of $S_0$, into a schema $\StS(C,S)$ without dynamic references,
and that maps every pair $(C,K)$ to a keyword $\StK(C,K)$ without dynamic references.
This function maps references as specified below, and acts as a homomorphism on
all the other operators, as exemplified here with $\qany$.
$$
\begin{array}{lllllll}
\StK(C,\qddref: \key{absURI}\catHash\key{f})
 = 
 \qdref: \fstURI(C+?\key{absURI},\key{f}) \catHash \ECC\cat\key{f} \\[\NL]
\StK(C,\qdref: \key{absURI}\catHash\key{f})
 =   \qdref: \key{absURI} \catHash \ECC\cat\key{f} \\[\NL]
\StK(\qany: \JArr{S_1,\ldots,S_n}) 
 =  
 \qany: \JArr{\StS(C,S_1),\ldots,\StS(C,S_n)} \\[\NL]
\ldots
\end{array}
$$

Consider now a schema $S_0$ and the set $\CCC$ of all possible contexts, that is,
of all lists with no repetitions of absolute URIs of resources inside $S_0$;
a \emph{fragment} of $S_0$ is any subschema that is identified by a static or a dynamic anchor
(e.g., the subschema identified by $\qkw{urn:phi\#phi}$ is a fragment).
The static translation of $S_0$, $\StG(S_0)$, is obtained by substituting, 
in $S_0$,  each fragment 
$S_f$ identified by $\key{absURI}\catHash\key{f}$ with many fragments $\ECC\cat\key{f}$, 
one for any context $C\in\CCC$, where the schema identified by each $\key{absURI}\lcatHash\ECC\lcat\key{f}$ is
$\StS(C,S_f)$, as exemplified in the {\FV}. \hide{
We say that a translated fragment $\key{absURI}\catHash\ECC\cat\key{f}$ refers 
another translated fragment if the schema of the first contains a reference to the second;
when we translate a schema $S_0$, we do not actually need to generate every 
possible fragment $\key{absURI}\catHash\ECC\cat\key{f}$, but we only need to translate
those that are recursively reachable from the root fragment
$\key{b}\catHash\Encode{\List{b}}\cat S_0$; this is exemplified in the translation of our
example, in the {\FV}, where we only translate the context-fragment pairs that are reachable.}
If we have $n_U$ absolute URIs in $S_0$, we have
$\Sigma_{i\in\SetFromTo{0}{n_U}}(i!)$ lists of URIs without repetitions, hence, if we have $n_f$ fragments, 
the possible $(C,S_f)$ pairs are $(\Sigma_{i\in\SetFromTo{0}{n_U}}(i!))\times n_f$, which is
included between $n_U!\times n_f$ and $(n_U+1)!\times n_f$.
This exponential expansion was to be expected, since this transformation can be used to reduce the validation problem
of $J$ using $S_0$, 
that is PSPACE-complete with respect to $|J|+|S_0|$, to validation using $\StG(S_0)$, which is P with respect to
$|J|+|\StG(S_0)|$.

We can now prove that this process preserves the schema behaviour. 

\begin{theoremrep}[Encoding correctness]
Let $S$ be a closed schema with base URI $\key{b}$.
Then:
$$\SJudg{\List{\key{b}}}{J}{\StG(S)}\Ret{r}{(r,\ps)}\ \ \Iff\ \ \SJudg{\List{\key{b}}}{J}{S}\Ret{r}{(r,\ps)}$$
\end{theoremrep}

\begin{appendixproof}

For any schema $S_0$ with base URI $b$, we want to prove that

$$\SJudg{\List{\key{b}}}{J_0}{\StG(S_0)}\Ret{r}{(r_0,\ps_0)}\ \ \Iff\ \ \SJudg{\List{\key{b}}}{J_0}{S_0}\Ret{r}{(r_0,\ps_0)}$$

Our validation rules are deterministic, hence for each triple $C,J,S$ or $C,J,K$ we can define its 
``standard proof'', that is just the only proof obtained by the application of the rules.

We then consider the standard proof for 
$$\SJudg{\List{\key{b}}}{J_0}{S_0}\Ret{r}{(r_0,\ps_0)}$$
and we prove that, for any judgment
$\SJudg{C}{J}{S}\Ret{r}{(r,\ps)}$
in that proof, we have that:
$$\SJudg{C}{J}{S}\Ret{r}{(r,\ps)}\ \Iff\ \SJudg{C}{J}{\StS(C,S)}\Ret{r}{(r,\ps)}$$
and similarly that,  for any judgment
$\KJudg{C}{J}{K}\Ret{r}{(r,\pk)}$
in that proof, we have that:
$$\KJudg{C}{J}{K}\Ret{r}{(r,\pk)}\ \Iff\ \KJudg{C}{J}{\StK(C,K)}\Ret{r}{(r,\pk)}.$$

We prove it by mutual induction on the size of the proof.
The base cases are trivial, since $\StK(C,K)$ is equal to $K$ when $K$ is not a reference operator and does not
contain any subschema.
All the applicators different from references, such as $\qany$, are immediate by induction.

The only interesting cases are those for $\qdref$ and $\qddref$.
 
$$
\begin{array}{lllllll}
\StK(C,\qddref: \key{absURI}\catHash\key{f})
& = & \qdref: \fstURI(C+?\key{absURI},\key{f}) \catHash \ECC\cat\key{f} \\[\NL]
\StK(C,\qdref: \key{absURI}\catHash\key{f})
& = & \qdref: \key{absURI} \catHash \ECC\cat\key{f} \\[\NL]
\end{array}
$$

Let us consider the $(\qddref)$ case. Let this be the root of the proof
of $\KJudg{C}{J}{K}\Ret{r}{(r,\pk)}$:

\infrule[\rddref]
{
\DGet(\Load(\key{absURI}),\key{f}) \neq \bot  \andalso
\key{fURI} = \fstURI(C+?\key{absURI},f) \\[\NL]
S' = \DGet(\Load(\key{fURI}),\key{f}) \andalso
\SJudg{C  +? \key{fURI}\ }{J}{S'}\Ret{r}{(r,\ps)}
}
{
\KJudg{C}{J}{\addref: \key{absURI}\catHash\key{f}}
  \Ret{r}
 (r,\ps)
}

By definition, 
$$\StK(C,\qddref: \key{absURI}\catHash\key{f})
 =  \qdref: \fstURI(C+?\key{absURI},\key{f}) \catHash \ECC\cat\key{f}$$
That is
$$\StK(C,\qddref: \key{absURI}\catHash\key{f})
 =  \qdref: \key{fURI} \catHash \ECC\cat\key{f}$$

\medskip
Hence, the proof for $\KJudg{C}{J}{\StK(C,K)}\Ret{r}{(r,\pk)}$ is the following one.

\infrule[\rdref]
{
S'_1 = \Get(\Load(\key{fURI}),\key{\ECC\cat\key{f}}) \andalso
\SJudg{C+?\key{fURI} }{J}{S'_1}\Ret{r}{(r_1,\ps_1)}
}
{
\KJudg{C}{J}{ \adref: \key{fURI} \catHash \ECC\cat\key{f}}
  \Ret{r}
 (r_1,\ps_1)
}

By the definition of $\StG(S_0)$, if $S'$ is the schema identified by 
$\key{fURI} \catHash\key{f}$, then 
$\key{fURI} \catHash \ECC\cat\key{f}$ refers to $\StS(C,S')$,
hence $S_1'=\StS(C,S')$, and we conclude by induction.

The case for $(\rdref)$ is similar but simpler.

\end{appendixproof}


\section{Dynamic references in {\DNineteen}: {\qdrecA} and {\qdrRef}}


Dynamic references have first been introduced in {\DNineteen}, in a restricted form, 
where (1) dynamic anchors have no name, which means that they behave as if they all shared a
unique name, and (2) the initial reference of every dynamic reference is the root of its own resource.

In detail, the dynamic references of {\DNineteen} are based on two keywords, $\qdrecA : b$, with $b\in\Set{\atrue,\afalse}$,
and $\qdrRef: \qkw{\#}$.
The keyword $\qdrecA : \atrue$,\footnote{The
keyword $\qdrecA : \afalse$ has no effect at all.} placed at the top-level of its resource, has the same effect as 
$\qdda : \key{RecRoot}$ in {\DTwenty}, where $\key{RecRoot}$ is an arbitrary unique anchor name: it associates a dynamic anchor 
to the entire resource;
please note that a \emph{unique} anchor name $\key{RecRoot}$ must be used 
to interpret every $\qdrecA$ and every $\qdrRef$ in the schema and in all schemas that are reachable from it.
The keyword $\qdrRef: \qkw{\#}$ is equivalent to
$\qddref: \underline{\key{baseURI}}\catHash\key{RecRoot}$, where \key{baseURI} is the base URI of the current resource.
Hence, $\qdrRef: \qkw{\#}$ is a dynamic reference that \emph{initially} refers
to the root of the resource where the keyword is found; for this reason, it is called a dynamic \emph{recursive}
reference.
Here ``initially refers to...'' indicates the target of the static interpretation of $\underline{\key{baseURI}}\catHash\key{RecRoot}$, 
but, at validation time, a dynamic reference is resolved to the outermost resource that defines an anchor with the same name,
which will often not coincide with the static ``initial'' target.

If you consider the schemas of Figures \ref{fig2} and \ref{fig3}, their {\DNineteen} versions are obtained by replacing every
$\qdda : \qkw{tree}$ with $\qdrecA : \atrue$, and every $\qddref: \qkw{...\#tree}$ with $\qdrRef: \qkw{\#}$
--- \qkw{tree} plays the role of \key{RecRoot}.
Similarly, the {\DNineteen} version of the
metaschema in Figure \ref{fig:schema} is obtained by replacing every
$\qdda : \qkw{meta}$ with $\qdrecA : \atrue$, and every $\qddref: \qkw{...\#meta}$ with $\qdrRef: \qkw{\#}$.

Hence, the two restrictions of {\DNineteen} do not prevent dynamic references from being used for their most important application,
 the representation of JSON Schema metaschema as a collection of many fragments.
However, the presence of only one dynamic anchor name (which we indicate as \key{RecRoot}) means that every dynamic reference will be resolved,
in a fixed dynamic context, to the same target, the root of the first resource in the context that has $\qdrecA: \atrue$ at its root,
which makes it very complicated to mix different uses of dynamic references into a single project,
as we crucially do in our encoding of QBF.
\hide{For example,
if one validates refinable trees defined as in the {\DNineteen} version of Figure \ref{fig3} in a dynamic context where
composable metaschemas, as in the {\DNineteen} version of Figure \ref{fig:schema}, are already in scope, then
every dynamic reference $\qdrRef: \qkw{\#}$ found inside the definition of a refinable tree is actually interpreted as a
reference to a composable metaschema.}

%
The unique-name restriction drastically reduces the expressive power of the mechanism, but
it makes validation polynomial: since a schema in {\DNineteen} corresponds to a schema in {\DTwenty} 
with a single dynamic anchor name then, by Corollary \ref{cor:ptime}, its validation time is in P.

\begin{corollary}\label{cor:nineteen}
Validation of a JSON instance by a schema that respects {\DNineteen} is in P.
\end{corollary}

\section{Experiments}\label{sec:experiments}

We implemented Algorithm~\ref{alg:ptime} for the entire JSON Schema
language in Scala, applying the rules described here and in the {\FV}.

%

%


\subsection{Correctness of Formalization}

We applied our algorithm to the official JSON Schema test suite~\citep{testsuite}.\footnote{We only focus on \emph{main} schemas and do not consider the \emph{optional} ones, and use the version with git commit hash 6afa9b3.}
\hide{
We pass all tests apart from those that contain references to external files (concerning 24 schemas out of 345, at the time of writing).
These references must be resolved at runtime, a feature that we do not yet support. \ST{TODO}
This is a technical limitation that we plan to overcome within the upcoming weeks.
}
Out of a total of 1,210 tests, we pass all apart from 14 pertaining to schemas with the following unsupported features: special characters in patterns or in URI references, 
unknown keywords, a  vocabulary different from \jsonsch\/, a decimal with a high precision. 
\hide{We also ignore the 16 tests pertaining to a non-valid format problem: Our implementation, which follows the original Draft 2020-12, correctly identifies these tests as invalid while a recent update to the standard suggests format problems to generate an annotation only, while considering the instance to be valid. We are planning to integrate this modification into our rules and implementation.
}

This experiment shows that the rules that we presented, and which are faithfully reflected by our algorithm,
are correct and complete w.r.t.\ the standard test suite. 

\subsection{Complexity}

%
Validation for {\mJS} is PSPACE-complete in the presence of dynamic references, while it is in P
when dynamic references are bound by a constant; 
this is not something that may be proved by a finite set of experiments, but we already provided a formal proof for this.

In the upcoming experiment, we test a rich set of validators, 
in order to see
(1)~whether there exist test cases where the PSPACE-hardness result is reflected by 
considerable 
validation times with small schemas and instances
and (2) whether the difference between schemas with a variable number of variables and with a fixed number of variables
is visible, by this set of validators.

\paragraph{Schemas.} We define three families of artificial schemas, 
designed in order to stress-test a generic JSON validator, and we validate the 
instance \emph{null} against each of them; the schemas are satisfied by any instance.
The schemas can be inspected online at \protect\url{https://github.com/sdbs-uni-p/mjs-schemas}.

The \emph{dyn schema} family comprises schemas \qkw{dyn1.js} up to 
\qkw{dyn100.js} and generalizes our running example;
the file \qkw{dyn}$\lcat i\lcat$\qkw{.js} contains the encoding of 
\[
\begin{array}{llllll}
$$\forall x_1.\ \exists x_2.\ \ldots 
\forall x_{2i-1}.\ \exists x_{2i}.\ 
 ((x_{1} \And x_2) \Or (\Not x_{1} \And \Not x_2)) 
\And \ldots \And 
 ((x_{2i-1} \And x_{2i}) \Or (\Not x_{2i-1} \And \Not x_{2i}))
\end{array}
\]
as defined in Section~\ref{sec:hardness}; schema \qkw{dyn}$\lcat i\lcat$\qkw{.js}, hence, contains $i$ pairs of variables.

The corresponding  \emph{stat schema} family comprises
 {\DFour} schemas \qkw{stat1.js} up to \qkw{stat100.js}, where each keyword $\qddref$
is just substituted with the keyword $\qdref$, without applying the expansion we described in Section~\ref{sec:elimination}.

The \emph{dyn\_bounded schema} family, from \qkw{dyn.bounded1.js} up to \qkw{dyn.bounded100.js},
encodes
$$\forall x_1.\ \exists x_2.\ \ldots 
\forall x_{2i-1}.\ \exists x_{2i}.\ 
((x_{1} \And x_{2i}) \Or (\Not x_{1} \And \Not x_{2i})).$$
Hence, the  \emph{dyn\_bounded} schemas only contain four dynamic references, a fact that, according to Corollary \ref{cor:ptime}, allows an optimized algorithm to run in polynomial time.
Observe that the size of \qkw{dyn}$\lcat i\lcat$\qkw{.js}, \qkw{dyn.bonded}$\lcat i\lcat$\qkw{.js}, and \qkw{stat}$\lcat i\lcat$\qkw{.js} schemas grows linearly with $i$. 

\paragraph{Validators.} 

For third-party validators, we employ the meta-validator Bowtie~\citep{bowtie}, which invokes validators encapsulated in Docker containers. 
We tested all 16 different open-source validators currently\footnote{As of July 2023, the time of writing this paper.} provided by Bowtie that support Draft~4 or Draft~2020.
They are written in 11 different programming languages, as detailed 
in Table~3 of the {\FV}.
We also integrated within Bowtie the academic
validator from~\citep{DBLP:conf/www/PezoaRSUV16}, as well as our own implementation.

\paragraph{Execution environment.}
Our execution environment is a 40-core Debian server with 384GB of RAM.\footnote{The 40 cores allow us to run experiments in parallel. However, the experiment may just as well run on a commodity laptop. In fact, in our reproduction package, we evaluate all experiments sequentially, assuming that 2 CPUs and 6MB of memory are assigned to the virtual machine.} Each core runs with with 3.1Gz and CPU frequency set to performance mode.
We are running Docker version 20.10.12, Bowtie version~0.67.0, and Scala version~2.12.

All runtimes were measured as GNU time, averaged over five runs, and include the overhead of invoking Bowtie and Docker.
We overrode the default timeout setting in Bowtie, to allow for longer-running experiments.
%

In the figures showing the measured runtimes, plotted lines terminate when the validator produces a logical validation error, a runtime exception (most commonly, a stack overflow),
or when Bowtie reports \enquote{no response} by the validator.

\begin{figure*}
\centering
\begin{subfigure}[t]{0.195\textwidth}
\includegraphics[scale=.85]{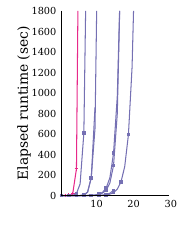}
\caption{\emph{dyn} schemas.} 
\label{fig:exp_dyn}
\end{subfigure}
\hfill
\begin{subfigure}[t]{0.39\textwidth}
\includegraphics[scale=.85]{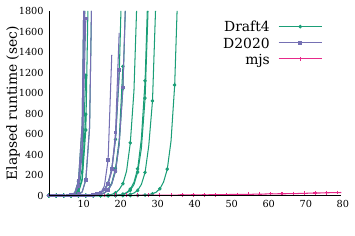}
\caption{\emph{stat} schemas (Draft 4 compatible).} 
\label{fig:exp_stat}
\end{subfigure}
\hfill
\begin{subfigure}[t]{0.39\textwidth}
\includegraphics[scale=.85]{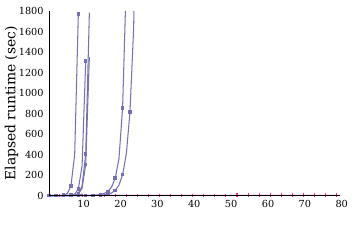}
\begin{tikzpicture}[overlay]
    \node at (-1.5, 2) {
     \resizebox{0.5\textwidth}{!}{
	 \renewcommand{\arraystretch}{0.8} 
	  \sffamily
      \begin{tabular}{@{}rrrr@{}}
          \multicolumn{4}{c}{Size of schema file $i$ in kB} \\
        \toprule

        $i$  & \emph{dyn} & \emph{stat} & \emph{dyn\_bounded} \\
        \hline
        1 & 4  & 4 & 4 \\
        10 & 24 & 24 & 20\\
        20 & 44 & 48 & 40\\
        30 & 64 & 72 & 60\\
        40 & 84 & 96 & 76\\
        50 & 104 & 116 & 96\\
        60 & 128 & 140 & 116\\
        70 & 148 & 164 & 132\\
        80 & 168 & 188 & 152\\
        \bottomrule
      \end{tabular} } 
    };
  \end{tikzpicture}
\caption{\emph{dyn\_bounded} schemas.}
\label{fig:exp_bounded}
\end{subfigure}
\caption{Runtimes of validators on the three schema families, where the tics on the horizontal axes denote the index of the schema,
(e.g, 10 is the index of schema  $\qkw{dyn}\cat 10\cat\qkw{.js}$);
the table in Subfigure~\ref{fig:exp_bounded} maps the schema indexes to the actual file sizes in kB.
The tested validators support {\DFour} (green lines) or {\DTwenty} (blue lines). \enquote{mjs} (red line) refers to our implementation.}
\label{fig:all}
\end{figure*}

\paragraph{Results.} 
In Figures~\ref{fig:exp_dyn}-\ref{fig:exp_bounded} we show the results of our evaluation. In all figures, the x-axis indicates the $i$ index of schemas, while the y-axis reports the runtime.
We distinguish the runtimes for the {\DFour} and {\DTwenty} validators, as well as our own prototype implementation, by different line styles.
The tics denote the data points, where data points can lie outside the plotted area.


Results on the \emph{dyn} and \emph{stat} schemas show
that, on this specific example, the difference in the asymptotic complexity of the static and the dynamic versions
is extremely visible for our validator (red line): we see that validation with dynamic references can
become impractical even with reasonably-sized  files (e.g., schema \qkw{stat5.js} counts fewer than 250 lines 
when pretty-printed), while the runtime remains very modest when dynamic references are substituted with static references.
The runtime on the \emph{dyn\_bounded} family reflects Corollary~\ref{cor:ptime}, which shows the effectiveness of the
proposed optimization on this specific example.

The results on the \emph{stat} family strongly suggest that all other validators have chosen to
implement an algorithm that is exponential even when there is no dynamic reference present.
This is not surprising for the validators designed for {\DTwenty}, since we have been the first to describe an algorithm
(Algorithm \ref{alg:ptime}) that runs in polynomial time over the static fragment of {\DTwenty}.
It is somewhat more surprising for the validators for {\DFour}, especially for the one
published as additional material for~\citep{DBLP:conf/www/PezoaRSUV16}: 
for {\DFour}, as for {\cJS} in general, the validation problem is in P, as proved
for the first time in that same paper~\citep{DBLP:conf/www/PezoaRSUV16}.

\paragraph{Discussion.}
This experiment shows that there are families of schemas where the PSPACE-hardness of the problem is
visible (Figure \ref{fig:exp_dyn}), and that the algorithm we describe in~Section \ref{sec:ptime} is extremely effective when 
dynamic references are replaced with static references (Figure \ref{fig:exp_stat}), or limited in number (Figure \ref{fig:exp_bounded}).
In this paper, we focus on worst-case asymptotic complexity, and we do not make claims
about real-world relevance of our algorithm, which is an important issue, but is not in the scope of this paper.



\section{Related Work}\label{sec:relwork}



To the best of our knowledge, {\mJS} has not been formalized before, nor has validation in the presence of dynamic references been studied.

Overviews over schema languages for {\json} can be found 
in~\citep{DBLP:conf/www/PezoaRSUV16,DBLP:conf/edbt/BaaziziCGS19,DBLP:conf/pods/BourhisRSV17,DBLP:conf/sigmod/BaaziziCGS19}.
In \citep{DBLP:conf/www/PezoaRSUV16} Pezoa et al. proposed the first formalization of {\cJS} {\DFour} and studied the complexity of validation. They proved that {\JS} {\DFour} expressive power goes beyond MSO and tree automata, and showed that validation is PTIME-complete. 
They also described and experimentally analyzed a Python validator that exhibits good performance and scalability. 
Their formalization of semantics and validation, however, cannot be extended to modern {\JS} due to the presence of dynamic references and annotation-dependent validation.

In~\citep{DBLP:conf/pods/BourhisRSV17} Bourhis et al.\ refined the analysis of Pezoa et al. They mapped {\cJS} onto an equivalent modal logic, called recursive JSL, and studied the complexity of validation and satisfiability. In particular, they proved that validation for recursive JSL and {\cJS} is PTIME-complete and that it can be solved in $O(|J|^{2}|S|)$ time; then they showed that satisfiability for {\cJS}, i.e., checking whether 
there exists at least one instance that is validated by the input schema,  is EXPTIME-complete for schemas without {\xuniqIt} and is in 2EXPTIME otherwise. Again, their approach does not seem very easy to extend to modern {\JS}, as it relies on modal logic and a very special kind of alternating tree automata. 

While  we are not aware of any other formal study about {\JS} validation, dozens of validators have been designed and implemented in the past (please, see \citep{vallist} for a rather complete list of about 50 implementations). Only some of them (about 21),  like ajv \citep{ajv} and Hyperjump \citep{hyperjump}, support modern {\JS} and dynamic references. These validators 
usually compile schemas to an efficient internal representation, that is later used for validation purposes. ajv, for instance, uses modern code generation techniques and compiles a schema into a specialized validator, designed to support advanced v8 optimization.


Validation has been widely studied in the context of XML data (see \citep{DBLP:journals/tods/MartensNSB06,DBLP:journals/siamcomp/MartensNS09}, for instance). 
However, schema languages for XML are based on regular expressions, while {\JS} exploits record types, recursion, and full boolean logics, and this
makes it very difficult to import techniques from one field to the other.


Schema languages such as {\JS} and type systems for functional languages are clearly related, and
a lot of work has been invested in the analysis of the computational complexity of type checking and type inference
for programming languages and for module systems (we will only cite \citep{DBLP:journals/jfp/HengleinM94}, as an example).
We are well aware of this research field, but we do 
not think that it is related to this specific work, since in that case the focus is on the analysis of \emph{code} while
{\JS} validation analyzes \emph{instances of data structures}.

\iflong{In \citep{DBLP:conf/erlang/EarleFHM14}, Benac Earle et al. present a systematic approach to testing behavioral aspects of Web Services that communicate using {\json} data. In particular, this approach builds a finite state machine capturing the schema describing the exchanged data, but this machine is only used for generating data and is restricted to atomic values, objects and to some form of boolean expressions.}

%
%

\section{Conclusions and Open Problems}\label{sec:conclusion}

{\MJS} introduced annotation-dependent validation and dynamic references, whose exact interpretation
is regarded as difficult to understand \citeoneone, \citefiveseven.
The changes to the evaluation model invalidate the theory developed for {\cJS}.

Here we provide the first published formalization for {\mJS}. This formalization provides a language
to unambiguously describe and discuss the standard, and a tool to understand its subtleties, and
it has been discussed with the community of {\jsonsch} tools developers.
The formalization has been expressed as a Scala program, which passes the tests of the standard {\jsonsch}
validation test suite and is available in the {\FV}.

We use our formalization to study the complexity of validation of {\mJS}. We proved that the problem is 
PSPACE-complete, and that a very small fragment of the language is already PSPACE-hard.
We proved that this increase in asymptotic complexity is caused by dynamic references, while
annotation-dependent validation without dynamic references can be decided in polynomial time. We have
implemented and experimented with an explicit algorithm to this aim.

We defined a technique to eliminate dynamic references, at the price of a potential exponential increase
in the schema size. 

Many interesting problems remain open, such as the definition of a new notion of schema equivalence and inclusion that is compatible
with annotation-dependent validation, the study of its properties, and the study of the computational complexity of the problems of
satisfiability, validity, inclusion, and example generation.

\section*{Acknowledgments}
This work is partly funded by \emph{Deutsche For\-schungs\-gemein\-schaft} (DFG, German Research Foundation) grant \#385808805.

This work is partly supported by the European Union under the scheme HORIZON-INFRA-2021-DEV-02-01 --- Preparatory phase of new ESFRI research infrastructure projects, Grant Agreement n.101079043, ``SoBigData RI PPP: SoBigData RI Preparatory Phase Project''.

This work is partly supported by \emph{Ministero dell'Università e della Ricerca} (MUR, Ministry of University and Research) under the PRIN Project ``BioConceptum'' (grant \#2022AEEKXS).
 
The authors thank Stefan Klessinger and Sajal Jain for integrating academic validators with the Bowtie framework, 
and Thomas Kirz for assisting with the gnuplot visualizations.
The authors thank Julian Bergman, author of the Bowtie framework, for making timeouts configurable for our experiments.

\section*{Data Availability Statement}

We provide our code, scripts, and data within a reproduction package hosted on Zenodo \citep{lyes_attouche_2023_10019663}.

For reuse purposes, the code of our validator is available here \citep{mjsvalidator}. 


\bibliographystyle{ACM-Reference-Format}
\bibliography{references}


\begin{thebibliography}{34}


\ifx \showCODEN    \undefined \def \showCODEN     #1{\unskip}     \fi
\ifx \showDOI      \undefined \def \showDOI       #1{#1}\fi
\ifx \showISBNx    \undefined \def \showISBNx     #1{\unskip}     \fi
\ifx \showISBNxiii \undefined \def \showISBNxiii  #1{\unskip}     \fi
\ifx \showISSN     \undefined \def \showISSN      #1{\unskip}     \fi
\ifx \showLCCN     \undefined \def \showLCCN      #1{\unskip}     \fi
\ifx \shownote     \undefined \def \shownote      #1{#1}          \fi
\ifx \showarticletitle \undefined \def \showarticletitle #1{#1}   \fi
\ifx \showURL      \undefined \def \showURL       {\relax}        \fi
\providecommand\bibfield[2]{#2}
\providecommand\bibinfo[2]{#2}
\providecommand\natexlab[1]{#1}
\providecommand\showeprint[2][]{arXiv:#2}

\bibitem[\protect\citeauthoryear{??}{jsd}{2022}]%
        {jsdev}
 \bibinfo{year}{2022}\natexlab{}.
\newblock \bibinfo{title}{jschon.dev}.
\newblock
\newblock
\urldef\tempurl%
\url{https://jschon.dev}
\showURL{%
\tempurl}
\newblock
\shownote{Online tool. Retrieved 14 October 2022.}


\bibitem[\protect\citeauthoryear{??}{jse}{2022}]%
        {jsever}
 \bibinfo{year}{2022}\natexlab{}.
\newblock \bibinfo{title}{json-everything validator}.
\newblock
\newblock
\urldef\tempurl%
\url{https://json-everything.net/json-schema/}
\showURL{%
\tempurl}
\newblock
\shownote{Online tool. Retrieved 14 October 2022.}


\bibitem[\protect\citeauthoryear{??}{ajv}{2023}]%
        {ajv}
 \bibinfo{year}{2023}\natexlab{}.
\newblock \bibinfo{title}{Ajv JSON Schema validator}.
\newblock
\newblock
\urldef\tempurl%
\url{https://ajv.js.org}
\showURL{%
\tempurl}
\newblock
\shownote{Retrieved 10 January 2023.}


\bibitem[\protect\citeauthoryear{??}{hyp}{2023}]%
        {hyperjump}
 \bibinfo{year}{2023}\natexlab{}.
\newblock \bibinfo{title}{Hyperjump JSON Schema Validator}.
\newblock
\newblock
\urldef\tempurl%
\url{https://json-schema.hyperjump.io/}
\showURL{%
\tempurl}
\newblock
\shownote{Online tool. Retrieved 10 January 2023.}


\bibitem[\protect\citeauthoryear{??}{val}{2023}]%
        {vallist}
 \bibinfo{year}{2023}\natexlab{}.
\newblock \bibinfo{title}{JSON Schema validators}.
\newblock
\newblock
\urldef\tempurl%
\url{https://json-schema.org/implementations.html#validators}
\showURL{%
\tempurl}
\newblock
\shownote{Retrieved 10 January 2023.}


\bibitem[\protect\citeauthoryear{Andrews}{Andrews}{2023}]%
        {modern}
\bibfield{author}{\bibinfo{person}{Henry Andrews}.}
  \bibinfo{year}{2023}\natexlab{}.
\newblock \bibinfo{title}{Modern {JSON} {S}chema}.
\newblock
\newblock
\newblock
\shownote{Available online at {\url{https://modern-json-schema.com/}}.}


\bibitem[\protect\citeauthoryear{Attouche, Baazizi, Colazzo, Ghelli, Sartiani,
  and Scherzinger}{Attouche et~al\mbox{.}}{2022}]%
        {DBLP:journals/pvldb/AttoucheBCGSS22}
\bibfield{author}{\bibinfo{person}{Lyes Attouche},
  \bibinfo{person}{Mohamed~Amine Baazizi}, \bibinfo{person}{Dario Colazzo},
  \bibinfo{person}{Giorgio Ghelli}, \bibinfo{person}{Carlo Sartiani}, {and}
  \bibinfo{person}{Stefanie Scherzinger}.} \bibinfo{year}{2022}\natexlab{}.
\newblock \showarticletitle{Witness Generation for {JSON} Schema}.
\newblock \bibinfo{journal}{\emph{Proc. {VLDB} Endow.}} \bibinfo{volume}{15},
  \bibinfo{number}{13} (\bibinfo{year}{2022}), \bibinfo{pages}{4002--4014}.
\newblock
\urldef\tempurl%
\url{https://www.vldb.org/pvldb/vol15/p4002-sartiani.pdf}
\showURL{%
\tempurl}


\bibitem[\protect\citeauthoryear{Attouche, Baazizi, Colazzo, Ghelli, Sartiani,
  and Scherzinger}{Attouche et~al\mbox{.}}{2023a}]%
        {mjsvalidator}
\bibfield{author}{\bibinfo{person}{Lyes Attouche},
  \bibinfo{person}{Mohamed-Amine Baazizi}, \bibinfo{person}{Dario Colazzo},
  \bibinfo{person}{Giorgio Ghelli}, \bibinfo{person}{Carlo Sartiani}, {and}
  \bibinfo{person}{Stefanie Scherzinger}.} \bibinfo{year}{2023}\natexlab{a}.
\newblock \bibinfo{booktitle}{\emph{{ModernJSONSchemaValidator}}}.
\newblock
\urldef\tempurl%
\url{https://gitlab.lip6.fr/jsonschema/modernjsonschemavalidator}
\showURL{%
\tempurl}


\bibitem[\protect\citeauthoryear{Attouche, Baazizi, Colazzo, Ghelli, Sartiani,
  and Scherzinger}{Attouche et~al\mbox{.}}{2023b}]%
        {lyes_attouche_2023_10019663}
\bibfield{author}{\bibinfo{person}{Lyes Attouche},
  \bibinfo{person}{Mohamed-Amine Baazizi}, \bibinfo{person}{Dario Colazzo},
  \bibinfo{person}{Giorgio Ghelli}, \bibinfo{person}{Carlo Sartiani}, {and}
  \bibinfo{person}{Stefanie Scherzinger}.} \bibinfo{year}{2023}\natexlab{b}.
\newblock \bibinfo{booktitle}{\emph{{Reproduction Package for: Validation of
  Modern JSON Schema: Formalization and Complexity}}}.
\newblock
\urldef\tempurl%
\url{https://doi.org/10.5281/zenodo.10019663}
\showDOI{\tempurl}


\bibitem[\protect\citeauthoryear{Attouche, Baazizi, Colazzo, Ghelli, Sartiani,
  and Scherzinger}{Attouche et~al\mbox{.}}{2023c}]%
        {attouche2023validation-arXiv}
\bibfield{author}{\bibinfo{person}{Lyes Attouche},
  \bibinfo{person}{Mohamed-Amine Baazizi}, \bibinfo{person}{Dario Colazzo},
  \bibinfo{person}{Giorgio Ghelli}, \bibinfo{person}{Carlo Sartiani}, {and}
  \bibinfo{person}{Stefanie Scherzinger}.} \bibinfo{year}{2023}\natexlab{c}.
\newblock \bibinfo{title}{Validation of Modern JSON Schema: Formalization and
  Complexity}.
\newblock
\newblock
\showeprint[arxiv]{2307.10034}~[cs.DB]


\bibitem[\protect\citeauthoryear{Baazizi, Colazzo, Ghelli, and
  Sartiani}{Baazizi et~al\mbox{.}}{2019a}]%
        {DBLP:conf/edbt/BaaziziCGS19}
\bibfield{author}{\bibinfo{person}{Mohamed~Amine Baazizi},
  \bibinfo{person}{Dario Colazzo}, \bibinfo{person}{Giorgio Ghelli}, {and}
  \bibinfo{person}{Carlo Sartiani}.} \bibinfo{year}{2019}\natexlab{a}.
\newblock \showarticletitle{{Schemas And Types For {JSON} Data}}. In
  \bibinfo{booktitle}{\emph{Proc.\ {EDBT}}}. \bibinfo{pages}{437--439}.
\newblock


\bibitem[\protect\citeauthoryear{Baazizi, Colazzo, Ghelli, and
  Sartiani}{Baazizi et~al\mbox{.}}{2019b}]%
        {DBLP:conf/sigmod/BaaziziCGS19}
\bibfield{author}{\bibinfo{person}{Mohamed~Amine Baazizi},
  \bibinfo{person}{Dario Colazzo}, \bibinfo{person}{Giorgio Ghelli}, {and}
  \bibinfo{person}{Carlo Sartiani}.} \bibinfo{year}{2019}\natexlab{b}.
\newblock \showarticletitle{Schemas and Types for {JSON} Data: From Theory to
  Practice}. In \bibinfo{booktitle}{\emph{Proc.\ {SIGMOD} Conference}}.
  \bibinfo{pages}{2060--2063}.
\newblock


\bibitem[\protect\citeauthoryear{Bergman}{Bergman}{2023a}]%
        {bowtie}
\bibfield{author}{\bibinfo{person}{Julian Bergman}.}
  \bibinfo{year}{2023}\natexlab{a}.
\newblock \bibinfo{title}{Bowtie JSON Schema Meta Validator}.
\newblock
\newblock
\urldef\tempurl%
\url{https://github.com/bowtie-json-schema/bowtie}
\showURL{%
\tempurl}
\newblock
\shownote{Online tool. Version 0.67.0.}


\bibitem[\protect\citeauthoryear{Bergman}{Bergman}{2023b}]%
        {testsuite}
\bibfield{author}{\bibinfo{person}{Julian Bergman}.}
  \bibinfo{year}{2023}\natexlab{b}.
\newblock \bibinfo{title}{JSON-Schema-Test-Suite (draft2020-12)}.
\newblock
\newblock
\urldef\tempurl%
\url{https://github.com/json-schema-org/JSON-Schema-Test-Suite/tree/main/tests/draft2020-12}
\showURL{%
\tempurl}


\bibitem[\protect\citeauthoryear{Berners-Lee, Fielding, and
  Masinter}{Berners-Lee et~al\mbox{.}}{2005}]%
        {RFC3986}
\bibfield{author}{\bibinfo{person}{T. Berners-Lee}, \bibinfo{person}{R.
  Fielding}, {and} \bibinfo{person}{L. Masinter}.} \bibinfo{year}{January
  2005}\natexlab{}.
\newblock \bibinfo{booktitle}{\emph{{U}niform {R}esource {I}dentifier ({URI}):
  Generic Syntax}}.
\newblock \bibinfo{type}{{T}echnical {R}eport}. \bibinfo{institution}{Internet
  Engineering Task Force}.
\newblock
\urldef\tempurl%
\url{https://datatracker.ietf.org/doc/html/rfc3986}
\showURL{%
\tempurl}


\bibitem[\protect\citeauthoryear{Blackler}{Blackler}{2022}]%
        {bl}
\bibfield{author}{\bibinfo{person}{Jim Blackler}.}
  \bibinfo{year}{2022}\natexlab{}.
\newblock \bibinfo{title}{JSON Generator}.
\newblock
\newblock
\newblock
\shownote{Available at \url{https://github.com/jimblackler/jsongenerator}.
  Retrieved 19 September 2022.}


\bibitem[\protect\citeauthoryear{Bourhis, Reutter, Su{\'{a}}rez, and
  Vrgoc}{Bourhis et~al\mbox{.}}{2017}]%
        {DBLP:conf/pods/BourhisRSV17}
\bibfield{author}{\bibinfo{person}{Pierre Bourhis}, \bibinfo{person}{Juan~L.
  Reutter}, \bibinfo{person}{Fernando Su{\'{a}}rez}, {and}
  \bibinfo{person}{Domagoj Vrgoc}.} \bibinfo{year}{2017}\natexlab{}.
\newblock \showarticletitle{{{JSON:} Data model, Query languages and Schema
  specification}}. In \bibinfo{booktitle}{\emph{Proc.\ {PODS}}}.
  \bibinfo{pages}{123--135}.
\newblock
\urldef\tempurl%
\url{https://doi.org/10.1145/3034786.3056120}
\showDOI{\tempurl}


\bibitem[\protect\citeauthoryear{Bourhis, Reutter, and Vrgoc}{Bourhis
  et~al\mbox{.}}{2020}]%
        {DBLP:journals/is/BourhisRV20}
\bibfield{author}{\bibinfo{person}{Pierre Bourhis}, \bibinfo{person}{Juan~L.
  Reutter}, {and} \bibinfo{person}{Domagoj Vrgoc}.}
  \bibinfo{year}{2020}\natexlab{}.
\newblock \showarticletitle{{JSON:} Data model and query languages}.
\newblock \bibinfo{journal}{\emph{Inf. Syst.}}  \bibinfo{volume}{89}
  (\bibinfo{year}{2020}), \bibinfo{pages}{101478}.
\newblock
\urldef\tempurl%
\url{https://doi.org/10.1016/j.is.2019.101478}
\showDOI{\tempurl}


\bibitem[\protect\citeauthoryear{Bryan, Zyp, and Nottingham}{Bryan
  et~al\mbox{.}}{2013}]%
        {RFC6901}
\bibfield{author}{\bibinfo{person}{P. Bryan}, \bibinfo{person}{K. Zyp}, {and}
  \bibinfo{person}{M. Nottingham}.} \bibinfo{year}{Aprile 2013}\natexlab{}.
\newblock \bibinfo{booktitle}{\emph{{JavaScript} {O}bject {N}otation ({JSON})
  {P}ointer}}.
\newblock \bibinfo{type}{{T}echnical {R}eport}. \bibinfo{institution}{Internet
  Engineering Task Force}.
\newblock
\urldef\tempurl%
\url{https://www.rfc-editor.org/info/rfc6901}
\showURL{%
\tempurl}


\bibitem[\protect\citeauthoryear{Earle, Fredlund, Herranz{-}Nieva, and
  Mari{\~{n}}o}{Earle et~al\mbox{.}}{2014}]%
        {DBLP:conf/erlang/EarleFHM14}
\bibfield{author}{\bibinfo{person}{Clara~Benac Earle},
  \bibinfo{person}{Lars{-}{A}ke Fredlund}, \bibinfo{person}{{\'{A}}ngel
  Herranz{-}Nieva}, {and} \bibinfo{person}{Julio Mari{\~{n}}o}.}
  \bibinfo{year}{2014}\natexlab{}.
\newblock \showarticletitle{Jsongen: A quickcheck based library for testing
  {JSON} web services}. In \bibinfo{booktitle}{\emph{Proceedings of the
  Thirteenth {ACM} {SIGPLAN} workshop on Erlang, Gothenburg, Sweden, September
  5, 2014}}, \bibfield{editor}{\bibinfo{person}{Laura~M. Castro} {and}
  \bibinfo{person}{Hans Svensson}} (Eds.). \bibinfo{publisher}{{ACM}},
  \bibinfo{pages}{33--41}.
\newblock
\showISBNx{978-1-4503-3038-1}
\urldef\tempurl%
\url{https://doi.org/10.1145/2633448.2633454}
\showDOI{\tempurl}


\bibitem[\protect\citeauthoryear{Galiegue and Zyp}{Galiegue and Zyp}{2013}]%
        {Draft04}
\bibfield{author}{\bibinfo{person}{Francis Galiegue} {and}
  \bibinfo{person}{Kris Zyp}.} \bibinfo{year}{2013}\natexlab{}.
\newblock \bibinfo{booktitle}{\emph{JSON Schema: interactive and non
  interactive validation - draft-fge-json-schema-validation-00}}.
\newblock \bibinfo{type}{{T}echnical {R}eport}. \bibinfo{institution}{Internet
  Engineering Task Force}.
\newblock
\urldef\tempurl%
\url{https://tools.ietf.org/html/draft-fge-json-schema-validation-00}
\showURL{%
\tempurl}


\bibitem[\protect\citeauthoryear{Henglein and Mairson}{Henglein and
  Mairson}{1994}]%
        {DBLP:journals/jfp/HengleinM94}
\bibfield{author}{\bibinfo{person}{Fritz Henglein} {and}
  \bibinfo{person}{Harry~G. Mairson}.} \bibinfo{year}{1994}\natexlab{}.
\newblock \showarticletitle{The Complexity of Type Inference for Higher-Order
  Typed lambda Calculi}.
\newblock \bibinfo{journal}{\emph{J. Funct. Program.}} \bibinfo{volume}{4},
  \bibinfo{number}{4} (\bibinfo{year}{1994}), \bibinfo{pages}{435--477}.
\newblock
\urldef\tempurl%
\url{https://doi.org/10.1017/S0956796800001143}
\showDOI{\tempurl}


\bibitem[\protect\citeauthoryear{Jacobson}{Jacobson}{2021}]%
        {discussion57}
\bibfield{author}{\bibinfo{person}{Mark Jacobson}.}
  \bibinfo{year}{2021}\natexlab{}.
\newblock \bibinfo{title}{The meaning of "additionalProperties" has changed}.
\newblock
\newblock
\newblock
\shownote{Available online at
  {\url{https://github.com/orgs/json-schema-org/discussions/57}}.}


\bibitem[\protect\citeauthoryear{Martens, Neven, and Schwentick}{Martens
  et~al\mbox{.}}{2009}]%
        {DBLP:journals/siamcomp/MartensNS09}
\bibfield{author}{\bibinfo{person}{Wim Martens}, \bibinfo{person}{Frank Neven},
  {and} \bibinfo{person}{Thomas Schwentick}.} \bibinfo{year}{2009}\natexlab{}.
\newblock \showarticletitle{Complexity of Decision Problems for {XML} Schemas
  and Chain Regular Expressions}.
\newblock \bibinfo{journal}{\emph{{SIAM} J. Comput.}} \bibinfo{volume}{39},
  \bibinfo{number}{4} (\bibinfo{year}{2009}), \bibinfo{pages}{1486--1530}.
\newblock
\urldef\tempurl%
\url{https://doi.org/10.1137/080743457}
\showDOI{\tempurl}


\bibitem[\protect\citeauthoryear{Martens, Neven, Schwentick, and Bex}{Martens
  et~al\mbox{.}}{2006}]%
        {DBLP:journals/tods/MartensNSB06}
\bibfield{author}{\bibinfo{person}{Wim Martens}, \bibinfo{person}{Frank Neven},
  \bibinfo{person}{Thomas Schwentick}, {and} \bibinfo{person}{Geert~Jan Bex}.}
  \bibinfo{year}{2006}\natexlab{}.
\newblock \showarticletitle{Expressiveness and complexity of {XML} Schema}.
\newblock \bibinfo{journal}{\emph{{ACM} Trans. Database Syst.}}
  \bibinfo{volume}{31}, \bibinfo{number}{3} (\bibinfo{year}{2006}),
  \bibinfo{pages}{770--813}.
\newblock
\urldef\tempurl%
\url{https://doi.org/10.1145/1166074.1166076}
\showDOI{\tempurl}


\bibitem[\protect\citeauthoryear{Neal}{Neal}{2022}]%
        {discussion1172}
\bibfield{author}{\bibinfo{person}{Oliver Neal}.}
  \bibinfo{year}{2022}\natexlab{}.
\newblock \bibinfo{title}{Ambiguous behaviour of ``additionalProperties'' when
  invalid}.
\newblock
\newblock
\newblock
\shownote{Available online at
  {\url{https://github.com/json-schema-org/json-schema-spec/issues/1172}}.}


\bibitem[\protect\citeauthoryear{Org}{Org}{2022}]%
        {jsonschema}
\bibfield{author}{\bibinfo{person}{JSON~Schema Org}.}
  \bibinfo{year}{2022}\natexlab{}.
\newblock \bibinfo{title}{JSON Schema}.
\newblock
\newblock
\newblock
\shownote{Available at \url{https://json-schema.org}.}


\bibitem[\protect\citeauthoryear{Pezoa, Reutter, Su{\'{a}}rez, Ugarte, and
  Vrgoc}{Pezoa et~al\mbox{.}}{2016}]%
        {DBLP:conf/www/PezoaRSUV16}
\bibfield{author}{\bibinfo{person}{Felipe Pezoa}, \bibinfo{person}{Juan~L.
  Reutter}, \bibinfo{person}{Fernando Su{\'{a}}rez},
  \bibinfo{person}{Mart{\'{\i}}n Ugarte}, {and} \bibinfo{person}{Domagoj
  Vrgoc}.} \bibinfo{year}{2016}\natexlab{}.
\newblock \showarticletitle{Foundations of {JSON} Schema}. In
  \bibinfo{booktitle}{\emph{Proc.\ {WWW}}}. \bibinfo{pages}{263--273}.
\newblock


\bibitem[\protect\citeauthoryear{Sipser}{Sipser}{2012}]%
        {sipser}
\bibfield{author}{\bibinfo{person}{Michael Sipser}.}
  \bibinfo{year}{2012}\natexlab{}.
\newblock \bibinfo{booktitle}{\emph{Introduction to the Theory of Computation -
  {T}hird {E}dition}}.
\newblock \bibinfo{publisher}{Cengage}.
\newblock


\bibitem[\protect\citeauthoryear{Stockmeyer and Meyer}{Stockmeyer and
  Meyer}{1973}]%
        {DBLP:conf/stoc/StockmeyerM73}
\bibfield{author}{\bibinfo{person}{Larry~J. Stockmeyer} {and}
  \bibinfo{person}{Albert~R. Meyer}.} \bibinfo{year}{1973}\natexlab{}.
\newblock \showarticletitle{Word Problems Requiring Exponential Time:
  Preliminary Report}. In \bibinfo{booktitle}{\emph{Proceedings of the 5th
  Annual {ACM} Symposium on Theory of Computing, April 30 - May 2, 1973,
  Austin, Texas, {USA}}}, \bibfield{editor}{\bibinfo{person}{Alfred~V. Aho},
  \bibinfo{person}{Allan Borodin}, \bibinfo{person}{Robert~L. Constable},
  \bibinfo{person}{Robert~W. Floyd}, \bibinfo{person}{Michael~A. Harrison},
  \bibinfo{person}{Richard~M. Karp}, {and} \bibinfo{person}{H.~Raymond Strong}}
  (Eds.). \bibinfo{publisher}{{ACM}}, \bibinfo{pages}{1--9}.
\newblock
\urldef\tempurl%
\url{https://doi.org/10.1145/800125.804029}
\showDOI{\tempurl}


\bibitem[\protect\citeauthoryear{Vardi}{Vardi}{1982}]%
        {DBLP:conf/stoc/Vardi82}
\bibfield{author}{\bibinfo{person}{Moshe~Y. Vardi}.}
  \bibinfo{year}{1982}\natexlab{}.
\newblock \showarticletitle{The Complexity of Relational Query Languages
  (Extended Abstract)}. In \bibinfo{booktitle}{\emph{Proceedings of the 14th
  Annual {ACM} Symposium on Theory of Computing, May 5-7, 1982, San Francisco,
  California, {USA}}}, \bibfield{editor}{\bibinfo{person}{Harry~R. Lewis},
  \bibinfo{person}{Barbara~B. Simons}, \bibinfo{person}{Walter~A. Burkhard},
  {and} \bibinfo{person}{Lawrence~H. Landweber}} (Eds.).
  \bibinfo{publisher}{{ACM}}, \bibinfo{pages}{137--146}.
\newblock
\urldef\tempurl%
\url{https://doi.org/10.1145/800070.802186}
\showDOI{\tempurl}


\bibitem[\protect\citeauthoryear{Wright, Andrews, and Hutton}{Wright
  et~al\mbox{.}}{2019}]%
        {Version09}
\bibfield{author}{\bibinfo{person}{A. Wright}, \bibinfo{person}{H. Andrews},
  {and} \bibinfo{person}{B. Hutton}.} \bibinfo{year}{2019}\natexlab{}.
\newblock \bibinfo{booktitle}{\emph{{JSON} {S}chema Validation: A Vocabulary
  for Structural Validation of JSON -
  draft-handrews-json-schema-validation-02}}.
\newblock \bibinfo{type}{{T}echnical {R}eport}. \bibinfo{institution}{Internet
  Engineering Task Force}.
\newblock
\urldef\tempurl%
\url{https://tools.ietf.org/html/draft-handrews-json-schema-validation-02}
\showURL{%
\tempurl}
\newblock
\shownote{Retrieved 19 September 2022.}


\bibitem[\protect\citeauthoryear{Wright, Andrews, Hutton, and Dennis}{Wright
  et~al\mbox{.}}{2022}]%
        {specs2020}
\bibfield{author}{\bibinfo{person}{A. Wright}, \bibinfo{person}{H. Andrews},
  \bibinfo{person}{B. Hutton}, {and} \bibinfo{person}{G. Dennis}.}
  \bibinfo{year}{2022}\natexlab{}.
\newblock \bibinfo{booktitle}{\emph{{JSON} {S}chema: A {M}edia {T}ype for
  {D}escribing {JSON} {D}ocuments - draft-bhutton-json-schema-01}}.
\newblock \bibinfo{type}{{T}echnical {R}eport}. \bibinfo{institution}{Internet
  Engineering Task Force}.
\newblock
\urldef\tempurl%
\url{https://json-schema.org/draft/2020-12/json-schema-core.html}
\showURL{%
\tempurl}
\newblock
\shownote{Retrieved 15 October 2022.}


\bibitem[\protect\citeauthoryear{Wright, Luff, and Andrews}{Wright
  et~al\mbox{.}}{2017}]%
        {Draft06}
\bibfield{author}{\bibinfo{person}{A. Wright}, \bibinfo{person}{G. Luff}, {and}
  \bibinfo{person}{H. Andrews}.} \bibinfo{year}{2017}\natexlab{}.
\newblock \bibinfo{booktitle}{\emph{JSON Schema Validation: A Vocabulary for
  Structural Validation of JSON - draft-wright-json-schema-validation-01}}.
\newblock \bibinfo{type}{{T}echnical {R}eport}. \bibinfo{institution}{Internet
  Engineering Task Force}.
\newblock
\urldef\tempurl%
\url{https://tools.ietf.org/html/draft-wright-json-schema-validation-01}
\showURL{%
\tempurl}
\newblock
\shownote{Retrieved 19 September 2022.}


\end{thebibliography}


\onecolumn

\begin{toappendix}
\section{Complete formalization of {\mJS}}\label{sec:complete}

We report here a formalization that describes that entire grammar of {\mJS}, {\DTwenty}, and
of all of its typing rules.\subsection{Complete grammar}

\begin{figure}[h!]

$$
\begin{array}{lllllll}
\multicolumn{3}{l}{
q\in \Num, i \in \Int, k \in \Str, \key{absURI} \in \Str, f \in \Str, \key{format} \in \Str, p \in \Str, J \in \semt
} \\[3\NL]
\key{Tp} &::=& \qobject \Mm \qnumber \Mm \qinteger \Mm \qstr \Mm \qarray \Mm \qboolean \Mm \qnull  \\[\NL]
\key{S} &::=
&  \JObjOpen\   \key{IKOrT}
             \ (, \key{IKOrT})^*
             \ (, \key{FLD})^*
             \ (, \key{SLD})^*  
            \ \JObjClose \\[\NL]
          && \ \M\ \JObjOpen\  \key{FLD}
             \ (, \key{FLD})^*
             \ (, \key{SLD})^*  
            \ \JObjClose 
            \ \M\ \JObjOpen\  \key{SLD}
              \ (, \key{SLD})^*  
            \ \JObjClose
            \ \M\  \JObjOpen\   \JObjClose \\[\NL]
           & &  \ \M\  \xtrue \ \M\ \xfalse  \\[\NL]
\key{IKOrT} & ::= & \key{IK} \Mm 
                 \Bb  \key{ITETriple} \Cc \Mm \Bb \key{ContainsTriple} \Cc  \Mm \Bb \key{Other} \Cc \\[\NL]
\key{IK} & ::= &
\Bb \aexmin: q  \Cc \Mm   
\Bb \aexmax: q  \Cc \\ && \Mm
\Bb \amin: q   \Cc \Mm
\Bb \amax: q   \Cc \Mm
\Bb \amof: q   \Cc \\ && \Mm
\Bb \apatt: p   \Cc \Mm
\Bb \aminL: i    \Cc \Mm
\Bb \amaxL: i  \Cc \\ && \Mm
\Bb \aminP: i   \Cc \Mm
\Bb \amaxP: i \Cc \\ && \Mm
\Bb  \areq: \JArr{k_1,\ldots,k_n}\Cc \\ &&\Mm
\Bb  \auniqIt: \xtrue  \Cc \Mm
\Bb \auniqIt: \xfalse \Cc \\ &&\Mm
\Bb \aminIt: i  \Cc \Mm
\Bb \amaxIt: i  \Cc \Mm
\Bb \aformat: \key{format}  \Cc \\ && \Mm
\Bb \adepR:\JObjOpen\  k_1: \JArr{k^1_1,\ldots,k^1_{o_1}}\ldots,k_n: \JArr{k^n_1,\ldots,k^n_{o_n}}\ \JObjClose \Cc \\ && \Mm
\Bb  \aenum: [J_1,\ldots,J_n] \Cc \Mm
\Bb  \aconst: J_c \Cc \\ && \Mm
\Bb  \atype: \key{Tp} \Cc \Mm
\Bb  \atype: [\key{Tp}_1,\ldots,\key{Tp}_n]\Cc \\ && \Mm
\Bb \adid: \key{absURI}  \Cc \Mm   
\Bb  \adref: \key{absURI}\catHash\key{f}  \Cc \Mm
\Bb  \addref: \key{absURI}\catHash\key{f}\Cc \\ && \Mm
\Bb \addefs :  \JObj{k_1: S_1,\ldots,k_n:S_n} \Cc \Mm
\\ && \Mm
\Bb \ada: \emph{plain-name}  \Cc \Mm
\Bb \adda: \emph{plain-name}  \Cc \\ && \Mm
\Bb  \aany: \JArr{S_1,\ldots,S_n} \Cc \Mm
\Bb  \aall: \JArr{S_1,\ldots,S_n} \Cc \\ && \Mm
\Bb  \aone: \JArr{S_1,\ldots,S_n} \Cc \Mm
\Bb  \anot: S \Cc \\ && \Mm
\Bb \aprefIts: \JArr{S_1,\ldots,S_n} \Cc \Mm
\Bb \acont: S  \Cc \\ && \Mm
\Bb \apattProps: \JObj{ p_1 : S_1,\ldots,p_m : S_m }  \Cc \\ && \Mm
\Bb \aprops: \JObj{ k_1 : S_1,\ldots,k_m : S_m }  \Cc \\ && \Mm
\Bb \apropN: S \Cc \Mm
\Bb \adepS:  \JObj{k_1: S_1,\ldots,k_n:S_n}  \Cc \\[\NL]
\key{ITETriple}\!\! & ::= & (\aif : S,)^?\   \athen : S,\  \aelse : S  \\[\NL]
\key{ContainsTriple} & ::= & (\acont : S,)^?\  \aminC : i\  (, \amaxC : i)^?  \\[\NL]
\key{Other} & ::= &   
\Bb \adschema: k \Cc \Mm
\Bb \advoc: k \Cc  \Mm
\Bb \adcomm: k  \Cc \\ && \Mm
\Bb \atitle: k \Cc \Mm
\Bb \adescr: k \Cc \\ &&  \Mm
\Bb \adepr: b \Cc \Mm
\Bb \areadOnly: b \Cc \Mm
\Bb \awriteOnly: b \Cc \\ && \Mm
\Bb \adefault: J \Cc \Mm
\Bb \aexamples: [J_1,\ldots,J_n] \Cc \\ && \Mm
\Bb k: J \Cc \text{\ \ (with $k$ not cited as keyword in any other production)}
 \\[\NL]
\key{FLD}  & ::= &   
\Bb \aaddProps: S \Cc \Mm
\Bb \ait: S \Cc \\
\key{SLD}  & ::= & 
\Bb \aunProps: S  \Cc \Mm
\Bb \aunIts: S \Cc
\end{array}
$$

\caption{Complete grammar of normalized JSON Schema Draft 2020-12.}
\label{fig:fullgrammar}
\end{figure}

{The complete grammar is reported in Figure \ref{fig:fullgrammar}.

This grammar groups the keywords $\qif$-$\qthen$-$\qelse$,
and specifies that the presence of any keyword among
$\aif$-$\athen$-$\aelse$ implies the presence of $\athen$ and $\aelse$,
which is enforced by adding a trivial $\qthen : \{\}$, or $\qelse : \{\}$, when one or both are missing;
this presentation reduces the number of rules needed to formalize $\qif$-$\qthen$-$\qelse$.
In the same way, the grammar groups $\acont$-$\aminC$-$\amaxC$ and imposes the presence
of $\aminC$ when any of the three is present, which is enforced by adding the default
 $\aminC: 1$ when $\aminC$ is missing.
}


\hide{
\adcomm
\adescr
\adefault
\adefs  -----------
\addefs
\atitle
\aname
\atags
\acomment
\aid
\adid
\adschema
\ada
\addref
\adda
}

\subsection{Terminal keywords}\label{sec:typecond}

%
%
%
%


The types and the conditions of the terminal keywords are specified in Table \ref{tab:completetab}.
There, the length operator $|J|$ counts the number of characters of a string, the number of fields of an object,
and the number of elements of an array. \kw{names}(J) extracts the names of an object.
When $p$ is a pattern or a format, we use $\rlan{p}$ to indicate 
the corresponding set of strings.

When  \TypeOf(\key{kw}) is \emph{no type} then the assertion does not have the (\key{kw}\TTriv) rule and does not have the condition
$\TypeOf(J) = \TypeOf(\key{kw})$ in the (\key{kw}) rule. \\ 
\medskip

\begin{table}
\begin{tabular}{| l | l | l |}
\hline 
assertion \key{kw}:J' & \TypeOf(\key{kw}) & \rkw{cond}(J,\key{kw}:J') \\ \hline 
$\aenum: [J_1,\ldots,J_n]$   & \emph{no type} &  $J \in \Set{J_1,\ldots,J_n}$ \\ \hline
$\aconst: J_c $  & \emph{no type} &  $J = J_c$ \\ \hline
$\atype: \rkw{Tp}$ & \emph{no type} &   $\TypeOf(J)= \rkw{Tp}$ \\ \hline
$\atype: [\rkw{Tp}_1,\ldots,\rkw{Tp}_n]$  & \emph{no type} &   $\TypeOf(J) \in \Set{\rkw{Tp}_1,\ldots,\rkw{Tp}_n}$ \\ \hline
\aexmin: q   &  \rnumber &  $J > q$ \\ \hline
\aexmax: q   &  \rnumber &  $J < q$ \\ \hline
\amin: q   &  \rnumber &  J $\geq$ q \\ \hline
\amax: q   &  \rnumber &  J $\leq$ q \\ \hline
\amof: q   &  \rnumber &  $\exists i\in\Int.\ J = i\times q$ \\ \hline
\apatt: p  &  \rstr &  $J \in \rlan{p}$ \\ \hline
\aminL: i   &  \rstr &  $|J| \geq i$ \\ \hline
\amaxL: i   &  \rstr &  $|J| \leq i$ \\ \hline
\aminP: i  & \robject &  $|J| \geq i$  \\ \hline
\amaxP: i  & \robject &  $|J| \leq i$   \\ \hline
$\areq: \JArr{k_1,\ldots,k_n}$  & \robject &  $\forall i.\ k_i\in \kw{names}(J)$   \\ \hline
\auniqIt: \xtrue  & \rarray &  $J=[J_1,\ldots,J_n]\mbox{\ with\ } n\geq 0 $  \\ 
&&  $\And\ \forall i,j.\ 1 \leq i\neq j \leq n \Implies J_i\neq J_j  $  \\ 
\hline
\auniqIt: \xfalse  & \rarray & T  \\ 
\hline
\aminIt: i  & \rarray &  $|J| \geq i$    \\ \hline
\amaxIt: i  & \rarray &  $|J| \leq i$  \\ \hline
\aformat: \key{format}  & \rstr &  $J \in \rlan{\key{format}}  $  \\ \hline
$\adepR:$
      &  &  $\forall i\in\SetTo{n}.\ $  \\ 
 $\quad\JObjOpen\  k_1: \JArr{k^1_1,\ldots,k^1_{o_1}}$ &\robject  & $\qquad k_i \in   \kw{names}(J)$  \\ 
 $\quad\ \ \ldots,k_n: \JArr{k^n_1,\ldots,k^n_{o_n}}\ \JObjClose$ & & $\qquad \Implies \Set{k^i_1,\ldots,k^i_{o_i}} \subseteq \kw{names}(J)$ \\ \hline
\end{tabular}
\caption{Types and conditions for terminal keywords.}
\label{tab:completetab}
\end{table}

\hide{
\subsection{Terminal unconditional assertions}

\medskip
\begin{tabular}{| l | l |}
\hline 
assertion \akw{kw}:J'  & \rkw{cond}(J,\akw{kw}:J') \\ \hline 
$\aenum: [J_1,\ldots,J_n]$   &   $J \in \Set{J_1,\ldots,J_n}$ \\ \hline
$\aconst: J_c $  &  $J = J_c$ \\ \hline
$\atype: \rkw{Tp}$ &   $\TypeOf(J)= \rkw{Tp}$ \\ \hline
$\atype: [\rkw{Tp}_1,\ldots,\rkw{Tp}_n]$  &   $\TypeOf(J) \in \Set{\rkw{Tp}_1,\ldots,\rkw{Tp}_n}$ \\ \hline
\end{tabular}

\infrule[\rkw{kw}]
{
\TypeOf(J) = \TypeOf(\akw{kw}) \andalso
r = \rkw{cond}(J,\akw{kw}\!:\!J')
}
{\KJudgCJ{\akw{kw}\!:\!J'}
 \Ret{r}
 \rkw{kw}(\_)
}
}

\newcommand{\CP}[1]{C^+_{#1}}
\renewcommand{\CP}[1]{C}

\subsection{The boolean applicators}

\infrule[\rany]
{
\forall i\in \SetTo{n}.\ \ \SJudg{\CP{.,i}}{J}{S_i}\Ret{r_i}{(r_i,\ps_i)} \andalso
r=\Or(\SetIIn{r_i}{i}{\SetTo{n}}) 
}
{\KJudg{C}{J}{\aany:[S_1,...,S_n]}
 \Ret{r}
 (r,\bigcup_{i\in\SetTo{n}}\ps_i)
}

\infrule[\rall]
{
\forall i\in \SetTo{n}.\ \SJudg{\CP{.,i}}{J}{S_i}\Ret{r_i}{(r_i,\ps_i)} \andalso
r=\And(\SetIIn{r_i}{i}{\SetTo{n}}) 
}
{\KJudgCJ{\aall:[S_1,...,S_n]}
 \Ret{r}
 (r,\bigcup_{i\in\SetTo{n}}\ps_i)
}

\infrule[\rone]
{
\forall i\in \SetTo{n}.\ \SJudg{\CP{.,i}}{J}{S_i}\Ret{r_i}{(r_i,\ps_i)} \andalso
r = (\ |\SetST{i}{r_i=\btrue}| = 1 \ )
}
{\KJudgCJ{\aone:[S_1,...,S_n]}
 \Ret{r}
 (r,\bigcup_{i\in\SetTo{n}}\ps_i)
}

\infrule[\rnot]
{
\SJudg{\CP{.,.}}{J}{S}\Ret{r}{(r,\ps)} 
}
{\KJudgCJ{\anot:S}
 \Ret{\Not r}
 (\Not r,\ps)
}

\subsection{References}

\infrule[\rdref]
{
S' = \Get(\Load(\key{absURI}),\key{f}) \andalso
\SJudg{C  +? \key{absURI}\ }{J}{S'}\Ret{r}{(r,\ps)}
}
{
 \KJudg{C}{J}{\adref: \key{absURI}\cat\qkw{\#}\cat\key{f}}
 \Ret{r}
 (r,\ps)
}

\infrule[\rddref]
{
\DGet(\Load(\key{absURI}),\key{f}) \neq \bot  \andalso
\key{fURI} = \fstURI(C+?\key{absURI},f) \\[\NL]
S' = \DGet(\Load(\key{fURI}),\key{f}) \andalso
\SJudg{C  +? \key{fURI}\ }{J}{S'}\Ret{r}{(r,\ps)}
}
{
 \KJudg{C}{J}{\addref: \key{absURI}\cat\qkw{\#}\cat\key{f}}
 \Ret{r}
 (r,\ps)
}

\infrule[\rddref\rkw{AsRef}]
{
\DGet(\Load(\key{absURI}),\key{f}) = \bot  \\[\NL]
S' = \Get(\Load(\key{absURI}),\key{f}) \andalso
\SJudg{C  +? \key{absURI}\ }{J}{S'}\Ret{r}{(r,\ps)}
}
{
 \KJudg{C}{J}{\addref: \key{absURI}\cat\qkw{\#}\cat\key{f}}
  \Ret{r}
 (r,\ps)
}

\subsection{If-then-else}\label{sec:iterulescomplete}

\infrule[\rkw{if-true}]
{
\SJudg{\CP{.,.}}{J}{S_i}\Ret{T}{(\btrue,\ps_i)} \andalso
\SJudg{\CP{.,.}}{J}{S_t}\Ret{r_t}{(r,\ps_t)}
}
{
\KLJudgCJ{(
  \aif: S_i \plus
  \athen: S_t \ \plus
  \aelse: S_e\ )}
  \Ret{\rl \plus T \plus r_t}
  {(r,\ps_i \cup \ps_t)}
}

\infrule[\rkw{if-false}]
{
\SJudg{\CP{.,.}}{J}{S_i}\Ret{F}{(F,\ps_i)} \andalso
\SJudg{\CP{.,.}}{J}{S_e}\Ret{r_e}{(r,\ps_e)}
}
{
\KLJudgCJ{(
  \aif: S_i \plus
  \athen: S_t \ \plus
  \aelse: S_e\ )}
  \Ret{\rl \plus T \plus r_e}
  {(r,\ps_i \cup \ps_e)}
}

\infax[\rkw{missing-if}]
{
\KLJudgCJ{(
  \athen: S_t \ \plus
  \aelse: S_e\ )}
  \Ret{}
  {(\btrue,\ES)}
}

\subsection{Object and array applicators: independent applicators}\label{sec:indeprules}

Independent applicators are defined by a set of two rules: (\key{kw}\TTriv) and (\key{kw});
the (\key{kw}\TTriv) rules are specified by the following table; the  (\key{kw}) rules are specified after the table.

\medskip

\begin{tabular}{| l | l | l |}
\hline 
assertion $\key{kw}:J'$ & $\TypeOf(\key{kw})$  \\ \hline 
$\apattProps: \JObj{ p_1 : S_1,\ldots,p_m : S_m }$  & \robject   \\ \hline
$\aprops: \JObj{ k_1 : S_1,\ldots,k_m : S_m }$  & \robject   \\ \hline
$\apropN: S$  & \robject   \\ \hline
$\aprefIts: S$  & \rarray   \\ \hline
$\acont: S$  & \rarray   \\ \hline
$\adepS:  \JObj{k_1: S_1,\ldots,k_n:S_n}$  & \robject   \\ \hline
\end{tabular}

\infrule[\rpattProps]
{
J = \JObj{k'_1:J_1,\ldots,k'_n:J_n} \andalso
\Set{(i_1,j_1),\ldots,(i_l,j_l)} = \SetST{(i,j)}{k'_i \in \rlan{p_j}} \\[\NL]
\forall q\in \SetTo{l}.\ 
\SJudg{\CP{k'_{i_q},k_{j_q}}}{J_{i_q}}{S_{j_q}}\Ret{r_q}{(r_q,\ps_q)} \andalso
r=\And(\SetIIn{r_q}{q}{\SetTo{l}}) 
}
{
\KJudgCJ{\apattProps: \JObj{ p_1 : S_1,\ldots,p_m : S_m }} 
 \Ret{r}
 {(r,\Set{k'_{i_1},\ldots,k'_{i_l}})}
}

\infrule[\rprops]
{
J = \JObj{k'_1:J_1,\ldots,k'_n:J_n} \andalso
\Set{(i_1,j_1),\ldots,(i_l,j_l)} = \SetST{(i,j)}{k'_i=k_j} \\[\NL]
\forall q\in \SetTo{l}.\ 
\SJudg{\CP{k'_{i_q},k_{j_q}}}{J_{i_q}}{S_{j_q}}\Ret{r_q}{(r_q,\ps_q)} \andalso
r=\And(\SetIIn{r_q}{q}{\SetTo{l}})
}
{
\KJudgCJ{\aprops: \JObj{ k_1 : S_1,\ldots,k_m : S_m }} 
 \Ret{r}
 {(r,\Set{k'_{i_1},\ldots,k'_{i_l}})}
}

\infrule[\rpropN]
{
J = \JObj{k_1:J_1,\ldots,k_n:J_n} \andalso
\forall i\in \SetTo{n}.\ 
\SJudg{\CP{k_i,.}}{k_i}{S}\Ret{r_i}{(r_i,\ps_i)} \andalso
r=\And(\SetIIn{r(\ps_i)}{i}{\SetTo{n}}) 
}
{
\KJudgCJ{\apropN:S}\Ret{r}
 {(r,\ES)}
}

\infrule[\rprefIts]
{
J = \JArr{J_1,\ldots,J_m} \andalso
\forall i\in \SetTo{\Min(n,m)}.\ \SJudg{\CP{i,i}}{J_i}{S_i}\Ret{r_i}{(r_i,\ps_i)} \andalso
r=\And(\SetIIn{r_i}{i}{\SetTo{\Min(n,m)}}) 
}
{\KJudgCJ{\aprefIts: \JArr{S_1,\ldots,S_n}}
 \Ret{r}
 {(r,\Set{1,\ldots,\Min(n,m)})}
}

\infrule[\rcont\rkw{-max}]
{
J=\JArr{J_1,...,J_n} \\[\NL]
\forall i\in \SetTo{n}.\ \SJudg{\CP{i,.}}{J_i}{S}\Ret{r_i}{(r_i,\ps_i)} \andalso
\pk_c = \SetST{i}{r_i = \btrue} \andalso
r_c = (i \leq |\pk_c| \leq j) 
}
{\KJudgCJ{(\acont : S\plus\aminC : i\plus\amaxC : j)}
\Ret{\rl\plus r}{(r_c,\pk_c)}
}

\infrule[\rcont\rkw{-no-max}]
{
\text{rule (\rcont\rkw{-max}) does not apply and} \andalso
J=\JArr{J_1,...,J_n} \\[\NL]
\forall i\in \SetTo{n}.\ \SJudg{\CP{i,.}}{J_i}{S}\Ret{r_i}{(r_i,\ps_i)} \andalso
\pk_c = \SetST{i}{r_i = \btrue} \andalso
r_c = (i \leq |\pk_c|) 
}
{\KJudgCJ{(\acont : S\plus\aminC : i)}
\Ret{\rl\plus r}{(r_c,\pk_c)}
}

\infrule[\rkw{missing-contains}]
{
\text{rules (\rcont\rkw{-max}) and (\rcont\rkw{-no-max}) do not apply}
}
{\KJudgCJ{(\aminC : i \ (\plus \amaxC i )^?)}
\Ret{\rl\plus r}{(\btrue,\ES)}
}

\infrule[\rdepS]
{
J = \JObj{k'_1: J_1,\ldots,\, k'_m: J_m} \andalso
\Set{i_1,\ldots,i_l} = \SetST{i}{i\in\SetTo{n},\ k_i \in \Set{k'_1,\ldots,\, k'_m}}\\[\NL]
\forall q\in\SetTo{l}.\ 
                       \SJudg{\CP{.,k_{i_q}}}{J}{S_{i_q}}\Ret{r_q}{\ps_q}\andalso
r=\And(\SetIIn{r_q}{q}{\SetTo{l}}) 
}
{\KJudgCJ{\adepS: \JObj{k_1: S_1,\ldots,k_n: S_n}}
 \Ret{r}
 {(r,\bigcup_{q\in\SetTo{l}}\ps_q)}
}

\hide{ OLD VERSION
\infrule[\rdepR]
{
J = \JObj{k'_1: J_1,\ldots,\, k'_m: J_m} \andalso
\Set{i_1,\ldots,i_l} = \SetST{i}{i\in\SetTo{n},\ k_i \in \Set{k'_1,\ldots,\, k'_m}}\\[\NL]
\forall q\in\SetTo{l}.\ 
r_q = ( \forall j\in \SetTo{o_q}.\ k^p_j \in \Set{k'_1,\ldots,\, k'_m} \andalso
r=\And(\SetIIn{r_q}{q}{\SetTo{l}})
}
{\KJudgCJ{\adepR: \JObj{k_1: \JArr{k^1_1,\ldots,k^1_{o_1}},\ldots,k_n: \JArr{k^n_1,\ldots,k^n_{o_n}}}}
\qquad\qquad\qquad\qquad
\\[\NL]
 \qquad\qquad\qquad\qquad
 \Ret{r}
 \rdepR(\_,\Set{k_{i_1},\ldots k_{i_l}},\Set{r_1,\ldots,r_l})   
}
}

\subsection{Dependent keywords}

Dependent keywords are defined by a set of two rules: (\key{kw}\TTriv) and (\key{kw});
the (\key{kw}\TTriv) rule is specified by the following table.
\medskip

\begin{tabular}{| l | l | l |}
\hline 
assertion $\key{kw}:J'$ & $\TypeOf(\key{kw})$  \\ \hline 
$\aunProps: S$  & \robject   \\ \hline
$\aaddProps: S$  & \robject   \\ \hline
$\aunIts: S$  & \rarray   \\ \hline
$\ait: S$  & \rarray   \\ \hline
\end{tabular}
\mbox{\ \ }\\

$$
\begin{array}{llll}
\akw{propsOf}(\qprops: \JObj{ k_1 : S_1,\ldots,k_m : S_m }) &=& \keykey{k_1} \cat \qkw{|}\cat\ldots \cat \qkw{|}\cat \keykey{k_n} \\[\NL]
\akw{propsOf}(\qpattProps: \JObj{ p_1 : S_1,\ldots,p_m : S_m }) &=& p_1 \cat \qkw{|}\cat\ldots \cat \qkw{|}\cat p_m \\[\NL]
\akw{propsOf}(K) &=& \ES \qquad\qquad\qquad\qquad \mbox{otherwise} \\[\NL]
\akw{propsOf}(\List{K_1,\ldots,K_n}) &=& \akw{propsOf}(K_1) \cat \qkw{|}\cat\ldots \cat \qkw{|}\cat \akw{propsOf}(K_n) 
\end{array}
$$

\infrule[\raddProps]
{
J=\JObj{k_1 : J_1,\ldots,k_n:J_n} \andalso
\KLJudgCJ{\Kl}\Ret{}{(r,\pk)} \\[\NL]
\Set{{i_1},\ldots,{i_l}} = \SetST{i}{1\leq i \leq n \And\ k_i \not\in \rlan{\akw{propsOf}(\Kl)}} \\[\NL]
\forall q\in \SetTo{l}.\ 
\SJudgC{J_{i_q}}{S}\Ret{r_q}{(r_q,\ps_q)} \andalso
r'= \And(\SetIIn{r_q}{q}{\SetTo{l}}) 
}
{
\KLJudgCJ{(\Kl\plus\aaddProps : S)}
\Ret{\rl\plus r}{(r\And r',\Set{k_{1}\ldots,k_{n}})}
}

\infrule[\runProps]
{
J=\JObj{k_1 : J_1,\ldots,k_n:J_n} \andalso
\KLJudgCJ{\Kl}\Ret{\rl}{(r,\pk)} \\[\NL]
\Set{{i_1},\ldots,{i_l}} = \SetST{i}{1\leq i \leq n \And\ k_i \not\in \pk} \\[\NL]
\forall q\in \SetTo{l}.\ 
\SJudgC{J_{i_q}}{S}\Ret{r_q}{(r_q,\ps_q)} \andalso
r'=\And(\SetIIn{r_q}{q}{\SetTo{l}}) 
}
{
\KLJudgCJ{(\Kl\plus\aunProps : S)}
 \Ret{\rl\plus r}{(r\And r',\Set{k_{1}\ldots,k_{n}})}
}

$$
\begin{array}{llll}
\akw{maxPrefixOf}({\Kl}) = 
& m &\mbox{if}\ \ (\qprefIts: \JArr{ S_1,\ldots, S_m }) \in \Kl \ \ \mbox{for some}\ \  S_1,\ldots,S_n \\[\NL]
\akw{maxPrefixOf}({\Kl}) = 
& 0 &\mbox{if}\ \ (\qprefIts: \JArr{ S_1,\ldots, S_m }) \not\in \Kl \ \ \mbox{for any}\ \  S_1,\ldots,S_n \\[\NL]
\end{array}
$$

\infrule[\rits]
{
J=\JArr{J_1,...,J_n} \andalso
\KLJudgCJ{\Kl}\Ret{\rl}{(r,\pk)} \andalso
\Set{i_1,\ldots,i_l} = \SetTo{n} \setminus \SetTo{\key{maxPrefixOf}(\Kl)} \\[\NL]
\forall q\in \SetTo{l}.\ \SJudgC{J_{i_q}}{S}\Ret{r_q}{\ps_q} \andalso
r'=\And(\SetIIn{r_q}{q}{\SetTo{l}}) 
}
{\KLJudgCJ{(\Kl\plus\aits : S)}
\Ret{\rl\plus r}
{(r \And r',\Set{1\ldots,n})}
}

\infrule[\runIts]
{
J=\JArr{J_1,...,J_n} \andalso
\KLJudgCJ{\Kl}\Ret{\rl}{(r,\pk)} \andalso
\Set{i_1,\ldots,i_l} = \SetTo{n} \setminus \pk \\[\NL]
\forall q\in \SetTo{l}.\ \SJudgC{J_{i_q}}{S}\Ret{r_q}{\ps_q} \andalso
r'=\And(\SetIIn{r_q}{q}{\SetTo{l}}) 
}
{\KLJudgCJ{(\Kl\plus\aunIts : S)}
\Ret{\rl\plus r}
{(r \And r',\Set{1\ldots,n})}
}

\infrule[\rklist-(n+1)]
{
k:J'\in\key{IndKeyOrT} \andalso
\KLJudgCJ{\Kl}\Ret{\rl}{(r_l,\pk_l)} \andalso
\KJudg{C}{J}{K}\Ret{r}{(r,\pk)} \\[\NL]
}
{\KLJudgCJ{(\Kl\plus K)}
\Ret{\rl\plus r}{(r_l \And r,\pk_l \cup \pk)}
}

\infax[\rklist-0]
{\KLJudgCJ{\List{}}
 \Ret{\List{}}{(T,\List{})}
}

\subsection{The trivial rule}

\infrule[\mbox{\em kw}-\rkw{trivial}]
{
kw \in \key{Other}
}
{\KJudgCJ{kw : J'}
 \Ret{T}
 {(\btrue,\ES)}
}

\subsection{Schema rules}

\medskip\medskip
\shortaxfirst{\rtrueS}
{\SJudgCJ{\atrue}
 \Ret{T}
 {(\btrue,\ES)}
}
\shortax{\rfalseS}
{\SJudgCJ{\afalse}
 \Ret{F}
 {(\bfalse,\ES)}
}
\medskip

\medskip
\shortrulefirst{\rschema\rkw{-true}}
{
\KLJudgCJ{\List{K_1,\ldots,K_n}}\Ret{\List{r_1,\ldots,r_n}}{(\btrue,\pk)} 
}
{\SJudgCJ{\JObj{K_1,\ldots,K_n}}
 \Ret{r}
 {(\btrue,\pk)}
}
\shortrule{\rschema\rkw{-false}}
{
\KLJudgCJ{\List{K_1,\ldots,K_n}}\Ret{\List{r_1,\ldots,r_n}}{(\bfalse,\pk)} 
}
{\SJudgCJ{\JObj{K_1,\ldots,K_n}}
 \Ret{r}
 {(\bfalse,\ES)}
}
\medskip

\hide{
\infrule[\rschema\rkw{\$id}]
{
\KLJudg{C+\key{absURI}}{J}{\List{K_1,\ldots,K_n}}\RetL{\List{r_1,\ldots,r_n}}{\List{\pk_1,\ldots,\pk_n}} \andalso
r=\And(\SetIIn{r(\pk_i)}{i}{\SetTo{n}}) 
}
{\SJudgCJ{\JObj{\qdid: \key{absURI}, K_1,\ldots,K_n}}
 \Ret{r}
 \rschema\rkw{\$id}(\_,\Set{\pk_1,\ldots,\pk_n})
}
}

\section{Functions $\Get(S,f)$ and $\DGet(S,f)$}\label{sec:get}

$\DGet(S,f)$ searches inside $S$ for a subschema $\JObj{\qdda : f, K_1, \ldots, K_n }$ 
and returns it. However, it only searches inside known keywords that contain a schema as a 
parameter, and the search is stopped by the presence of an internal $\qdid$ keyword, because
$\qdid$ indicates that the subschema is a separate resource, with a different URI.


We formalize this behavior by defining two functions:
\begin{enumerate}
\item $\DGetS(S,f)$, that searches $\qdda : f$ into $S$, and, if not found, invokes $\DGetK(k: J ,f)$ 
    on each keyword $k:J$ to search inside the schema;
\item $\DGetK(k:J,f)$ that invokes $\DGetS(S,f)$ to search $f$ inside the parts of $J$ that are known
    to contain a subschema.
\end{enumerate}

The function $\DGet(S,f)$ is defined as $\DGetS(\kw{StripId}(S),f)$, where 
$\kw{StripId}(S)$ removes any outermost $\qdid$ keyword from $S$. This is necessary since
$\DGetS$ interrupts its search when it meets a $\qdid$ keyword, but the presence of an
$\qdid$ in the outermost schema should not interrupt the search:
$$
\begin{array}{llllllll}
\kw{StripId}(\JObj{\qdid : URI, \Kl }) &=& \JObj{\Kl }   \\[\NL]
\kw{StripId}(S) &=& S  & & \mbox{otherwise}
\end{array}
$$

The functions $\DGetS(S,f)$ and $\DGetK(S,f)$ are defined as follows,
where $\setmax$ returns the maximum element of a set that contains either
schemas or $\bot$, according to the trivial order defined by $\bot \leq S$: that is,
$\setmax$ select the only element in the set that is different from $\bot$, if it exists and is unique.
If we
assume that no plain-name is used by two different anchors in the same schema, then,
in line 4, there exists at most one value of $i$ such that $\DGet(K_i,f)\neq\bot$,
hence the maximum is well defined, and the same holds for lines 6 and 7.

$$
\begin{array}{llllllll}
\begin{array}{llllllll}
0&\DGet(S,f) &=& \DGetS(\kw{StripId}(S),f) \\[\NL]
1& \DGetS(\xtrue/\xfalse,f) &=& \bot \\[\NL]
2& \DGetS(\JObj{ \Kl },f) &=& \bot &  \mbox{if } \qdid : URI \in \Kl \\[\NL]
3& \DGetS(\JObj{\Kl },f) &=& 
\JObj{\Kl } &  \mbox{if } \qdda : f \in \Kl \\
& & & &\mbox{\ and 2 does not apply}    \\[\NL]
4& \DGetS(\JObj{K_1,\ldots,K_n},f) &=& \setmax_{i\in \SetTo{n}} \DGetK(K_i,f)   & \mbox{if 2, 3 do not apply}\\[3\NL]
5& \DGetK(kw : S,f) &=& \DGetS(S,f) & kw\in\key{kwSimPar} \\[\NL]
6& \DGetK(kw : \JArr{S_1,\ldots,S_n},f) &=& \setmax_{i\in \SetTo{n}} \DGetS(S_i,f) & kw\in\key{kwArrPar} \\[\NL]
7& \DGetK(kw :  \JObj{k_1:S_1,\ldots,k_n:S_n},f) &=& \setmax_{i\in \SetTo{n}} \DGetS(S_i,f) & kw\in\key{kwObjPar}\\[\NL]
8& \DGetK(kw :  J,f) &=& \bot & kw\in\key{kwOther}\\[3\NL]
\end{array}
\\[3\NL]
\mbox{where}
\\[2\NL]
\begin{array}{llllllll}
\key{kwSimPar} &=&\SetOpen \anot , \acont, \apropN, \ait, \aaddProps, \\[\NL]
                         & & \ \ \ \aunProps, \aunIts \SetClose\\[\NL]
 \key{kwObjPar} &=&\Set{\qddefs, \qpattProps, \qprops, \qdepS}\\[\NL]
\key{kwArrPar}  &=& \Set{\aany, \qall, \qone, \qprefIts}\\[\NL]
 \key{kwOther} &=& \Str \setminus  \key{kwSimPar} \setminus\key{kwObjPar} \setminus \key{kwArrPar}
\end{array}
\end{array}
$$

The  four lines of $\DGetK$ definition specify that the search for a {\qdda} keyword is performed only inside keywords that are known
and whose parameter contains a schema object in a known position. For example, the search is not performed inside a user-defined
keyword or inside {\qconst} or {\qdefault}.

When $f$ is a plain-name,
then the function $\Get$ is identical to $\DGet$, but it extends case 3, since it matches both $\qda$ and $\qdda$:

$$
\begin{array}{llllllll}
0&\Get(S,f) &=& \GetS(\kw{StripId}(S),f) \\[\NL]
1& \GetS(\xtrue/\xfalse,f) &=& \bot \\[\NL]
2& \GetS(\JObj{ \Kl },f) &=& \bot &  \mbox{if } \qdid : URI \in \Kl \\[\NL]
3& \GetS(\JObj{\Kl },f) &=& 
\JObj{\Kl } &  \mbox{if } \qdda : f \in \Kl \\
& & & &\mbox{\ or\ } \qda : f \in \Kl    \\
& & & &\mbox{\ and 2 does not apply}    \\[\NL]
4& \GetS(\JObj{K_1,\ldots,K_n},f) &=& \setmax_{i\in \SetTo{n}} \GetK(K_i,f)   & \mbox{if 2, 3 do not apply}\\[3\NL]
5& \GetK(kw : S,f) &=& \GetS(S,f) & kw\in\key{kwSimPar} \\[\NL]
\ldots
\end{array}
$$

When $f$ is a JSON Pointer, it should be interpreted by $\Get(S,F)$ according to JSON Pointer specifications
of RFC6901 \citep{RFC6901}.
However, {\DTwenty} specs \citep{specs2020} (Section 9.2.1) specify that the behavior is undefined when the pointer crosses
resource boundaries.

\newpage
\section{The role of $\key{absURI}$ in $\qddref: \key{absURI}\cat\qkw{\#}\cat\key{f}$}\label{sec:nodynref}

In Remark \ref{rem:nodynref} we discuss the difference between looking for a dynamic reference in
$C$ or in $C+?\key{absURI}$. We report here a concrete test.

When the schema of Figure \ref{fig:dynref} is applied to the JSON instance $\JObj{ \qkw{children}: \JObj{}}$, the keyword 
$\qddref: \qkw{http://mjs.ex/tree\#node}$ (line 3) is applied in a dynamic context that only contains
$\qkw{http://mjs.ex/outer}$, which denotes a resources that does not contain
any $\qdda: \qkw{node}$ keyword. The only such keyword in the schema is at line 7, and is belongs to
the embedded resource $\qkw{http://mjs.ex/tree}$.
This specific example relies on the fact that an embedded resource, as identified by the $\qdid$ keyword, is not
part of the scope of the embedding resource, the one identified by $\qkw{http://mjs.ex/outer}$,
but it could be rephrased by putting the embedded resource in a separate file.
Hence, if $\qddref: \qkw{http://mjs.ex/tree\#node}$ were resolved using the dynamic context only,
it would raise a run-time failure. But our rule dictates that it is resolved in 
$C+?\key{absURI}$, that is in $$\qkw{http://mjs.ex/outer}+\qkw{http://mjs.ex/tree}$$
hence raising a validation error since the value of $\qkw{children}$ is not an array.
In Figure \ref{fig:dynref} we see how four different validators behaved at the date of 
10$^{th}$ Jan 2023: the first two exhibit an internal bug, and the other two use the approach that we 
formalized; none of them reports a resolution error due to the fact that 
$\qdda : \qkw{node}$ is not present in the dynamic scope.
Hence, they validate the choice to look for the fragment id $f$ inside $C+?\key{absURI}$, and not inside $C$ only.

\begin{figure}[hb]
\small
Schema:
\begin{querybox}{}
\begin{lstlisting}[style=query,escapechar=Z]
{ "$schema": "https://json-schema.org/draft/2020-12/schema","$id":
   "http://mjs.ex/outer",
   "$dynamicRef": "http://mjs.ex/tree#node",
   "$defs": {
       "http://mjs.ex/tree": {
         "$id": "http://mjs.ex/tree",
         "$dynamicAnchor": "node",
         "type": "object",
         "properties": {
             "data": true,
             "children": {
                "type": "array",
                "items": { "$dynamicRef": "#node" }
             }
          }
        }
    }
}
\end{lstlisting}
\end{querybox}

Instance:
\begin{querybox}{}
\begin{lstlisting}[style=querynonumbers]
{ "children" : {}}
\end{lstlisting}
\end{querybox}

Output:
\begin{querybox}{}
\begin{lstlisting}[style=query,escapechar=Z]
https://json-schema.hyperjump.io/:        TypeError: s[t] is undefined
https://tryjsonschematypes.appspot.com:   An error occurred - java.lang.NullPointerException
https://json-everything.net/json-schema/  "valid": false       
                                          "evaluationPath": "/$dynamicRef/properties/children"
https://jschon.dev/:                      "valid": false
                                          "keywordLocation": "/$dynamicRef/properties/children/type"
\end{lstlisting}
\end{querybox}
\caption{Different behavior of several {\JS} validators.}
\label{fig:dynref}
\end{figure}

\end{toappendix}

\newpage

\begin{toappendix}

\section{Algorithm \ref{alg:ptime}, part 2}\label{app:algptime}

From Section \ref{sec:ptime}.

\RestyleAlgo{ruled}
\begin{algorithm}
\footnotesize
\SetKwProg{Fn}{}{}{end}
\SetKw{kwWhere}{where}
\SetKw{Raise}{raise}

\caption{Polynomial Time Validation - part 2}
\tcc{Apply a keyword and return a new output based on PrevOutput}
\Fn{\KeywordValidate{Context,Instance, Keyword, PrevOutput, StopList, Up}}{ 
   \Switch{Keyword}{
      \Case{``anyOf'': List}{
          \Return (\AnyOf(Context,Instance,List,PrevOutput,StopList,Up))\;
      }
   \Case{``dynamicRef'': absURI ``\#'' fragmentId}{
          \Return (\DynRef(Context,Instance,absURI,fragmentId,PrevOutput,StopList,Up))\;
      }
  \ldots
   }
}
\mbox{\ \ }\\

\Fn{\AnyOf{Context, Instance, List, PrevOutput, StopList, Up}}{ 
(PrevResult, PrevEval, PrevDFragSet) := PrevOutput\;
 Result := \TrueK;
 Eval := \EmptySetK;
 DFragSet := \EmptySetK\;
 \For{Schema in List}{
  SchemaOutput := \SchemaValidateAS(Context, Instance, Keyword, StopList, Up) \;
  (SchemaRes,SchemaEval,SchemaDFragSet) := SchemaOutput \;
     Result := \OrK(Result,SchemaRes)\;
     Eval := \UnionK(Eval,SchemaEval)\;
     DFragSet := \UnionK(DFragSet,SchemaDFragSet);
  }
  {  \Return (\AndK(PrevResult,Result), \UnionK(PrevEval,Eval), \UnionK(PrevDFragSet,DFragSet))\;  }
} \mbox{\ \ }\\

\Fn{\DynRef{Context, Instance, AbsURI, fragmentId, PrevOutput, StopList, Up}}{ 
(PrevResult, PrevEval, PrevDFragSet) := PrevOutput\;
\lIf*{(dget(load(AbsURI),fragmentId) = bottom))}
                {\StaticRef(...)\;}
   fstURI := \FirstURI(Context+?AbsURI,fragmentId) \;
   fstSchema := get(load(fstURI),fragmentId)\;
  {  SchOutput := \SchemaValidateAS{\Saturate(Context,fstURI),Instance,fstSchema,StopList,Up}\; 
    (SchRes,SchEval,SchDT) := SchOutput\;     
    \Return (\AndK(PrevResult,SchRes), \UnionK(PrevEval,SchEval), 
    \UnionK(SchDT,\ \Singleton(fragmentId)))\;
  } 
}\mbox{\ \ }\\

\Fn{\FirstURI(context,fragmentId)}{
 \For{URI in context}{
   \lIf*{(dget(load(URI),fragmentId) != bottom)}{
          \{ \Return(URI); \}
       }
   }
   \Return(\Bottom)\;
} 

\end{algorithm}


{
\newpage

\section{Unfolding of the example}\label{sec:unfolding}

We report here the complete unfolding of the running example, using the procedure defined in Section
\ref{sec:elimination}. It is
the following document, where contexts are represented inside anchors 
using the abbreviations $\qkw{urn:psi}\rightarrow\qkw{p}$, 
$\qkw{urn:truex1}\rightarrow\qkw{t1}$, $\qkw{urn:truex2}\rightarrow\qkw{t2}$,
$\qkw{urn:falsex1}\rightarrow\qkw{f1}$, $\qkw{urn:falsex2}\rightarrow\qkw{f2}$,
and $\qkw{urn:phi}\rightarrow\qkw{ph}$. For space reasons, we split it into two parts. This is the first part.

\begin{querybox}{}
\begin{lstlisting}[style=query,escapechar=Z]
{ "id": "urn:psi",
  "$schema": "https://json-schema.org/draft/2020-12/schema",
  "allOf": [ { "$ref": "urn:truex1#p_t1_afterq1" }, { "$ref": "urn:falsex1#p_f1_afterq1" }],   
  "$defs": {
    "urn:truex1": {      
      "$id": "urn:truex1",
      "$defs": {
        "p_t1_t2_ph_x1": { "$dynamicAnchor": "p_t1_t2_ph_x1", "anyOf": [true] },
        "p_t1_f2_ph_x1": { "$dynamicAnchor": "p_t1_f2_ph_x1", "anyOf": [true] },
        "p_t1_t2_ph_not.x1":  { "$dynamicAnchor": "p_t1_t2_ph_not.x1", "anyOf": [false] },
        "p_t1_f2_ph_not.x1":  { "$dynamicAnchor": "p_t1_f2_ph_not.x1", "anyOf": [false] },
        "p_t1_afterq1":
           { "$anchor": "p_t1_afterq1",
              "anyOf": [ { "$ref": "urn:truex2#p_t1_t2_afterq2" }, 
                         { "$ref": "urn:falsex2#p_t1_f2_afterq2" } ]
    }}},   
    "urn:falsex1": {             
      "$id": "urn:falsex1",
      "$defs": {
        "p_f1_t2_ph_x1":  { "$dynamicAnchor": "p_f1_t2_ph_x1", "anyOf": [false] },
        "p_f1_f2_ph_x1":  { "$dynamicAnchor": "p_f1_f2_ph_x1", "anyOf": [false] },
        "p_f1_t2_ph_not.x1":  { "$dynamicAnchor": "p_f1_t2_ph_not.x1", "anyOf": [true] },
        "p_f1_f2_ph_not.x1":  { "$dynamicAnchor": "p_f1_f2_ph_not.x1", "anyOf": [true] },
        "p_f1_afterq1":
           { "$anchor": "p_f1_afterq1",
              "anyOf": [ { "$ref": "urn:truex2#p_f1_t2_afterq2" }, 
                         { "$ref": "urn:falsex2#p_f1_f2_afterq2" } ]
    }}},    
    "urn:truex2": {      
      "$id": "urn:truex2",
      "$defs": {
        "p_t1_t2_ph_x2":  { "$dynamicAnchor": "p_t1_t2_ph_x2", "anyOf": [true] },
        "p_f1_t2_ph_x2":  { "$dynamicAnchor": "p_f1_t2_ph_x2", "anyOf": [true] },
        "p_t1_t2_ph_not.x2":  { "$dynamicAnchor": "p_t1_t2_ph_not.x2", "anyOf": [false] },
        "p_f1_t2_ph_not.x2":  { "$dynamicAnchor": "p_f1_t2_ph_not.x2", "anyOf": [false] },
        "p_t1_t2_afterq2":  { "$anchor": "p_t1_t2_afterq2", "$ref": "urn:phi#p_t1_t2_ph_phi" },
        "p_f1_t2_afterq2":  { "$anchor": "p_f1_t2_afterq2", "$ref": "urn:phi#p_f1_t2_ph_phi" }
    }},      
    "urn:falsex2": {      
      "$id": "urn:falsex2",
      "$defs": {
        "p_t1_f2_ph_x2":  { "$dynamicAnchor": "p_t1_f2_ph_x2", "anyOf": [false] },
        "p_f1_f2_ph_x2":  { "$dynamicAnchor": "p_f1_f2_ph_x2", "anyOf": [false] },
        "p_t1_f2_ph_not.x2":  { "$dynamicAnchor": "p_t1_f2_ph_not.x2", "anyOf": [true] },
        "p_f1_f2_ph_not.x2":  { "$dynamicAnchor": "p_f1_f2_ph_not.x2", "anyOf": [true] },
        "p_t1_f2_afterq2":  { "$anchor": "p_t1_f2_afterq2", "$ref": "urn:phi#p_t1_f2_ph_phi" },
        "p_f1_f2_afterq2":  { "$anchor": "p_f1_f2_afterq2", "$ref": "urn:phi#p_f1_f2_ph_phi" }
    }},      
    "urn:phi": {   ...
\end{lstlisting}
\end{querybox}

Here is the second part;
since $\qkw{urn:phi\#phi}$ can be reached from four different contexts, we need four different definitions, as follows.
In the four cases, the way the dynamic variables $\qkw{x1}$ and $\qkw{x2}$ are resolved varies depending on the context
$\CC$, encoded in the anchor $\ECC\cat f$.

\begin{querybox}{}
\begin{lstlisting}[style=query,escapechar=Z]
    "urn:phi": {   
      "$id": "urn:phi",
      "$defs": {
        "p_t1_t2_ph_phi": {    
           "$anchor": "p_t1_t2_ph_phi", 
           "anyOf": [
             { "allOf":[{ "$ref": "urn:truex1#p_t1_t2_ph_x1" }, 
                        { "$ref": "urn:truex2#p_t1_t2_ph_x2" }]},
             { "allOf":[{ "$ref": "urn:truex1#p_t1_t2_ph_not.x1"}, 
                        { "$ref": "urn:truex2#p_t1_t2_ph_not.x2"}]}] },
        "p_f1_t2_ph_phi": {    
           "$anchor": "p_f1_t2_ph_phi", 
           "anyOf": [
             { "allOf":[{ "$ref": "urn:falsex1#p_f1_t2_ph_x1" }, 
                        { "$ref": "urn:truex2#p_f1_t2_ph_x2" }]},
             { "allOf":[{ "$ref": "urn:falsex1#p_f1_t2_ph_not.x1"}, 
                        { "$ref": "urn:truex2#p_f1_t2_ph_not.x2"}]}] },
        "p_t1_f2_ph_phi": {    
           "$anchor": "p_t1_f2_ph_phi", 
           "anyOf": [
             { "allOf":[{ "$ref": "urn:truex1#p_t1_f2_ph_x1" },
                        { "$ref": "urn:falsex2#p_t1_f2_ph_x2" }]},
             { "allOf":[{ "$ref": "urn:truex1#p_t1_f2_ph_not.x1"}, 
                        { "$ref": "urn:falsex2#p_t1_f2_ph_not.x2"}]}] },
        "p_f1_f2_ph_phi": {    
           "$anchor": "p_f1_f2_ph_phi", 
           "anyOf": [
             { "allOf":[{ "$ref": "urn:falsex1#p_f1_f2_ph_x1" }, 
                        { "$ref": "urn:falsex2#p_f1_f2_ph_x2" }]},
             { "allOf":[{ "$ref": "urn:falsex1#p_f1_f2_ph_not.x1"}, 
                        { "$ref": "urn:falsex2#p_f1_f2_ph_not.x2"}]}] }
}}}}
\end{lstlisting}
\end{querybox}

}

\newpage

\section{What does {\evaluated} mean?}\label{sec:ambiguity}

The keyword $\qunProps$ is applied to the instance properties that have not been {\evaluated} by adjacent keywords,
as discussed in Section \ref{sec:annotations}, but,
unfortunately, the definition in \citep{specs2020} of what counts as {\evaluated} presents some ambiguities.

The \emph{successful} application of a $\qprops$ keyword {\evaluates} all instance properties whose name appears
in the value of the $\qprops$ keyword; in Figure \ref{fig1}, these are the properties
named \qkw{data} or \qkw{children}. Moreover, a property is {\evaluated} by the successful application of a keyword that 
invokes a schema that {\evaluates} that
property, as happens in Figure \ref{fig1} with the keyword 
$\qdref: \qkw{http://mjs.ex/st\#tree}$.

If, however, the $\qprops$ keyword fails, then the specifications of {\DTwenty} (\citep{specs2020}) are ambiguous. They state without any ambiguity that 
validation will fail, but they give contradictory indications about which instance fields are {\evaluated},
which is a problem, since it affects the error messages and the annotations returned by the validation tool.
Consider the following example.

\begin{example}\label{ex:vlds}

The following schema expresses the fact that a property \qkw{a}, if present, must have type
\qkw{integer}, and that no other property, neither equal to \qkw{b} nor different from 
\qkw{a} and \qkw{b}, may appear.

    \begin{querybox}{}
        \begin{lstlisting}[style=query,escapechar=Z]
{
 "$id": "https://frml.edu/onlya_no_b.json",
 "type": "object",
 "properties":  { 
       "a": {"type": "integer"} ,
       "b": false
 },
 "unevaluatedProperties" : false
}
        \end{lstlisting}
    \end{querybox}

When applied to the instance $\JObj{ \qkw{a}: 0,\qkw{b}: 0,\qkw{c}: 0 }$, this schema fails, 
because of properties \qkw{b} and \qkw{c}.
While there is no ambiguity on the fact that the schema fails, it is not clear which properties 
count as {\evaluated} by the $\qprops$ keyword, 
hence,  which error messages should be produced
by {\qunProps}. This is testified by the discussion in \citeoneone, by that in~\citefiveseven, and by several more
that are cited therein.

We tested the above schemas with the  {\JE} validator~\citep{jsever},  {\BL} \citep{bl}, the {\JSD} validator~\citep{jsdev}, and {\HJ}~\citep{hyperjump},
in the online-versions on 07/07/2023.
These are established well-known validators that support {\DNineteen} and {\DTwenty}.
By analyzing their output, reported respectively in Figures~\ref{fig:evann}, \ref{fig:blann}, \ref{fig:jsann}, and~\ref{fig:hjann}, we observe that the first two validators apply 
 {\qunIts} to \qkw{c} only, hence they regard every matching field as {\evaluated} even when $\qprops$ fails;
 we call it the {\kfailuretol} interpretation: every property that matches is {\evaluated}, even if the keyword fails,
 and even if the subschema applied to that property fails, as happens here to $\qkw{b}$.
 However, {\JSD} applies $\qunProps$ to \qkw{a}, \qkw{b}, and \qkw{c}, 
 hence it considers no property as {\evaluated};  we call it the {\successonly} interpretation:
 when the keyword fails no property is regarded as {\evaluated}, neither $\qkw{a}$, whose value satisfies
 the $\qtype: \qinteger$ assertion, neither $\qkw{b}$. 
 Finally, {\HJ} does not apply $\qprops$ to any object property, because it adopts an interpretation where the failure
 of $\qprops$ produces a special \emph{null} annotation that completely prevents the execution of 
 $\qunProps$; we call it the {\nullannotation} interpretation.
\end{example}

In this paper we formalize the {\kfailuretol} interpretation, since it seems the one that is currently more
accepted by the community. Moreover, we provide a formal framework where these different interpretations
can be formalized and discussed.

It is important to clarify that these ambiguities do not affect validation, since they only arise with failing schemas, but 
they are still very relevant, because they affect the error messages and, more importantly,
the mental model of the specification reader. Whoever reads the specification document  builds a mental model about the internal logics of JSON Schema, and the way they define a schema, or the way they implement a validator, are guided by 
this mental model.
If the mental model is unclear, or ambiguous, their work becomes more complicated than it should be.


\section{List of downloaded validators}\label{app:validators}

List of the validators from Section \ref{sec:experiments}.

\begin{table}[ht]
\centering


\caption{Validators used in our experiments. Top: Listing active open-source projects, bottom:
Listing academic implementations. Stating programming language, supported drafts, and official release tag (if available).}
\label{tab:validators}

\begin{tabular}{llccll}
\toprule
Validator & Language & Draft4 & Draft 2020 & Origin/Version\\
\midrule 
clojure-json-schema &  clojure & $\times$ & & Bowtie, release tag 1ef0c \\
%
cpp-valijson & C\texttt{++} & $\times$ &  &Bowtie, release tag 1ef0c \\
%
go-gojsonschema & go & $\times$ & & Bowtie, release tag 1ef0c\\
%
go-jsonschema & go & $\times$ & $\times$ &  Bowtie, release tag 1ef0c\\
%
java-json-schema& Java & & $\times$ & Bowtie, release tag 1ef0c \\
%
js-ajv & javascript & $\times$ & $\times$ & Bowtie, release tag 1ef0c\\
%
%
js-hyperjump & javascript & $\times$ & $\times$ &  Bowtie, release tag 1ef0c\\
%
lua-jsonschema & lua & $\times$ & & Bowtie, release tag 1ef0c \\
%
ruby-json\_schemer & ruby & $\times$ & & Bowtie, release tag 1ef0c\\
%
rust-boon & python &  $\times$ & $\times$ & Bowtie, release tag 1ef0c\\
%
rust-jsonschema & rust & $\times$ & $\times$ &  Bowtie, release tag 1ef0c\\
%
python-fastjsonschema & python & $\times$ & & Bowtie, release tag 1ef0c\\
%
python-jschon & python & &  $\times$  & Bowtie, release tag 1ef0c\\
%
python-jsonschema & python & $\times$ & $\times$ &  Bowtie, release tag 1ef0c\\
%
ts-vscode-json-languageservice& typescript & $\times$ & $\times$ &  Bowtie, release tag 1ef0c\\
%
dotnet-jsonschema-net & dotnet & & $\times$ & Bowtie, release tag 1ef0c\\
\midrule 
foundations-jsonsonschema & python & $\times$ & & Code from~\citep{DBLP:conf/www/PezoaRSUV16}, integrated into Bowtie\\
modern-jsonschema & Scala & & $\times$ & Algorithm~\ref{alg:ptime}, integrated into Bowtie\\
\bottomrule
\end{tabular}

\end{table}

\newpage

\section{Output examples}\label{sec:figures}

\begin{figure}[h]
\footnotesize

Instance:
\begin{querybox}{}
\begin{lstlisting}[style=querynonumbers]
{ "a": 0, "b": 0, "c": 0 }
\end{lstlisting}
\end{querybox}

Schema:
\begin{querybox}{}
\begin{lstlisting}[style=query]
{
  "$id": "https://formalize.edu/onlya_no_b.json",
  "type": "object",
  "properties":  { 
        "a": {"type": "integer"} ,
        "b": false
  },
  "unevaluatedProperties" : false
}
\end{lstlisting}
\end{querybox}

Output:
\begin{querybox}{}
\begin{lstlisting}[style=query]
{
  "valid": false,
  "keywordLocation": "",
  "absoluteKeywordLocation": "https://formalize.edu/onlya_no_b.json#",
  "instanceLocation": "",
  "errors": [
    {
      "valid": true,
      "keywordLocation": "/properties/a",
      "absoluteKeywordLocation": "https://formalize.edu/onlya_no_b.json#/properties/a",
      "instanceLocation": "/a"
    },
    {
      "valid": false,
      "keywordLocation": "/properties/b",
      "absoluteKeywordLocation": "https://formalize.edu/onlya_no_b.json#/properties/b",
      "instanceLocation": "/b",
      "error": "All values fail against the false schema"
    },
    {
      "valid": false,
      "keywordLocation": "/unevaluatedProperties",
      "absoluteKeywordLocation": "https://formalize.edu/onlya_no_b.json#/unevaluatedProperties",
      "instanceLocation": "/c",
      "error": "All values fail against the false schema"
    }
  ]
}
\end{lstlisting}
\end{querybox}

\caption{Annotations returned by {\JE}, Example \ref{ex:vlds}.}
\label{fig:evann}
\end{figure}

\begin{figure}[h]
\footnotesize

Instance:
\begin{querybox}{}
\begin{lstlisting}[style=querynonumbers]
{ "a": 0, "b": 0, "c": 0 }
\end{lstlisting}
\end{querybox}

Schema:
\begin{querybox}{}
\begin{lstlisting}[style=query]
{
  "$id": "https://formalize.edu/onlya_no_b.json",
  "type": "object",
  "properties":  { 
        "a": {"type": "integer"} ,
        "b": false
  },
  "unevaluatedProperties" : false
}
\end{lstlisting}
\end{querybox}

Output:
\begin{querybox}{}
\begin{lstlisting}[style=query]
{
	"valid": false,
	"keywordLocation": "https://formalize.edu/onlya_no_b.json",
	"absoluteKeywordLocation": "",
	"instanceLocation": "",
	"errors": [
		{
			"valid": false,
			"error": "False",
			"keywordLocation": "https://formalize.edu/onlya_no_b.json#/properties/b",
			"absoluteKeywordLocation": "#/properties/b",
			"instanceLocation": "#/b"
		},
		{
			"valid": false,
			"error": "False",
			"keywordLocation": "https://formalize.edu/onlya_no_b.json#/unevaluatedProperties",
			"absoluteKeywordLocation": "#/unevaluatedProperties",
			"instanceLocation": "#/c"
		}
	]
}
\end{lstlisting}
\end{querybox}

\caption{Annotations returned by {\BL}, Example \ref{ex:vlds}.}
\label{fig:blann}
\end{figure}

\begin{figure}
\small
Instance:
\begin{querybox}{}
\begin{lstlisting}[style=querynonumbers]
{ "a": 0, "b": 0, "c": 0 }
\end{lstlisting}
\end{querybox}

Schema:
\begin{querybox}{}
\begin{lstlisting}[style=query]
{
  "$id": "https://formalize.edu/onlya_no_b.json",
  "type": "object",
  "properties":  { 
        "a": {"type": "integer"} ,
        "b": false
  },
  "unevaluatedProperties" : false
}
\end{lstlisting}
\end{querybox}

Output:
\begin{querybox}{}
\begin{lstlisting}[style=query]
{
    "valid": false,
    "instanceLocation": "",
    "keywordLocation": "",
    "absoluteKeywordLocation": "https://formalize.edu/onlya_no_b.json#",
    "errors": [
        {
            "instanceLocation": "/b",
            "keywordLocation": "/properties/b",
            "absoluteKeywordLocation": "https://formalize.edu/onlya_no_b.json#/properties/b",
            "error": "The instance is disallowed by a boolean false schema"
        },
        {
            "instanceLocation": "",
            "keywordLocation": "/unevaluatedProperties",
            "absoluteKeywordLocation": "https://formalize.edu/onlya_no_b.json#/unevaluatedProperties",
            "error": [
                "a",
                "b",
                "c"
            ]
        }
    ]
}
\end{lstlisting}
\end{querybox}
\caption{Annotations returned by {\JSD}, Example \ref{ex:vlds}.}
\label{fig:jsann}
\end{figure}

\begin{figure}
\small
Instance:
\begin{querybox}{}
\begin{lstlisting}[style=querynonumbers]
{ "a": 0, "b": 0, "c": 0 }
\end{lstlisting}
\end{querybox}

Schema:
\begin{querybox}{}
\begin{lstlisting}[style=query]
{
  "$id": "https://formalize.edu/onlya_no_b.json",
  "type": "object",
  "properties":  { 
        "a": {"type": "integer"} ,
        "b": false
  },
  "unevaluatedProperties" : false
}
\end{lstlisting}
\end{querybox}

Output:
\begin{querybox}{}
\begin{lstlisting}[style=query]
# fails schema constraint https://formalize.edu/onlya_no_b.json#/properties
#/b fails schema constraint https://formalize.edu/onlya_no_b.json#/properties/b
\end{lstlisting}
\end{querybox}
\caption{Annotations returned by {\HJ}, Example \ref{ex:vlds}.}
\label{fig:hjann}
\end{figure}

\end{toappendix}

\end{document}